\documentclass[12pt]{article}
\usepackage{epsfig} 
\usepackage{graphics}
\usepackage{xcolor}
\usepackage{a4wide}
\usepackage{amsfonts}
\usepackage{enumerate} 
\usepackage{amssymb,amsmath}
\usepackage{subfig}
\usepackage{extarrows}
\usepackage[all]{xy}  \SilentMatrices
\xyoption{pdf}
\usepackage{xfrac}

\usepackage[pdftex]{hyperref}

\newcommand{\comm}[1]{}

\newcommand{\crug}[1]{\raisebox{.6pt}{\textcircled{\raisebox{-.4pt} {$\mbox{\fontsize{9}{00}\selectfont $#1$}$}}}}

\newcommand{\h}{{h}}

\title{Phases of Five-dimensional Theories,\\ Monopole Walls, and Melting Crystals}
\author{
Sergey A.\ Cherkis\\
\ 
\it\small Department of Mathematics, University of Arizona, Tucson AZ 85721-0089, USA\\
\tt\small cherkis@math.arizona.edu
}
\date{}

\begin{document}
\begin{titlepage}

\renewcommand{\thepage}{ }

\vspace{1in}

\maketitle
\abstract{Moduli spaces of doubly periodic monopoles, also called monopole walls or monowalls, are hyperk\"ahler; thus, when four-dimensional, they are self-dual gravitational instantons.  We find all monowalls with lowest number of moduli. Their moduli spaces can be identified, on the one hand, with Coulomb branches of five-dimensional supersymmetric quantum field theories on $\mathbb{R}^3\times T^2$ and, on the other hand, with moduli spaces of local Calabi-Yau metrics on the canonical bundle of a del Pezzo surface.  We explore the asymptotic metric of these moduli spaces and compare our results with Seiberg's low energy description of the five-dimensional quantum theories.  We also give a natural description of the phase structure of general monowall moduli spaces in terms of triangulations of Newton polygons, secondary polyhedra, and associahedral projections of secondary fans.}

\vspace{-6in}
\parbox{\linewidth}
{\hfill \em  In memory of 
Andrei Zelevinsky
}

\end{titlepage}

\tableofcontents
\newpage

\section{Introduction} 
Most known self-dual gravitational instantons admit realizations as moduli spaces.  Moreover, usually a  gravitational instanton can be viewed as a moduli space in more then one way.  Such realizations are very useful in studying their geometry and topology.  In particular, they can be represented as moduli spaces of solutions of the self-duality equation for  Yang-Mills fields or its dimensional reductions.  This is a particularly convenient point of view, since antihermitian  connections on a hyperk\"ahler space (in particular on the Euclidean space with appropriate boundary conditions) form an infinite-dimensional affine hyperk\"ahler space.  This infinite-dimensional space of connections  carries the triholomorphic action of the group of gauge transformations and the self-dual Yang-Mills equations are the vanishing moment map conditions for this group action.  As a result, the moduli space of self-dual connections, up to gauge equivalence, is an infinite hyperk\"ahler quotient and thus, itself carries a hyperk\"ahler metric.  Whenever it is of real dimension four, its Riemann tensor is self-dual and it is a self-dual gravitational instanton.   

	The type of self-dual Yang-Mills solutions to consider is dictated by the desired asymptotic behavior of the moduli space.  This correspondence is presented in Table~\ref{Tab:Corresp}.
\begin{table}[htdp]
\small
\begin{center}
\begin{tabular}{c|ccc}
\begin{tabular}{@{}c@{}}Type of the\\ Moduli Space\\ \end{tabular} & Self-dual Yang-Mills Solution & &  Dual Equivalent Description  \\
\hline
ALE &  Instantons &  & ADHM Equations (Quivers)\\
	ALF & Monopoles & $\xleftrightarrow[\text{Transform}]{\text{ADHM-Nahm}}$ & Nahm Equations (Bows)\\
ALG & Periodic Monopoles &  &  Hitchin System (Slings)\\
ALH & Doubly-periodic Monopoles&  & Doubly-periodic Monopoles
\end{tabular}
\caption{Self-dual Gravitational Instantons as Moduli Spaces.}
\label{Tab:Corresp}
\end{center}
\end{table}%
The four types of the moduli spaces here are distinguished by their volume growth.  We distinguish these spaces by how fast the volume of a ball of geodesic radius $R$ centered at some fixed point $p$ grows with $R.$ A noncompact self-dual gravitational instanton space is of 1) ALE, 2) ALF, 3) ALG, or 4) ALH type if the volume growth is, respectively, 1) quartic, 2) cubic, 3) lower than cubic and no less than quadratic, and 4) lower than quadratic.

ALE spaces, such as Eguchi-Hanson space, are moduli spaces of four-dimensional objects: instantons or of zero-dimensional objects: quivers.  ALF spaces are moduli spaces of three-dimensional monopoles or of a system of ODEs  called the Nahm equations \cite{Atiyah:1985dv,Atiyah:1985fd,Cherkis:1997aa,Cherkis:1998xca}.  ALG spaces are moduli spaces of periodic monopoles or of two-dimensional Hitchin systems \cite{Cherkis:2000cj,Cherkis:2001gm}.  ALH spaces, in this view, appear as moduli spaces of doubly periodic monopoles. Thereby, in the pursuit of gravitational instantons we are led to doubly periodic monopoles, also called monopole walls, or {\em monowalls} for short. If in all previous cases (as indicated in Table~\ref{Tab:Corresp}) the Nahm transform produces a simpler, lower-dimensional object, in the case of ALH space the Nahm transform \cite{Nahm82,Nahm84}, when applied to a doubly periodic monopole, produces another doubly periodic monopole.    Thus we are destined to face the monowall.  

In a more extended view, not captured by Table \ref{Tab:Corresp} above, some ALG spaces appear as moduli spaces of ($\mathbb{Z}_n$ equivariant) doubly periodic instantons, or, equivalently, under the Nahm transform, of ($\mathbb{Z}_n$ equavariant) Hitchin systems on a two-torus \cite{AMSTalk}.  The possible values of $n$ in $\mathbb{Z}_n$ are $2,3,4,$ and $6$ and the corresponding instanton gauge groups are $SU(4), SU(3), SU(4),$ and $SU(6).$  If $\omega=\exp(2\pi i/n)$ and $(z,v)$ are linear coordinates on $\mathbb{R}^2\times T^2\simeq\mathbb{C}\times \left(\mathbb{C}/(\mathbb{Z}+\tau\mathbb{Z})\right),$ then the instanton equivariance condition is $A(z,v)=U^{-1}A(\omega z,\omega v)U,$ with $U$ given in terms of the $j\times j$  shift matrices $S_j$ by respectively $U=1_4, 1_3, 1_2\times S_2,$ and $1_1\times S_2\times S_3.$ 
On the Hitchin system side, on the other hand,  the $SU(n)$ Hitchin data $(\hat{A}=A_\zeta d\zeta+A_{\bar{\zeta}}d\bar{\zeta},\hat{\Phi})$ satisfy $\hat{\Phi}(\omega\zeta)=\omega^{-1}S_n \hat{\Phi}(\zeta)S^{-1}$ and $A_\zeta(\omega\zeta)=\omega^{-1} S A_\zeta(\zeta)S^{-1}.$   The intersection diagram of the compact two-cycles of one of these ALG spaces is respectively $D_4, E_6, E_7,$ and $E_8$ affine Dynkin diagram.

At least one case of an ALH space\footnote{Explored in collaboration with Marcos Jardim.}, a hyperk\"ahler deformation of $(T^3\times\mathbb{R})/\mathbb{Z}_2,$ appears as a moduli space of triply periodic $U(2)$ monopole with two positive and two negative Dirac singularities.  The orbifold limit is reached when one positive singularity is placed atop of a negative one, while the other pair of positive and negative singularities is placed on top of each other at the diametrically opposite point in $T^3.$

Monowalls were explored analytically and in terms of D-brane configurations in \cite{Lee:1998isa} and numerically in \cite{Ward:2006wt}.  More recently, the asymptotic metric on the moduli space of certain monowalls was computed in \cite{Hamanaka:2013lza}.

In \cite{Cherkis:2012qs} we associated to each monopole wall a decorated Newton polygon and found that dimension of the monopole wall moduli space is four times the number of internal points of its Newton polygon.  We also found that there is a  $GL(2,\mathbb{Z})$ action on monopole walls and their Newton polygons that is isometric on their moduli spaces.  In the study of gravitational instantons one is interested in four-dimensional moduli spaces.  Thus, after reviewing the monowall problem in Section~\ref{Sec:MMS}, we identify all monowalls with no moduli and all monowalls with four moduli  in Section~\ref{Sec:MWFM}. We find that all Newton polygons corresponding to monowalls with four moduli are reflexive. Furthermore, we find that some of these moduli spaces are isometric, ending with eight distinct moduli spaces. 
After discussing their significance in field theory and string theory in Section~\ref{Sec:GTHCY}, we conclude by  establishing the phase structure of these moduli spaces in Section~\ref{Sec:PhaseSpace}.

\section{Monowalls and their Moduli Spaces}\label{Sec:MMS}
We consider a monowall, also called a monopole wall, as defined in \cite{Cherkis:2012qs}. Namely, it is a Hermitian bundle $E\rightarrow T^2\times\mathbb{R}$ with a connection (the gauge field) $A$  and an endomorphism (the Higgs field) $\Phi$ satisfying the Bogomolny equation
\begin{equation} \label{Eq:Bogomolny}
dA+A\wedge A=-*(d\Phi+[A,\Phi]),
\end{equation}
and the asymptotic eigenvalues of $\Phi$ growing at most linearly along the $\mathbb{R}$ component.  We denote the linear coordinate along $\mathbb{R}$ by $z$, while the two periodic coordinates on the torus $T^2$ are $x$ and $y$ with respective periods $S$ and $R$, i.e. $x\sim x+S$ and $y\sim y+R.$

The Bogomolny equation can be viewed as the zero level moment map condition for the hyperk\"ahler reduction of the affine space of pairs $\{(A,\Phi)\}$ by the group of gauge transformations.  Thus the space of gauge equivalence classes of its solutions inherits a hyperk\"ahler metric from 
\begin{align}
|\delta(A,\Phi)|^2=-\int_{T^2\times\mathbb{R}} {\rm tr}\, \left(\delta A\wedge*\delta A+\delta\Phi\wedge *\delta\Phi\right).
\end{align}
Note, that this metric is the direct product of the gauge algebra center part and the rest.  As we shall see later, the center, trace $u(1),$ part will have no associated moduli, thus it is only the remaining $su(n)$ part that is of any significance.  In particular, two background solutions that differ only in the trace part will have exactly the same metric in their vicinity.  This fact will be significant for our classification below.

We demand that $\Phi$ is smooth everywhere except at a finite number of prescribed points in $T^2\times\mathbb{R},$ where it has positive or negative Dirac singularities, and that the asymptotic behavior of the eigenvalues of $\Phi$ is at most linear in $z.$  As in \cite{Cherkis:2012qs}, in order to introduce and motivate these conditions we first discuss some abelian solutions.

\subsection{Dirac Monowall}
Let us consider the rank one case, that is when the monowall fields are abelian.  We let $\Phi=i\phi$ and $A=i a$ so that the function $\phi$ and the one-form $a$ are real.  They satisfy the Bogomolny equation $*d\phi=-da.$  Geometrically it implies that $\phi$ is harmonic and, via the Stokes and Chern-Weil theorems, the flux of $\nabla\phi$ through any closed surface is proportional to $2\pi.$  The coefficient of proportionality $Q_+$ (and $Q_-$) as $z\rightarrow+\infty$ (and $z\rightarrow-\infty$) is called the right (and left) charge of the monowall.

The only harmonic function on $T^2\times\mathbb{R}$ that is at most linear at infinity is $\phi=2\pi(Qz+M)$ with corresponding one-form $a=2\pi\left(\frac{Q}{SR} ydx-\frac{p}{S}dx-\frac{q}{R}dy\right).$  Here the charge $Q$ has to be integer, and $M, p,$ and $q$ are arbitrary real constants.  This is the {\em constant energy density} solution.

Another way of constructing a monowall solution is by superimposing Dirac monopole solutions arranged along a doubly periodic array.  The Dirac solution of Eq.~\eqref{Eq:Bogomolny} on $\mathbb{R}^3$ is
\begin{align}
 \phi&=-\frac{1}{2 r},& a_\pm&=\frac{1}{2}\frac{ydx-xdy}{r(z\pm r)}.
\end{align}  
It satisfies the Bogomolny equation $*d\phi=-da$ and $\phi$ is the Green's function satisfying $\nabla^2\phi=2\pi\delta(z)\delta(y)\delta(z).$

Straightforward superposition of Dirac monopoles  arranged as a doubly periodic array at the lattice vertices $\mathbf{e}_{jk}=(j S,k R,0),$ with $ j,k\in\mathbb{Z}$ with the distance to the $\mathbf{e}_{jk}$ vertex denoted by $r_{jk}=|\mathbf{r}-\mathbf{e}_{jk}|,$ produces the Higgs field 
\begin{align}
-\frac{1}{2r}-\frac{1}{2}\sum_{(j,k)\neq(0,0)}\left(\frac{1}{r_{jk}}-\frac{1}{|\mathbf{e}_{jk}|}\right)=\pi\frac{|z|}{SR}-\Lambda+o(z^0),
\end{align}
with the constant \cite{Linton}
\begin{align}\nonumber
\Lambda&=\frac{1}{R}\left(\ln\frac{4\pi R}{S}-\gamma\right)-\frac{4}{R}\sum_{m,n} K_0\left(2\pi mn\frac{S}{R}\right)
  =\frac{1}{S}\left(\ln\frac{4\pi S}{R}-\gamma\right)-\frac{4}{S}\sum_{m,n} K_0\left(2\pi mn\frac{R}{S}\right).
\end{align}
Such a Higgs field does not have desired behavior as $|z|\rightarrow\infty,$ namely 
$$SR\frac{d}{dz} \phi=\pm\frac{1}{2}$$ and is not integer; thus there is no line bundle with a connection $a$ satisfying the Bogomolny equation for this Higgs field, since it would have to satisfy $\frac{1}{2\pi}\int_{T^2} da=SR\frac{d}{dz} \phi=\pm\frac{1}{2}.$

With this in mind, the basic Dirac monowall with $Q_-=0$ and $Q_+=1$ and $M_-=M_+=0$ has the following Higgs field 
\begin{align}
\phi&=\pi\frac{z}{SR}-\frac{1}{2r}-\frac{1}{2}\sum_{(j,k)\neq(0,0)}\left(\frac{1}{r_{jk}}-\frac{1}{|\mathbf{e}_{jk}|}\right)
+\Lambda
\end{align}
with asymptotic expansions \cite{Newman}
\begin{align}
      \label{Largez}
      \phi&=\pi\frac{z+|z|}{SR}-\frac{1}{2}\sum_{m,n}\frac{1}{\sqrt{S^2m^2+R^2n^2}}e^{-4\pi^2\left(\left(\frac{m}{S}\right)^2+\left(\frac{n}{R}\right)^2\right)|z|}e^{2\pi i\left(\frac{m}{S}x+\frac{n}{R}y\right)}\\
      \label{LargeR}
      &=\pi\frac{z}{SR}+\frac{1}{SR}\ln\left|2\sin\frac{\pi}{S}(x-iz)\right|-\frac{2}{R}\sum_{m,n} K_0\left(\frac{2\pi n}{R}\sqrt{z^2+(x-m S)^2}\right)\cos\left(\frac{2\pi}{R}n y\right)\\
      \label{Larger}
      &=\pi\frac{z}{SR}+\frac{1}{SR}\ln\left|2\sin\frac{\pi}{R}(y+iz)\right|-\frac{2}{S}\sum_{m,n} K_0\left(\frac{2\pi n}{S}\sqrt{z^2+(y-m R)^2}\right)\cos\left(\frac{2\pi}{S}n x\right).
\end{align}
Series \eqref{Largez} converges fast for large values of $|z|$, while  series \eqref{LargeR} and \eqref{Larger} can be used for large $(z^2+x^2)/R^2$ and $(z^2+y^2)/S^2$ respectively.  More details of various expansions of this function can be found in \cite{Newman} and \cite{Linton}.

\subsection{Moduli Problem}
In general we consider rank $n$ solutions of the Bogomolny equation $*D_A\Phi=-F_A$ on $T^2\times\mathbb{R}$ with asymptotic conditions on eigenvalues of the Higgs field 
\begin{align}
{\rm Eig\,Val}\ \Phi=\left\{2\pi i\left(Q_{\pm,l}z+M_{\pm,l}\right)+o(z^0)\, |\, l=1,\ldots,n \right\},
\end{align}
which split the bundle $E|_z\rightarrow T^2_z$ over the two-torus at large values of $|z|$ into eigen-bundles of $\Phi$:
\begin{align}
E|_z=\mathop{\oplus}_{j=1}^{f_+}E_{+j} &\ \text{for}\  z\rightarrow\infty&& \text{and}& E|_z=\mathop{\oplus}_{j=1}^{f_-}E_{-j} &\ \text{for}\  z\rightarrow-\infty.
\end{align}
Here $f_\pm$ are the numbers of distinct pairs $(Q_{\pm,l},M_{\pm,l})$. 
We also fix the conjugacy classes of the holonomy of the connection in each $E_{\pm j}$ by fixing the eigenvalues of the holonomy around the $x$-direction to be $p_{\pm,l}$ and around the $y$-direction to be $q_{\pm,l}.$ We also presume the holonomy conjugacy classes to be generic.

In addition, we choose points $\mathbf{r}_{+,\nu}$ and $\mathbf{r}_{-,\nu}$ at which one of the Higgs field eigenvalues has respectively positive and negative Dirac singularity, i.e. one of the eigenvalues of the Higgs field tends to imaginary positive or imaginary negative infinity, so that the Higgs field is gauge equivalent to:
\begin{align}
\Phi&=i\begin{pmatrix} \frac{1}{2|\mathbf{r}-\mathbf{r}_{+,\nu}|}& 0_{1\times(n-1)}\\0_{(n-1)\times1}&0_{(n-1)\times(n-1)}\end{pmatrix}+O(|\mathbf{r}-\mathbf{r}_{+,\nu}|^0),\\
\Phi&=i\begin{pmatrix} \frac{-1}{2|\mathbf{r}-\mathbf{r}_{-,\nu}|}& 0_{1\times(n-1)}\\0_{(n-1)\times1}&0_{(n-1)\times(n-1)}\end{pmatrix}+O(|\mathbf{r}-\mathbf{r}_{-,\nu}|^0).
\end{align}

The complete set of boundary data is thus $(Q_{\pm,l},M_{\pm,l},p_{\pm,l},q_{\pm,l},\mathbf{r}_{\pm,\nu})$.  Among the results of \cite{Cherkis:2012qs} is the statement that the space of solutions for generic boundary data is a smooth hyperk\"ahler manifold of dimension $4\times {\rm Int} N$, where $N$ is the Newton polygon (determined purely in terms of the charges $Q_{\pm,l}$ and the numbers of positive and negative singularities) and ${\rm Int} N$ is the number of  integer points in the interior of $N.$  Though the Newton polygon $N$ can be constructed directly from the charges  \cite[Sec. 4.1]{Cherkis:2012qs}, one gains more insight by considering how $N$ arises from the spectral curve of the monowall, that we now define.

\subsection{Spectral Description and Moduli Space Isometry}
As spelled out in \cite{Cherkis:2012qs}, a monowall has two spectral descriptions each corresponding to one of the periodic directions $x$ or $y$ of the torus $T^2.$  A spectral description consists of a spectral curve and a Hermitian holomorphic line bundle over it.  Singling out the $x$-direction for concreteness, the Bogomolny equation \eqref{Eq:Bogomolny} implies 
\begin{align}
[D_z-i D_y, D_x+i\Phi]=0,
\end{align}
where $D_x, D_y,$ and $D_z$ are the covariant derivatives $D_j=\partial_j+A_j$, with $j=x,y,$ or $z.$ As a consequence, the holonomy $V(y,z)$ of $D_x+i\Phi$ around the $x$ direction depends holomorphically on $z-i y$ (so long as we stay away from the monowall singularities).  As a result, he eigenvalues of $V(y,z)$ are locally meromorphic in $s=\exp(2\pi(z-iy)/R)$ (away from the singularities and branch points) with simple poles at $s=s_{+,\nu}:=\exp(2\pi (z_{+,\nu}-i y_{+,\nu})/R),$ at the positions of the positive Dirac singularities $\mathbf{r}_{+,\nu}=(x_{+,\nu},y_{+,\nu},z_{+,\nu}),$  and simple zeros at $s=s_{-,\nu}:=\exp(2\pi (z_{-,\nu}-i y_{-,\nu})/R),$ at the positions of the negative Dirac singularities $\mathbf{r}_{-,\nu}=(x_{-,\nu},y_{-,\nu},z_{-,\nu}),$.  The spectral curve 
\begin{align}
\Sigma_x: \left\{(s,t)\in\mathbb{C}^*\times\mathbb{C}^* \big|\, {\rm det}\, (V(z,y)-t)=0\right\}
\end{align}
of eigenvalues of the holonomy is an algebraic curve in $\mathbb{C}^*\times\mathbb{C}^*$ with cusps at infinity corresponding either to singularities or to the asymptotic eigenvalues of the Higgs field at $z\rightarrow\pm\infty.$

As $\Sigma_x$ is a curve of eigenvalues, it carries an associated eigensheaf over itself.  So long as the spectral curve $\Sigma_x$ is nondegenerate, this is an eigen line bundle ${\cal L}_x\rightarrow\Sigma_x.$  Since each fiber of this line bundle is a line in the Hermitian fiber of $E\rightarrow T^2\times\mathbb{R},$  the line bundle ${\cal L}_x$ is also Hermitian. 

The pair $(\Sigma_x, {\cal L}_x\rightarrow\Sigma_x)$ of the spectral curve and the Hermitian line bundle over it is equivalent to the monowall $(A,\Phi).$ This is a form of the Hitchin-Kobayashi correspondence, still to be proved in this particular setup.  It gives a view of the monowall moduli space as a Jacobian fibration over the moduli space of curves.  Namely, the base is the space of curves in $\mathbb{C}^*\times\mathbb{C}^*$ with fixed cusps going to infinity; these are determined in terms of the boundary data (charges, $s$ singularity positions, constant terms in Higgs asymptotics, and asymptotic $x$-holonomy) of the monowall problem.  The fiber over a given curve is a set of Hermitian line bundles over it with fixed holonomy around the cusps.  The holonomy around each cusp is fixed by the monowall asymptotic $y$-holonomy data and $x$-coordinates of the monowall singularities.

Since the curve $\Sigma_x$ is algebraic, it can be given by a polynomial equation $G(s,t)=0.$  Marking a lattice point $(m,n)$ for each monomial $s^m t^n$ with nonzero coefficient in $G(s,t)$, the minimal integer convex polygon containing all of these marked points is the Newton polygon $N_x.$  For a given monowall $(A,\Phi)$ its spectral curves $\Sigma_x$ and $\Sigma_y$ generally differ, and so do the line bundles ${\cal L}_x$ and ${\cal L}_y.$  However, the Newton polygon of $\Sigma_x$ is the same as that of $\Sigma_y$, thus, from now on, we denote $N_x$ by $N$.  $N$ is completely determined by the numbers of positive and negative singularities and  by the charges (with their multiplicities) of the monowall.

It is more elegant to take a toric view of the spectral curve $\Sigma_x$ not as a curve in $\mathbb{C}^*\times\mathbb{C}^*$, but as a curve $\overline{\Sigma}_x$ in its toric compactification given by the toric diagram $N.$  Then the cusps are the intersections of $\overline{\Sigma}_x$ with the `infinity divisor' and the positions of these points of intersection (together with the holonomy of ${\cal L}_x$ around them) are the asymptotic data of the monowall.  

An important observation for us is that it is the curve and line bundle that are important, and not any special coordinates $s$ and $t$ that $\mathbb{C}^*\times\mathbb{C}^*$ inherited from the monowall formulation.  As argued in \cite{Cherkis:2012qs}, the natural $GL(2,\mathbb{Z})$ action on $s$ and $t$ by $\big(\begin{smallmatrix} a&b\\c&d\end{smallmatrix}\big):(s,t)\mapsto(s^at^b,s^ct^d)$ is an isometry of the monowall moduli spaces.  The monowall changes drastically under such a transformation: its rank, charges, even the numbers of positive and negative singularities change.  The moduli spaces, nevertheless, remain isometric.

In terms of the spectral curve $\overline{\Sigma}_x$, its intersection with infinity divisor of the toric compactification is determined by the coefficients of the monomials in $G(s,t)$ that correspond to the perimeter points of the Newton polygon $N.$  Thus we can vary at will the coefficients in $G(s,t)$ corresponding to the internal points of $N$ while respecting the asymptotic conditions.  In other words, the coefficients of $G(s,t)$ at the internal points of $N$ are complex moduli.  They coordinatize the base of the monowall moduli space viewed as the Jacobian fibration.   

As for the dimension of the fiber,  Hermitian line bundles on a punctured Riemann surface of genus $g$ with fixed holonomy around its punctures are parameterized by a points in a $2g$-torus $T^{2g},$ its Jacobian.  The coordinates can be viewed as holonomies of the corresponding flat connection on ${\cal L}_x$ around the $2g$ generators of the fundamental group $\pi_1(\overline{\Sigma}_x).$  Thanks to the theorem of Khovanskii \cite{Khovanskii77}, the genus of $\Sigma_x$ equals to the number of integer internal points of $N.$  Thus the monowall moduli space is fibered by $2g$-dimensional real tori over a $2g$-real-dimensional base, with $g={\rm Int}\, N$ being the number of integer internal points of $N.$

\subsection{Monowalls Fusion}
Spectral description and Newton polygons in particular provide a good language for describing monowall fusion or concatenation.  A natural question to ask is the following.  Given a monowall $A$ and another monowall $B$ when can we arrange them back to back.  To begin with let us place  $A$ far to the left and $B$ far to the right on $T^2\times\mathbb{R}.$  If this can be done, then we view the result as another monowall $C$ and view this as a fusion or concatenation
$$A+B\rightarrow C.$$
We would like to know the requirements on $A$ and $B$ for this process to be possible.  We would also like to know the properties of the resulting monowall $C.$ 

With $A$ far to the left and $B$ far to the right, in the intermediate region eigenvales of $\Phi$ are linear.  Since away from the monowalls' nonabelian cores the eigenvalues with different charges are distinct, in the intermediate region between $A$ and $B,$ while still sufficiently far from both $A$ and $B,$ all $\Phi$ eigenvalues (of differing charges) associated with monowall $A$ are diverging from each other as $z$ increases, while those associated with monowall $B$ are converging.  

Thus there are two possibilities: no eigenvalue of $\Phi$ is associated with both $A$ and $B$ monowall or there is only one value of charge for which $Q^A_+=Q^B_-=\frac{\alpha}{\beta}$ and the corresponding eigenvalues of $\Phi$ associated with both $A$ and $B.$  The former possibility gives a monowall $C$ in the direct sum of vector bundles of $A$ and $B$, $E_A\times E_B\rightarrow T^2\times \mathbb{R}$ and the monowall $C$ configuration $(A,\Phi)$ is block-diagonal.  In this case there is no interaction whatsoever between $A$ and $B.$ It is a trivial case.  The latter case has in the intermediate, between-the-walls, region all eigenvalues of $A$ with a given charge equal to the eigenvalues of $B$ with that same charge, thus they are identified.  In terms of the Newton polygons it implies that they have antiparallel edges.
For polygons with a common edge  we obtain an associative operation $(N_A,e)+(N_B,-e)=N_C.$
It is defined  if $N_A$ has an edge $e=r \begin{pmatrix}\alpha\\-\beta\end{pmatrix}$ and $N_B$ has an edge $-e=r \begin{pmatrix}-\alpha\\ \beta\end{pmatrix},$ with $\beta>0.$  We orient the edges on a Newton polygon clockwise, as in \cite[Sec~4.1]{Cherkis:2012qs}, and $r$ is the integer length of the edge which equals to the multiplicity of the corresponding charge $Q=\alpha/\beta.$  The Newton polygon $N_C$ is obtained by joining $N_A$ and $N_B$ along these two edges and taking the minimal convex polygon with integer vertices containing $N_A$ and $N_B$.  This process is Viro's patchworking \cite{Viro06} (see \cite{IV} for an illustration of its power).

Next, we turn to monowall fission by identifying `elementary monowalls' and those with minimal number of independent constituents.

\section{Monopole Walls with Four Moduli}\label{Sec:MWFM}
We would like to identify all monopole walls with four moduli, moreover, we would like to know which of them have isometric moduli spaces.  To begin with we identify all  `elementary' monopole walls, these have no moduli at all.  Next, we identify all monopole walls with four moduli up to the action of $GL(2,\mathbb{Z})$ group.  This group, acting on the Newton polygon lattice, is generated by 
\begin{enumerate}
\item
reflection of one of the axes,
\item
T transformation of the lattice $(e_1,e_2)\rightarrow(e_1,e_1+e_2),$ and
\item
S transformation $(e_1,e_2)\rightarrow(-e_2,e_1)$,
\end{enumerate}
which in terms of the monopole wall correspond respectively to 
\begin{enumerate}
\item
the reflection of the noncompact and one of the periodic coordinates,
\item
adding a charge one constant energy solution in the center of the gauge group, and
\item
the Nahm transformation.
\end{enumerate}
Our first goal in this section is to classify $GL(2,\mathbb{Z})$ inequivalent convex integer polygons with only one internal point.  Considering how natural this question is, the answer is probably known since antiquity.  However, not finding a good reference, though there must be many, we proceed obtaining this classification in Section~\ref{Sec:FourModuli}.  The answer is sixteen reflexive polygons (Table~\ref{Tab:Relations}).

Once all $GL(2,\mathbb{Z})$ inequivalent monopole walls are identified we find pairs of these which are related by adding some abelian monopole wall in the gauge group center, and thus with isometric moduli spaces.  The final list of monopole walls with nonisometric moduli spaces, Table~\ref{FinalList}, is twice shorter.  

\subsection{Monopole Walls with no Moduli}
\label{Sec:Moduli}

The simplest monopole wall is the direct sum of a number of constant energy density solutions of the same charge.  It is translationally invariant in all directions and has no moduli.  Its Newton polygon is in fact not a polygon, but a single interval.  What are the other monopole walls without moduli.
Khovanskii proved in \cite{Khovanskii97} that, up to the $GL(2,\mathbb{Z})$ equivalence, the only polygons with no internal points that are not degenerate (i.e. not an interval) are 
\begin{enumerate}
\item a triangle with sides of integer length two.  A representative of this class is a triangle with vertices $(0,0), (2,0),$ and $(0,2).$  It corresponds to a $U(2)$ monopole wall with two negative singularities,
\item a trapezium (or a triangle if $k=0$) of integer height 1 and bases of integer lengths $k$ and $m$ with $k\leq m.$  A representative of this class has vertices $(0,0), (m,0), (1,k),$ and $(1,0).$  This corresponds to a $U(1)$ monopole wall with $k$ positive and $m$ negative singularities.
\end{enumerate}

The spectral curves of the corresponding monopole walls are 1) $t^2+(C_{11} s+C_{10})t+C_{02}s^2+C_{01}s+C_{00}=0$ and 2) $t=P_m(s)/Q_k(s).$

In some sense these can be viewed as `elementary walls' out of which other walls are composed.  

\subsection{Monopole Walls with Four Moduli}\label{Sec:FourModuli}
The smallest nonzero number of moduli that a monopole wall can have is four\footnote{Note that we have the regular linear growth asymptotic conditions on $\mathbb{R}\times T^2.$  In a theory with a boundary one might expect lower number of moduli.}.  These are particularly interesting since they deliver moduli spaces that are self-dual gravitational instantons.  The search for these is among the main motivations for this study.  

Since each monopole wall with given singularity structure and given boundary conditions determines a Newton polygon and its number of moduli is four times the number of internal internal points of its Newton polygon, we would like to list all Newton polygons with single integral point. To begin with, integer translations of this polygon do not change the spectral curve and are therefore immaterial.  Thus, for now, we choose our polynomial to have the origin as the end of one of its edges with longest integer length\footnote{Integer length of an edge is one short of the number of integer points on that edge.}. Then use $GL(2,\mathbb{Z})$ transformation to have this edge stretch along the positive horizontal axis, and to place the Newton polygon in the upper half-plane. If this edge had integer length $l$, then, after this transformation, it has end points $(0,0)$ and $(l,0)$.  

If the intersection of the Newton polygon $N$ with the horizontal line $\h$ passing through the point $(0,1)$ is empty then the whole Newton polygon is an interval $[(0,0),(l,0)]$.  In that case the monopole wall is $GL(2,\mathbb{Z})$ equivalent to the constant energy solution and $N$ has no internal points and no moduli.  Similarly, if this intersection $N\cap\h$ has no {\em internal} integer points, then the Newton polygon has no internal points at all.  As we are looking for a Newton polygon with a single integer internal point, we conclude that the $N\cap\h$ has exactly one integer internal point.  Now we use an $SL(2,\mathbb{Z})$ transformation  of the form $\left(\begin{smallmatrix} 1& q\\0 &1\end{smallmatrix}\right)$ to put the internal point at $(1,1).$  Since $N$ is by definition convex, contains $(1,1)$ as its only internal point and $[(0,0),(l,0)]$ as its side, we conclude that  $l\leq4.$  This conclusion follows from the fact that $N$ should be contained within the triangle bounded from above by the line containing $[(l,0),(2,1)]$ (otherwise, (2,1) is another internal integer point), bounded from the left by the vertical axis (otherwise, (0,1) is another integer internal point), from below by the horizontal axis, and has only integer vertices.  There are no such convex {\em integer polygons} with $(1,1)$ as their only internal point for $l>4.$
\subsubsection{Maximal Side of Integer Length Four}
\begin{figure}[h]
\begin{center}
\includegraphics[width=0.3\textwidth]{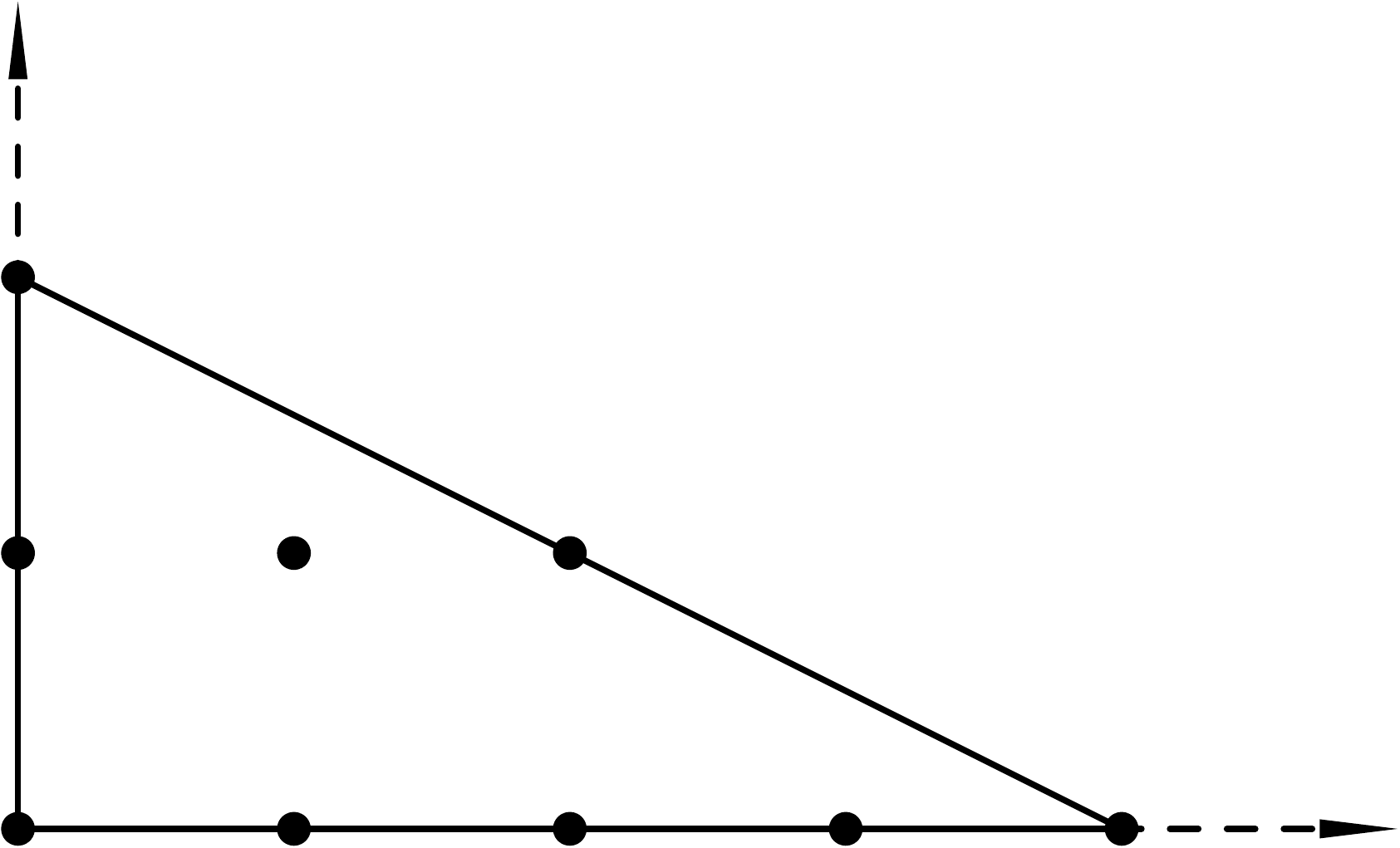}
\caption{The only Newton polygon with a single internal point and a side of integer length four.}
\label{Fig:l4}
\end{center}
\end{figure}

For $l=4$  there is exactly {\em one} such polygon with sides of integer lengths $4,2,2$ as in Figure~\ref{Fig:l4}.

\subsubsection{Maximal Side of Integer Length Three}

\begin{figure}[ht]
  \centering
   \subfloat[(3,3,3)]
   {\label{Fig:l3_333}\includegraphics[width=0.16\textwidth]{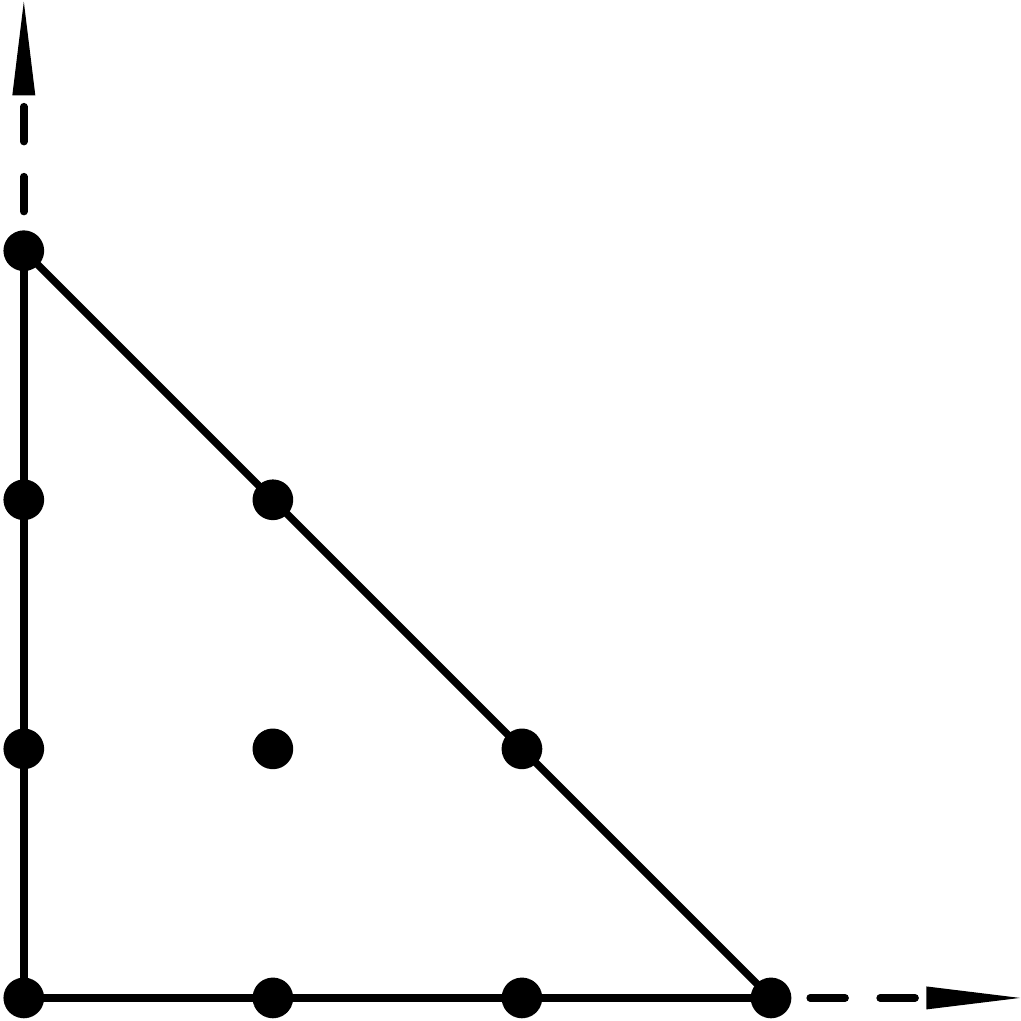}}
   \subfloat[(3,2,1,2)]{\label{Fig:l3_3212}\includegraphics[width=0.16\textwidth]{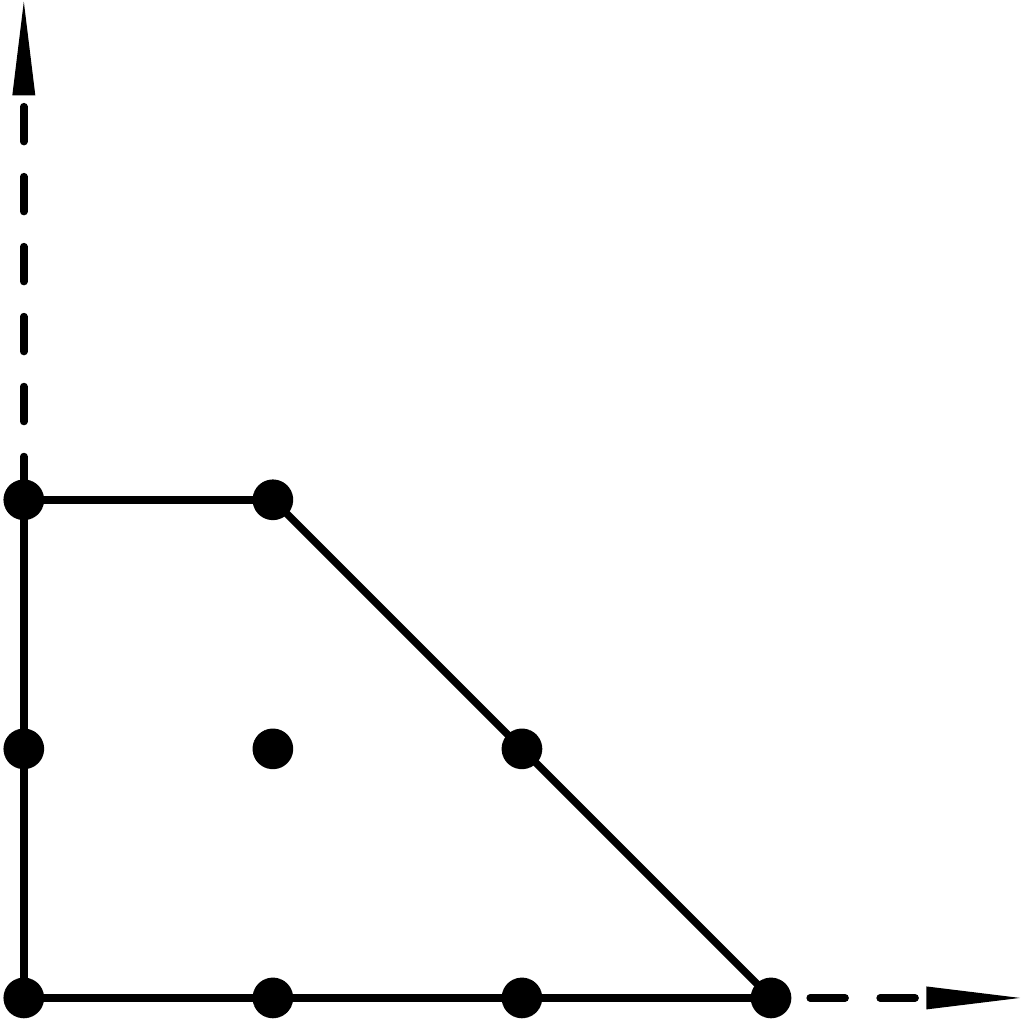}}
   \subfloat[(3,2,1,1)]{\label{Fig:l3_3211}\includegraphics[width=0.16\textwidth]{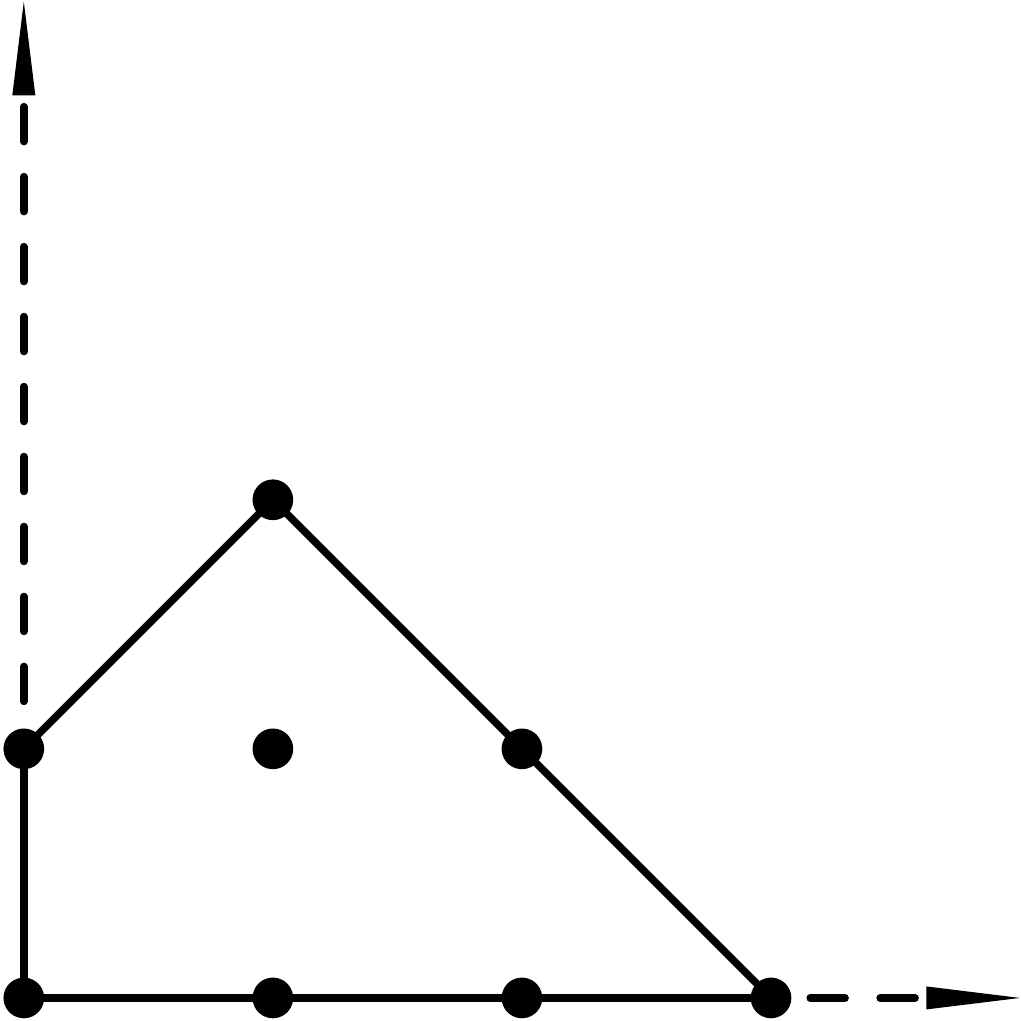}}
    \subfloat[(3,1,1,2)]{\label{Fig:l3_3112}\includegraphics[width=0.16\textwidth]{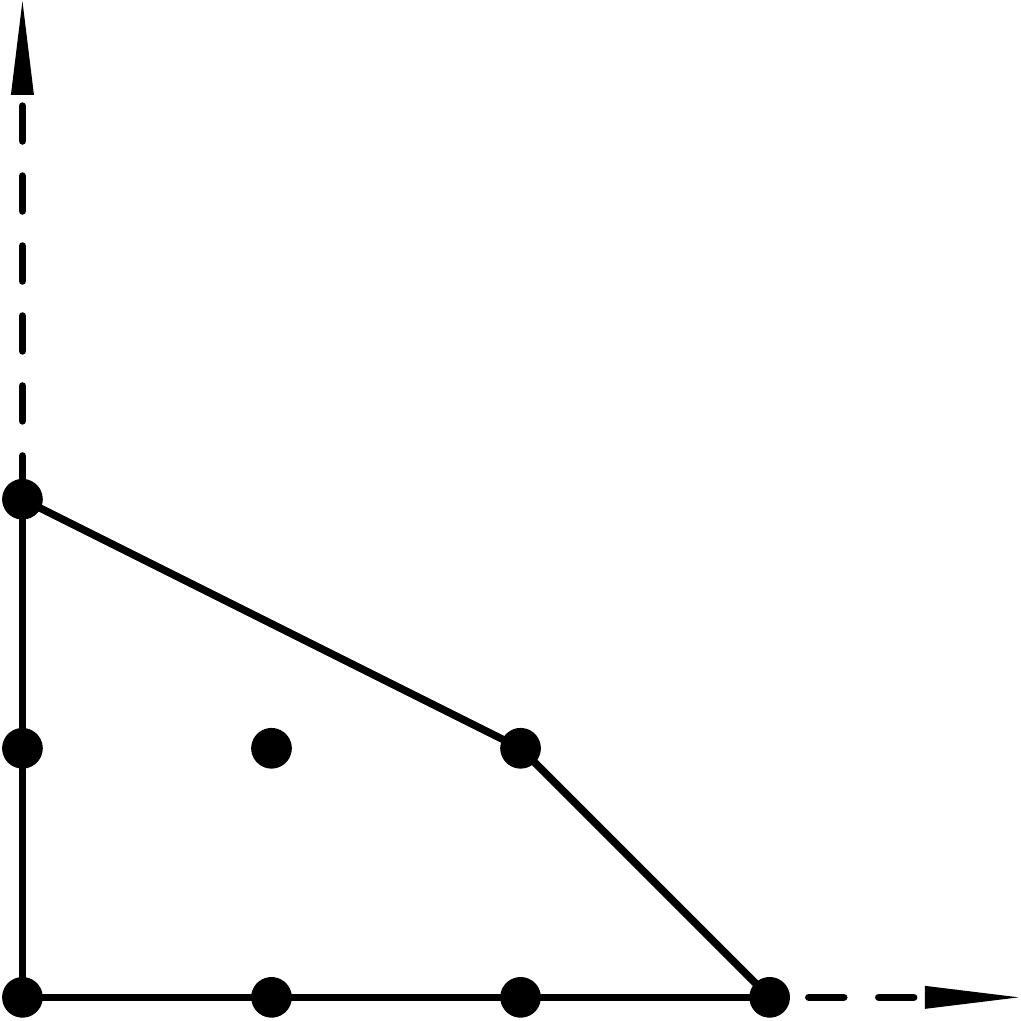}}
    \subfloat[(3,1,2)]{\label{Fig:l3_312}\includegraphics[width=0.16\textwidth]{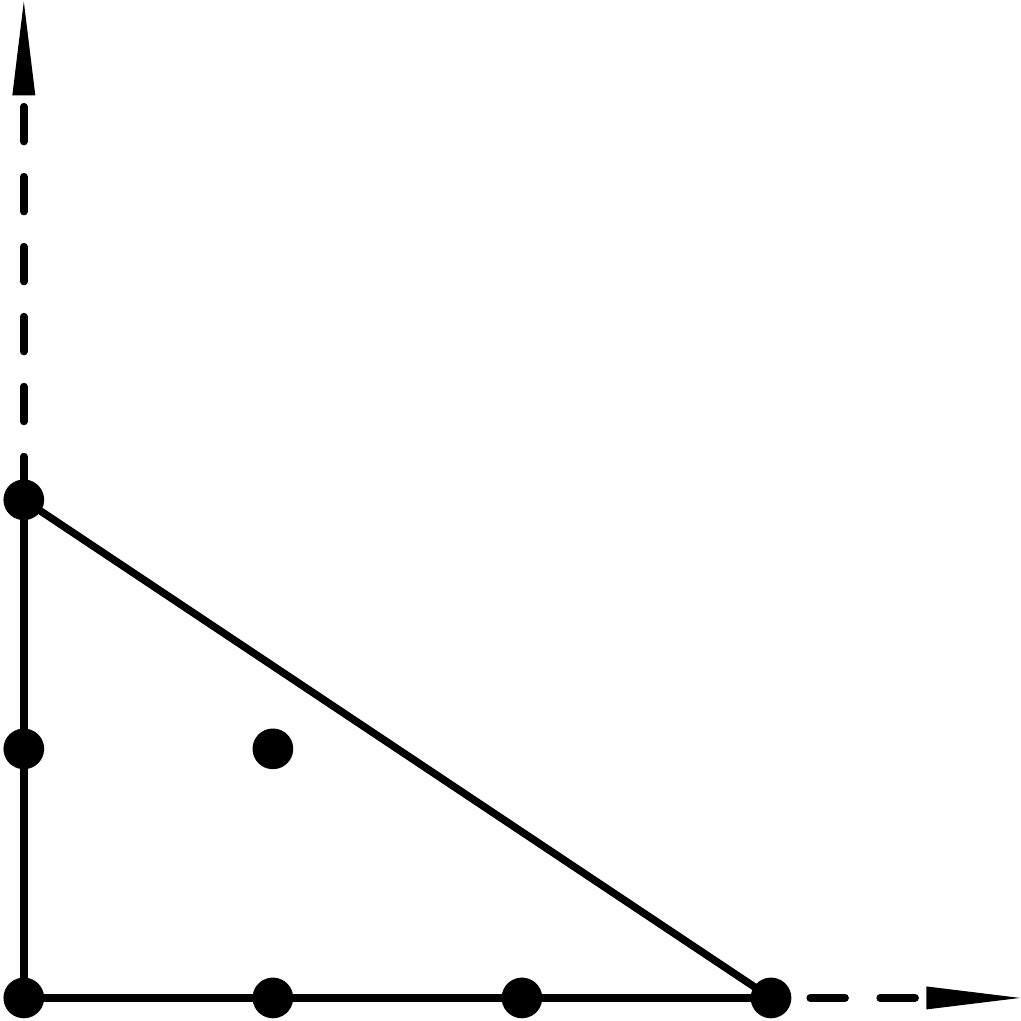}}
    \subfloat[(3,2,1)]{\label{Fig:l3_321}\includegraphics[width=0.16\textwidth]{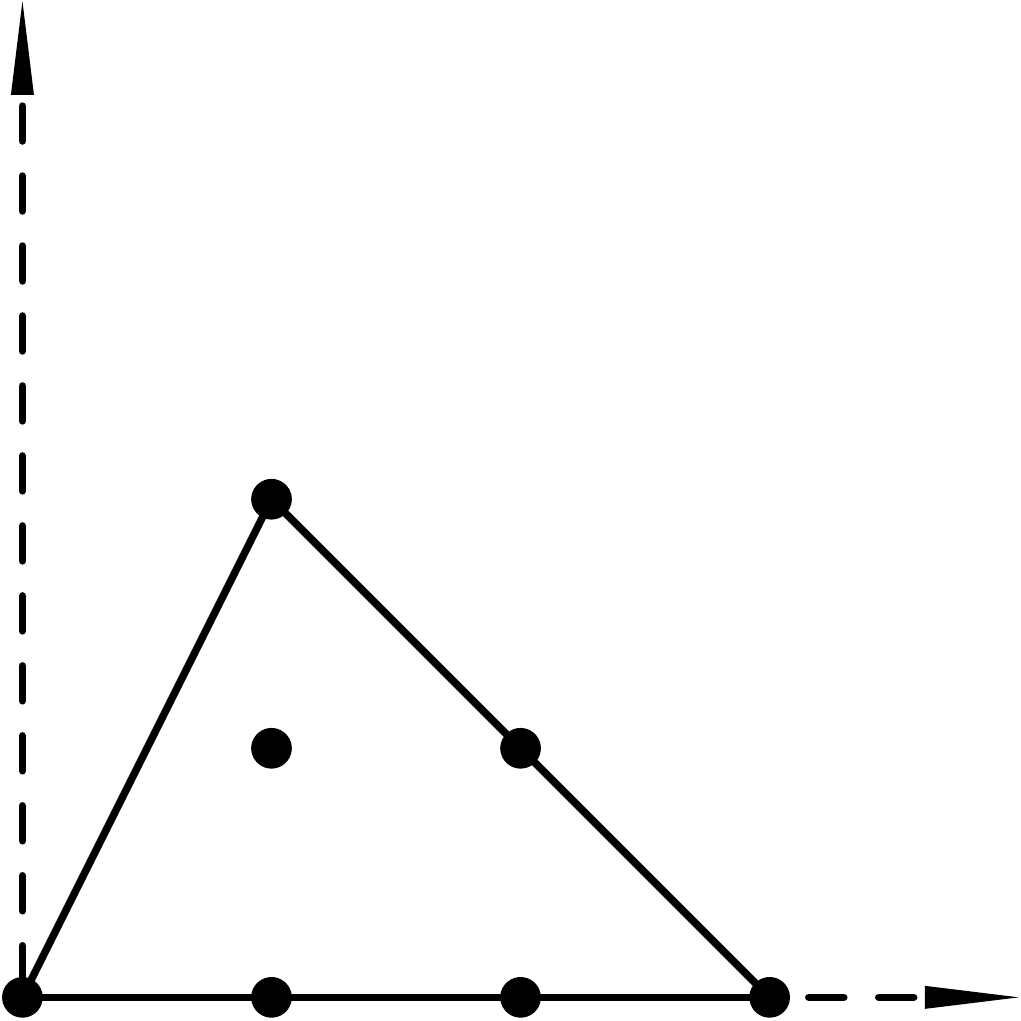}}
 \caption{Newton Polygons with single internal integer point and the longest side of length three.}
  \label{GI_l3}
\end{figure}

If $l=3$ then the Newton polygon is contained within the triangle $((0,0),(3,0),(0,3)).$  There are six such  Newton polygons that satisfy our conditions.  They are of integer sides lengths $(3,3,3), (3,2,1,2), (3,2,1,1), (3,1,1,2), (3,1,2),$ and $(3,2,1),$ presented in Figures~\ref{Fig:l3_333}, \ref{Fig:l3_3212}, \ref{Fig:l3_3211}, \ref{Fig:l3_3112}, \ref{Fig:l3_312}, and  \ref{Fig:l3_321} respectively.
The Newton polygon $(3,1,2)$ is related to $(3,2,1)$ by the transformation $\left(\begin{smallmatrix} -1&-1\\ 0&1 \end{smallmatrix}\right)$ followed by a right shift by three units;  the Newton polygon $(3,1,1,2)$ is related to $(3,2,1,1)$ by the same transformation.  The others are clearly $GL(2,\mathbb{Z})$ inequivalent, as they have either different number of sides or their side integer lengths spectra differ.

Thus there are {\em four distinct cases} of maximal side of length three.

\subsubsection{Maximal Side of Integer Length Two}

\begin{figure}[htbp]
  \centering
   \subfloat[(2,2,2,2)]
   {\label{Fig:l2_2222}\includegraphics[width=0.16\textwidth]{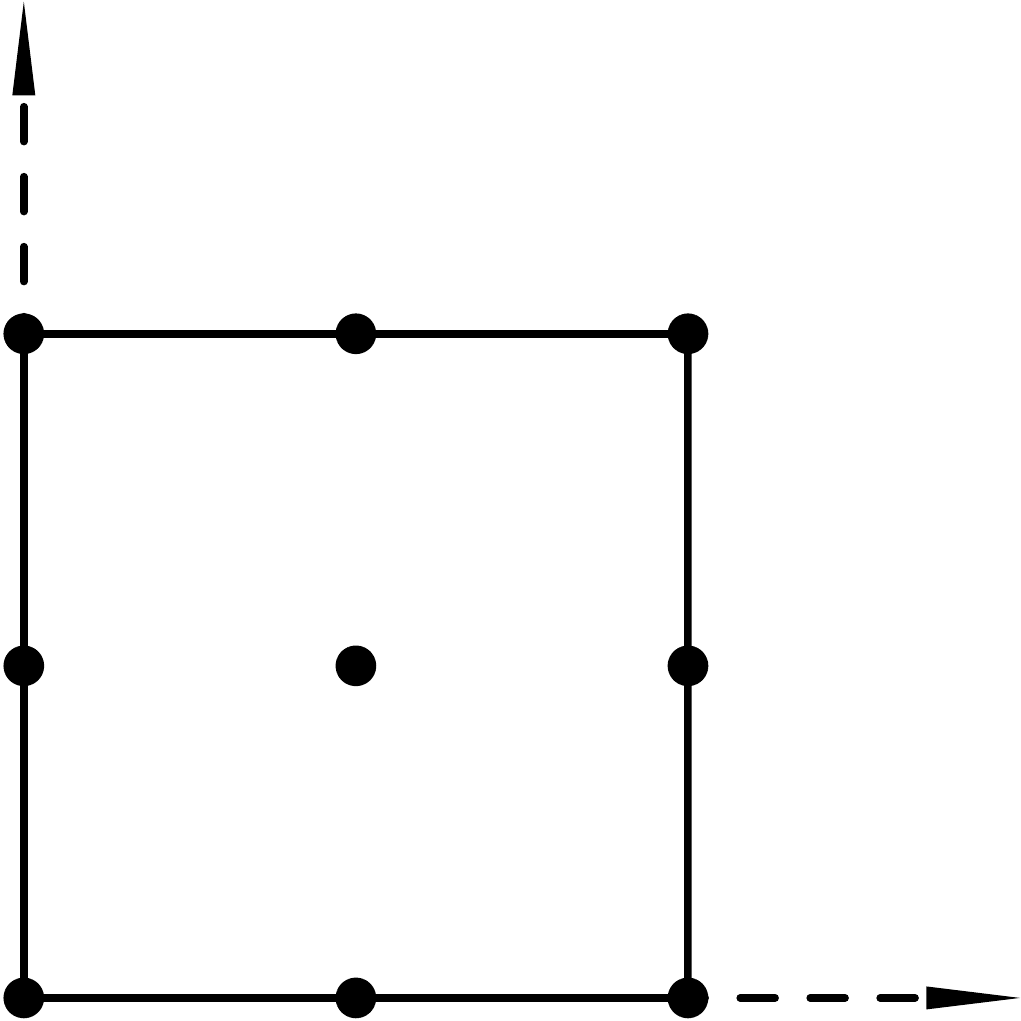}}
   \subfloat[(2,1,1,1,2)]{\label{Fig:l2_21112}\includegraphics[width=0.16\textwidth]{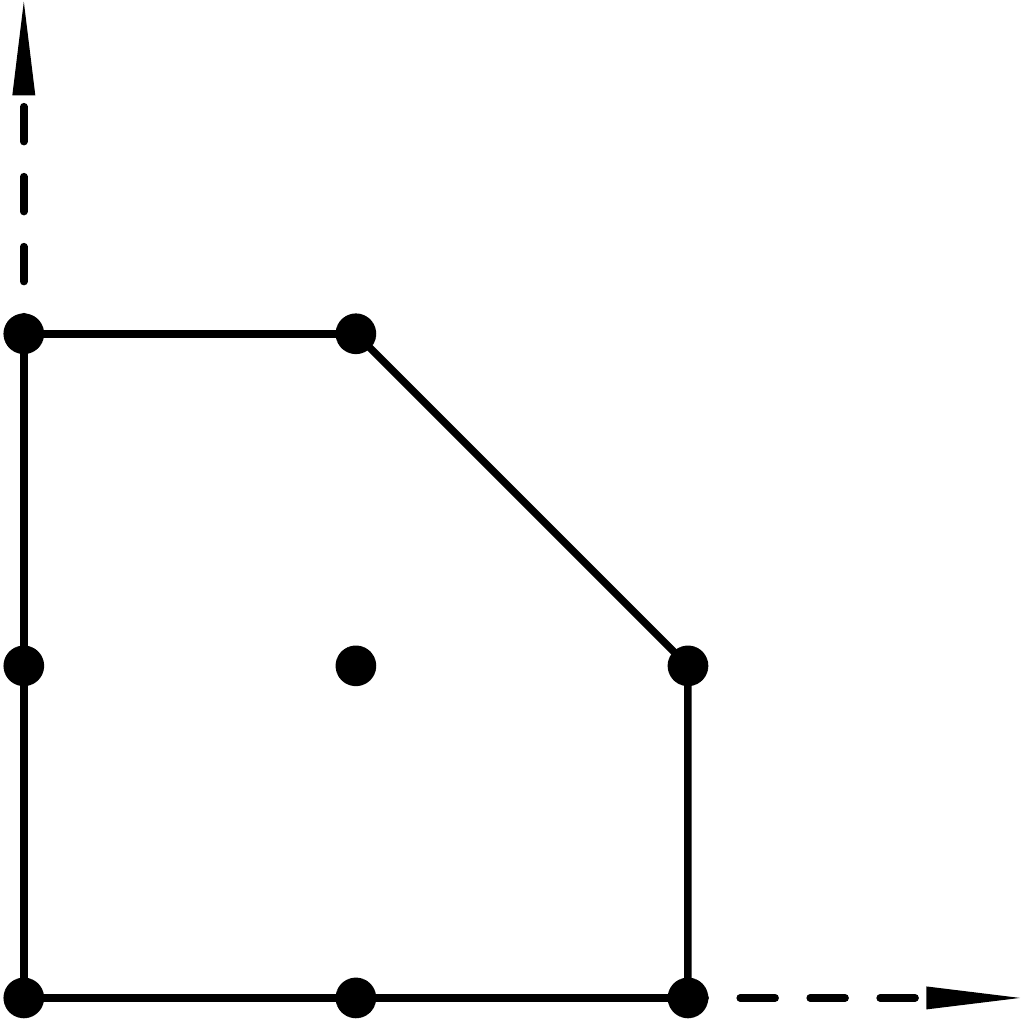}}
   \subfloat[(2,1,1,2)]{\label{Fig:l2_2112}\includegraphics[width=0.16\textwidth]{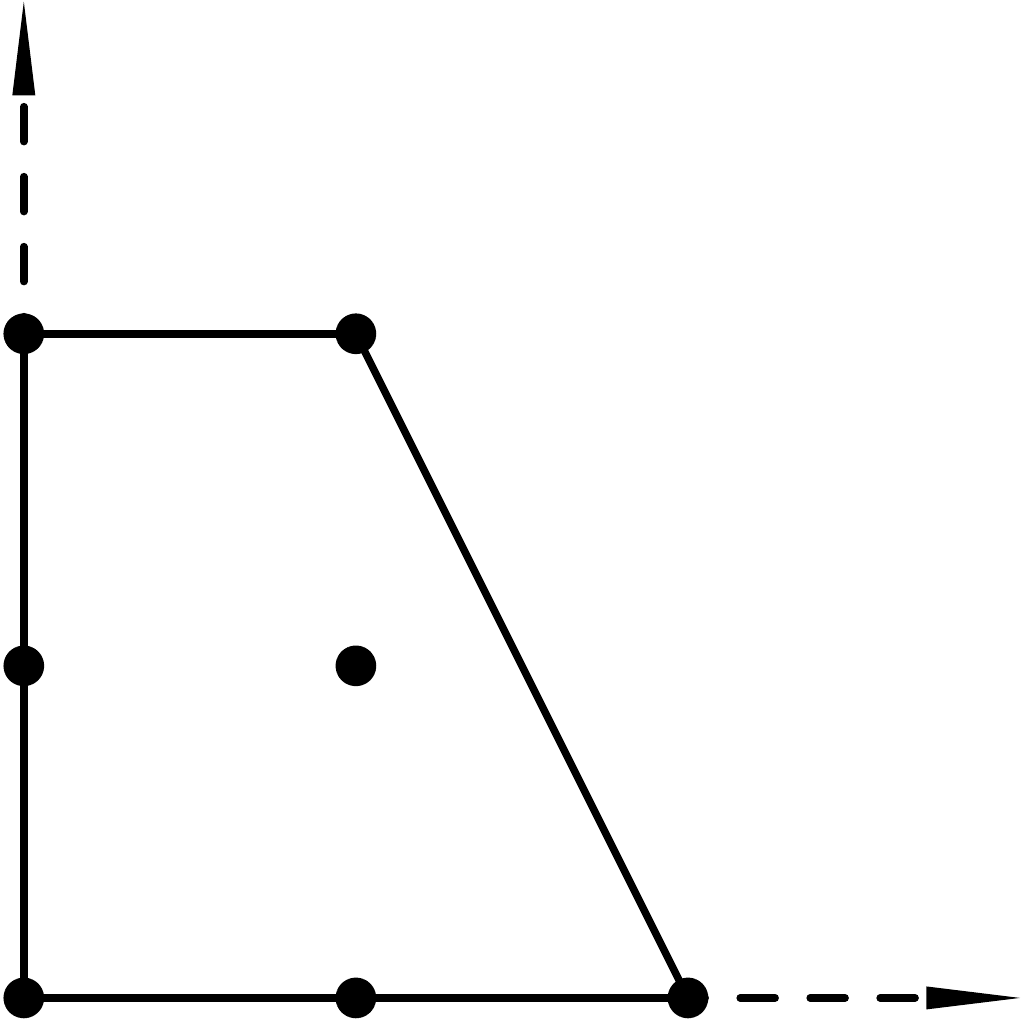}}
   \subfloat[(2,1,1,1,1)]{\label{Fig:l2_21111}\includegraphics[width=0.16\textwidth]{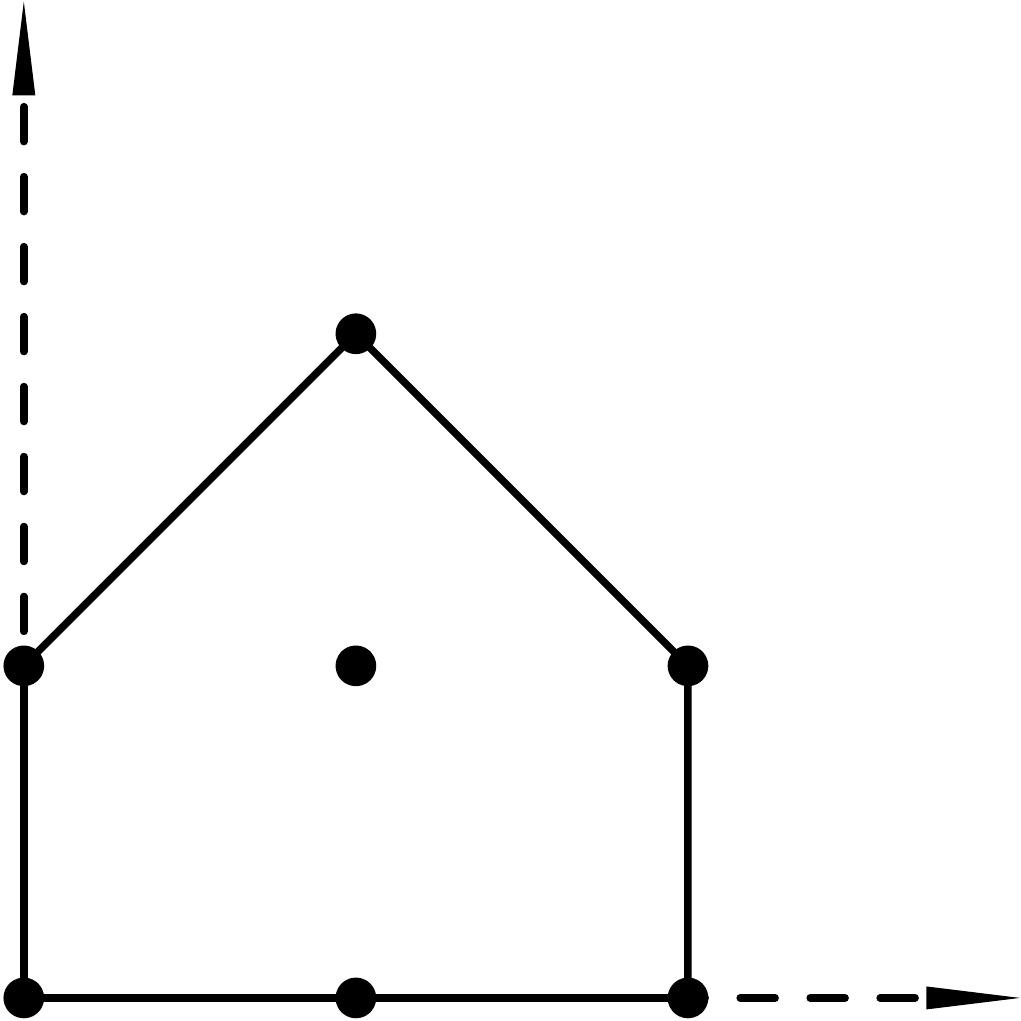}}
   \subfloat[(2,1,1,1)]{\label{Fig:l2_2111}\includegraphics[width=0.16\textwidth]{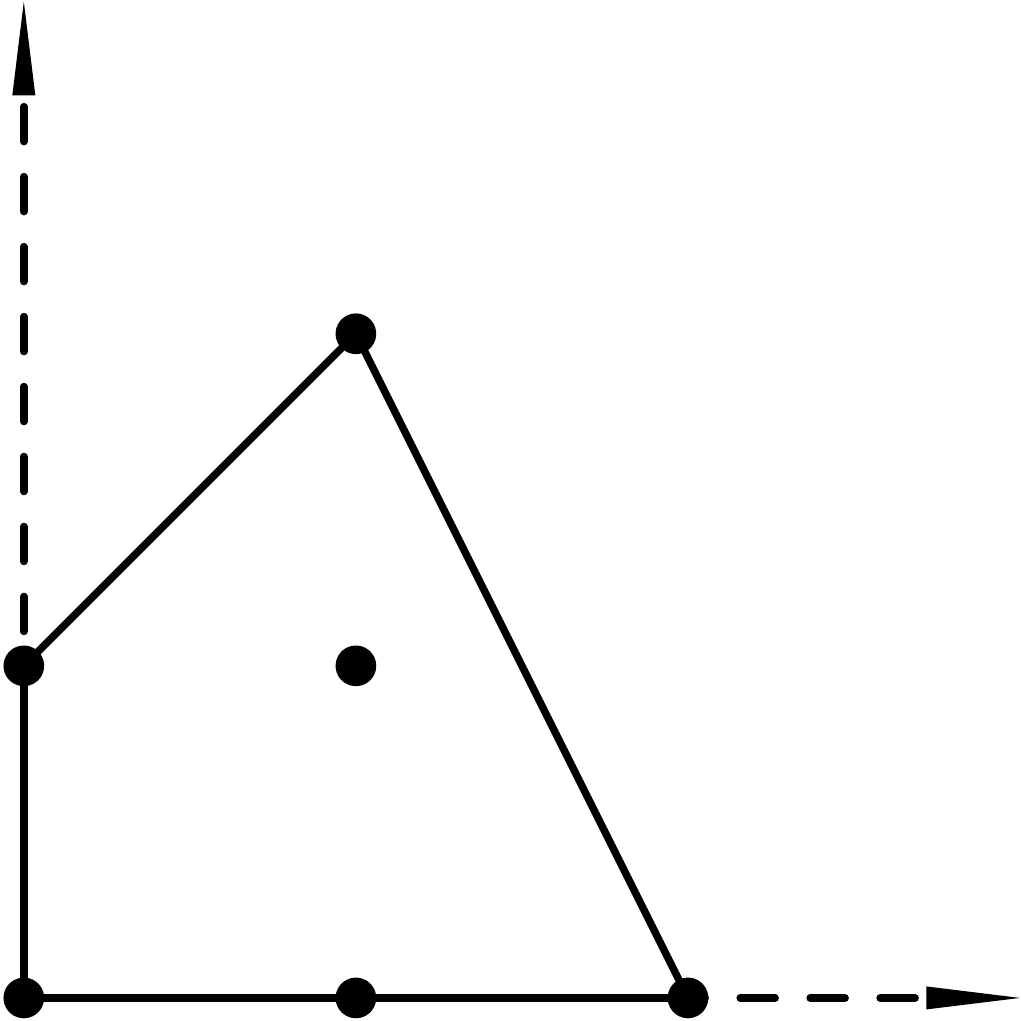}}
   \subfloat[(2,1,1)]{\label{Fig:l2_211}\includegraphics[width=0.16\textwidth]{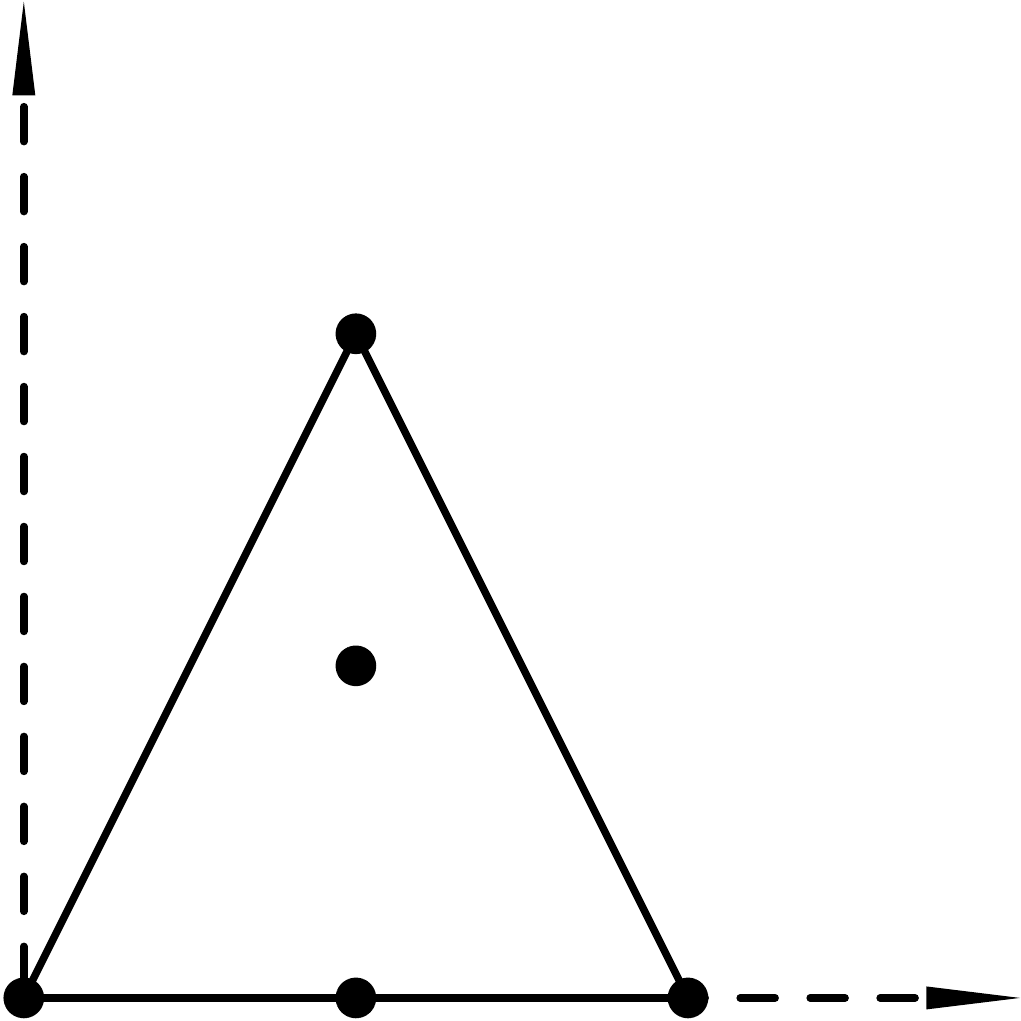}}
 \caption{Newton Polygons with single internal integer point and the longest side of length two.}
  \label{GI_l2}
\end{figure}

For $l=2$ $N$ lies in the strip between the vertical axis, the vertical line passing through the point $(2,0)$ and the horizontal line passing through $(0,2).$ There are six Newton polygons with one integer internal point (up to $GL(2,\mathbb{Z})$ equivalence).   The master polygon of Figure~\ref{Fig:l2_2222} is a square with sides of integer length 2.  All other cases result from truncating it, so that the result is convex and still contains the internal point $(1,1).$  Their integer side length spectra are $(2,1,1,1,2),$ $(2,1,1,2),$ $(2,1,1,1,1),$ $(2,1,1,1),$ and $(2,1,1).$  Since these spectra are all distinct, we have {\em six distinct cases} with $l=2.$

\subsubsection{Maximal Side of Integer Length One}
\begin{figure}[htbp]
  \centering
   \subfloat[Hexagon]{\label{Fig:Hex}\includegraphics[width=0.3\textwidth]{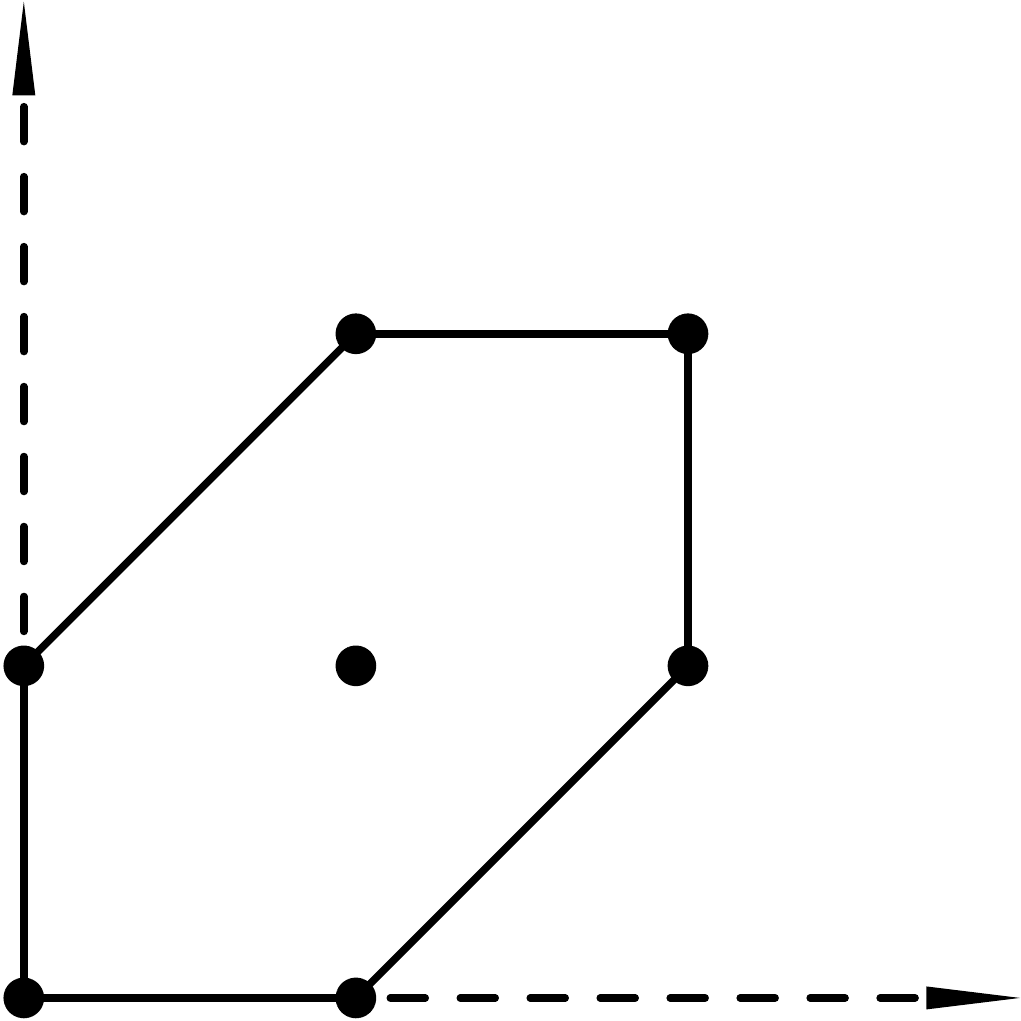}}
   \subfloat[First Pentagon]{\label{Fig:Pent}\includegraphics[width=0.3\textwidth]{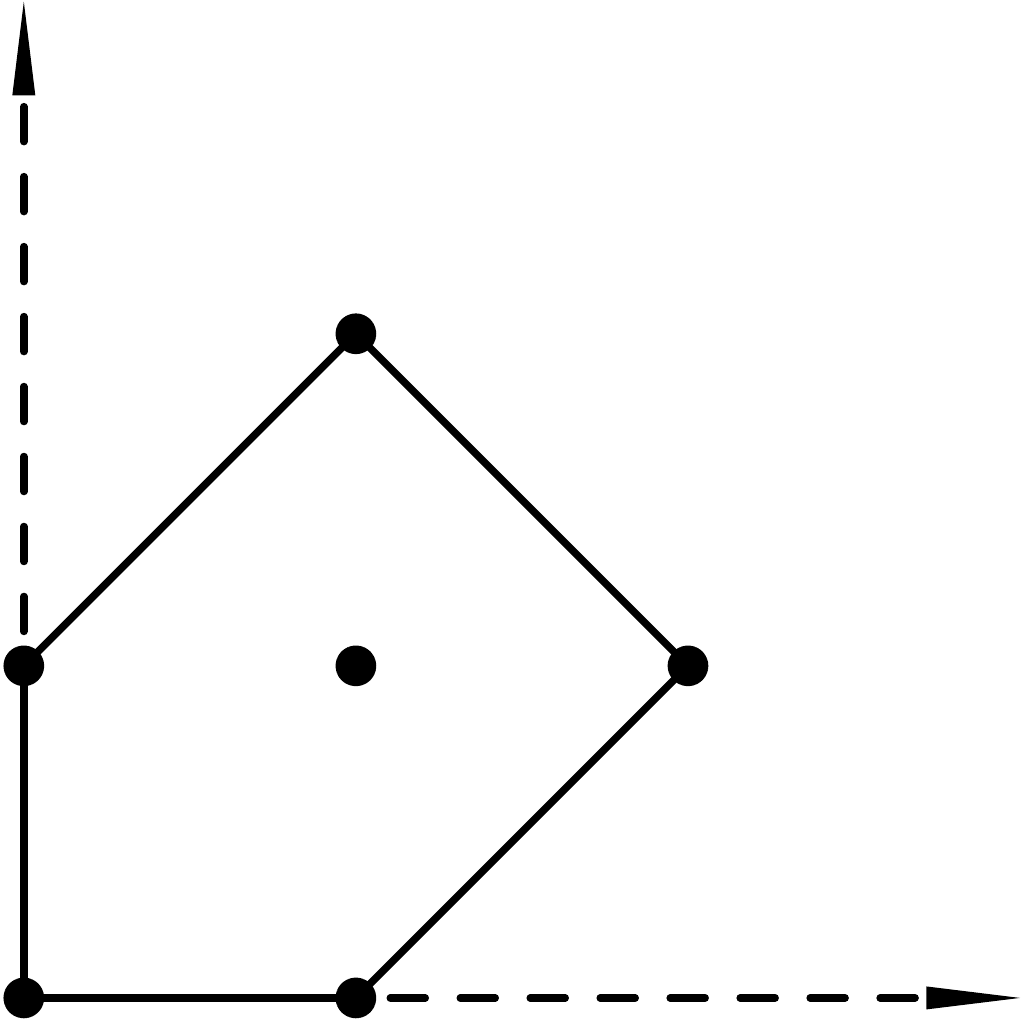}}
   \subfloat[Second Pentagon]{\label{Fig:PentS}\includegraphics[width=0.3\textwidth]{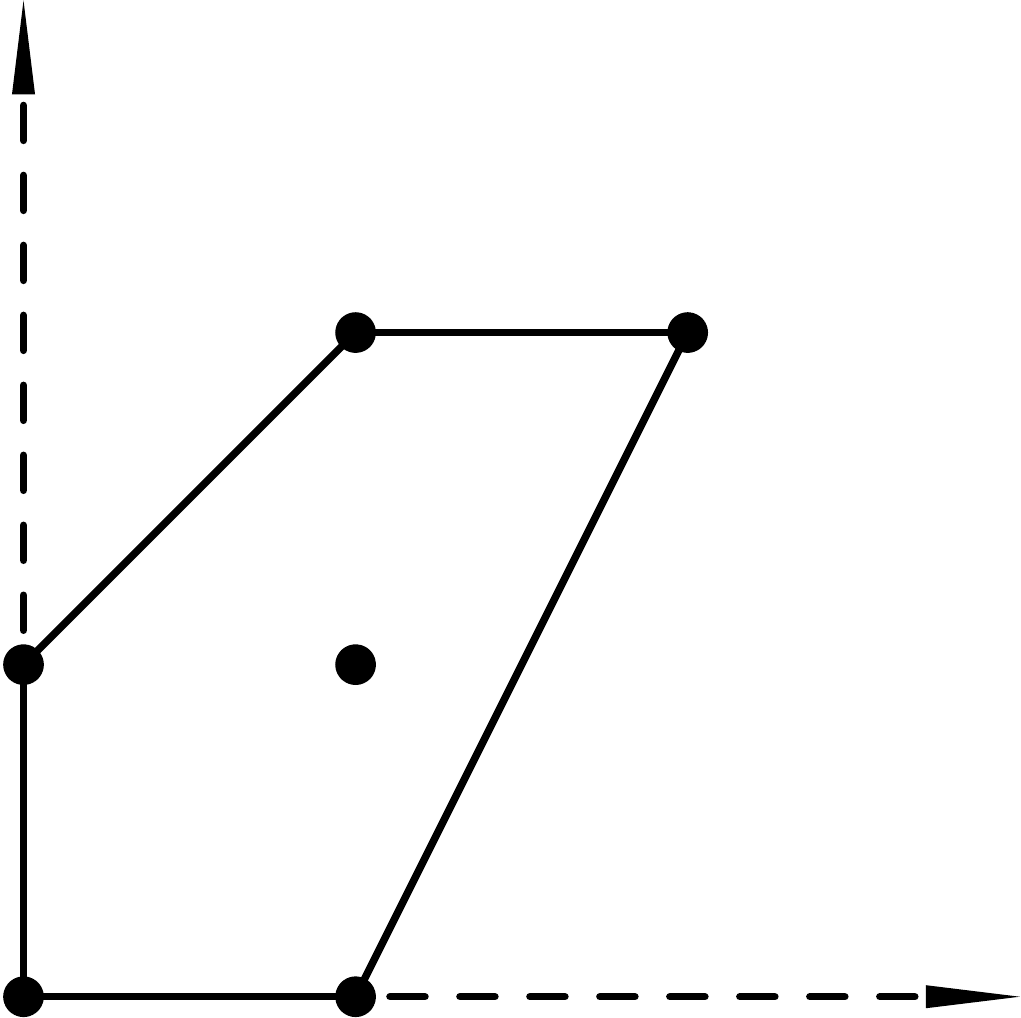}}
   
   \subfloat[Parallelogram or Rhombus]{\label{Fig:Pllg}\includegraphics[width=0.3\textwidth]{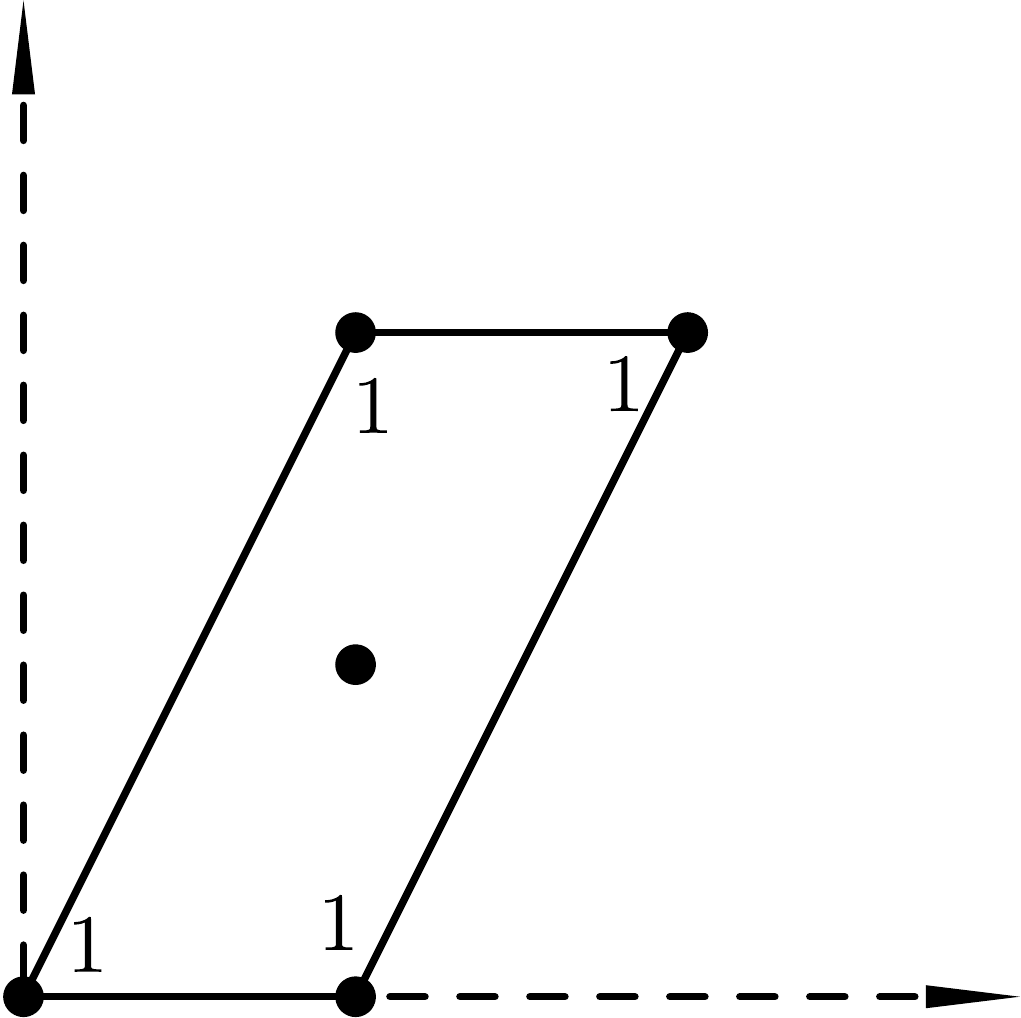}}
   \subfloat[Symmetric Quadrilateral]{\label{Fig:QuadSym}\includegraphics[width=0.3\textwidth]{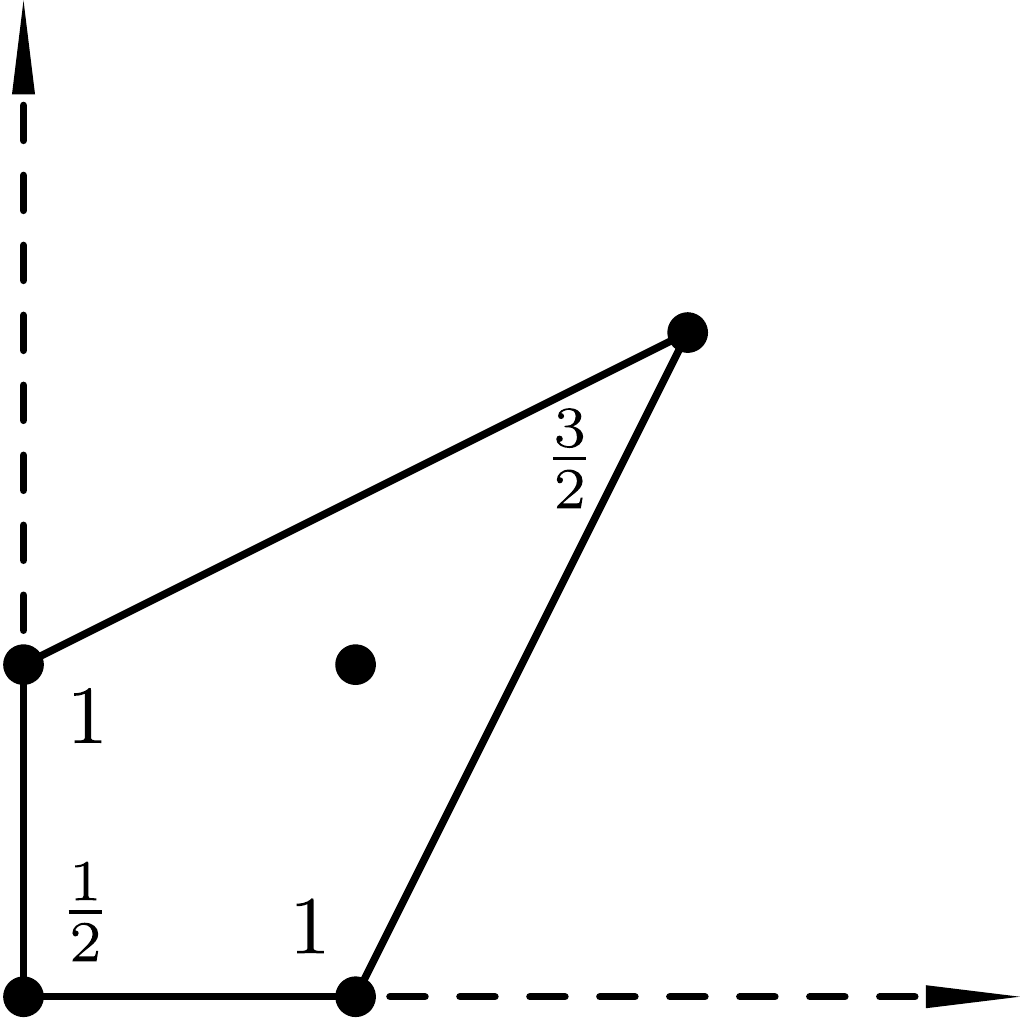}}
      \subfloat[Curious Quadrilateral]{\label{Fig:QuadCu}\includegraphics[width=0.3\textwidth]{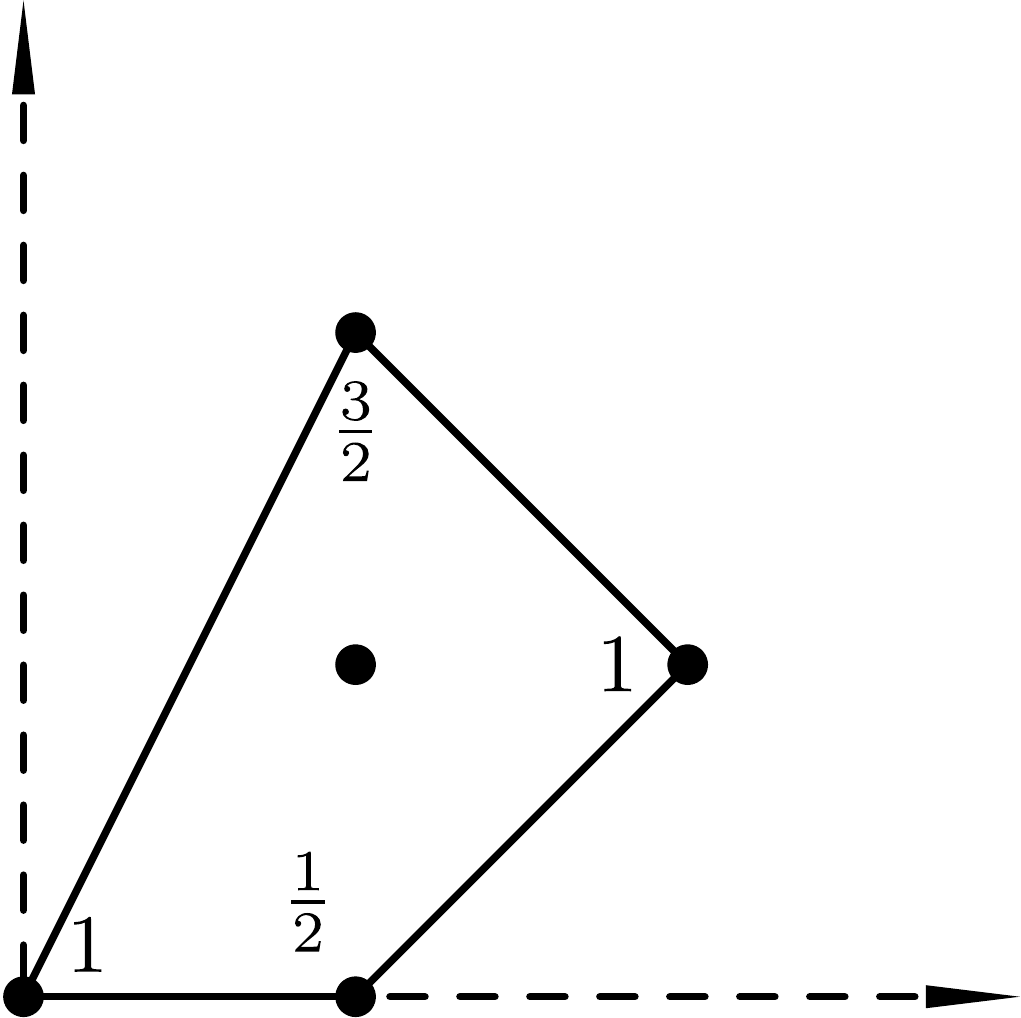}}
      
         \subfloat[Ambitious Quadrilateral]{\label{Fig:QuadAmb}\includegraphics[width=0.3\textwidth]{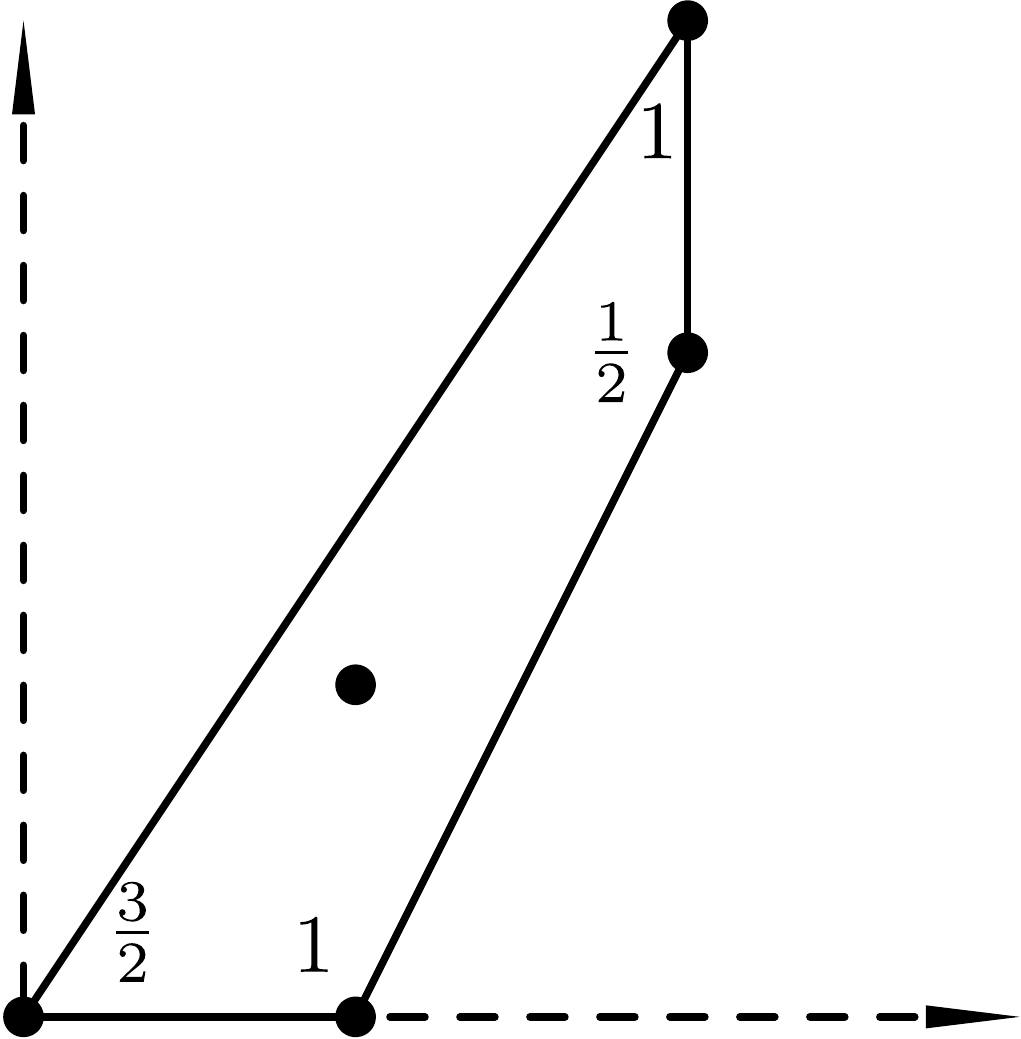}}
   \subfloat[Obnoxious Quadrilateral]{\label{Fig:QuadAbn}\includegraphics[width=0.3\textwidth]{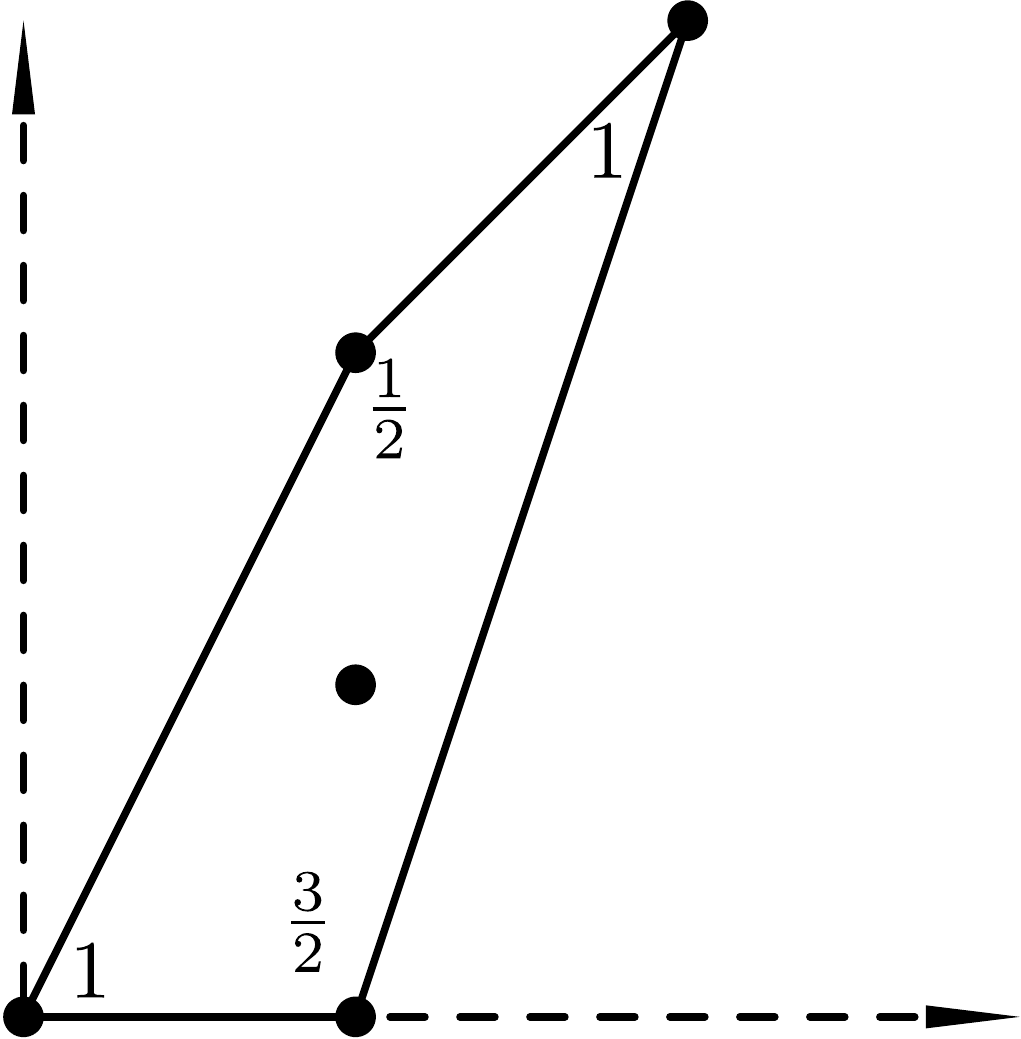}}
      \subfloat[Triangle]{\label{Fig:Triangle}\includegraphics[width=0.3\textwidth]{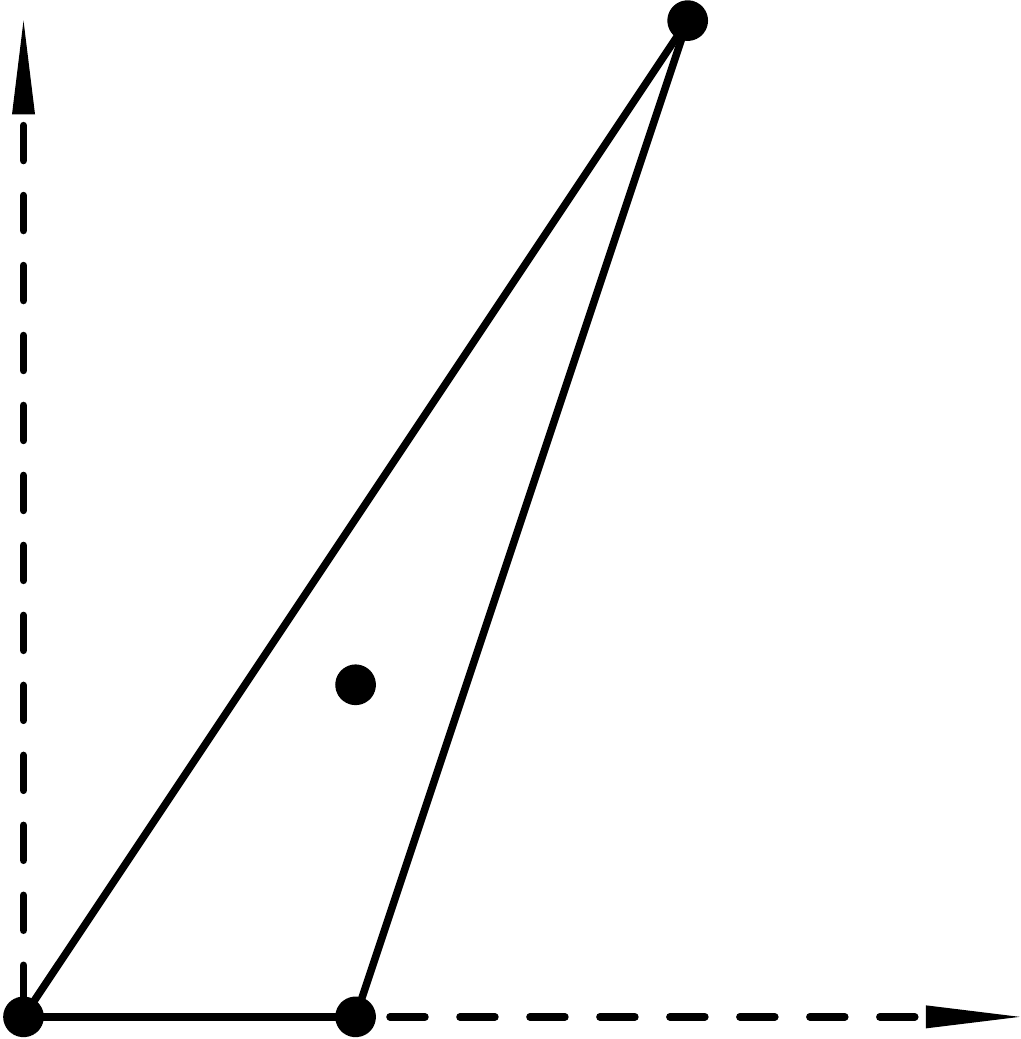}}
 \caption{Newton Polygons with single internal integer point and all sides of length one.}
  \label{GI_l1}   
\end{figure}

For $l=1$ all sides of the Newton polygon are of integer length 1, i.e. all integer points on the perimeter of the Newton polygon are vertices.  By our construction, one of the sides is $((0,0),(1,0)).$ Let us focus on the other side originating at $(0,0).$  Let us denote the coordinates of its other end by $(p,q).$  Using $\left(\begin{smallmatrix} 1& q\\0 &1\end{smallmatrix}\right)$ transformation of $GL(2,\mathbb{Z})$  we can make sure that $p\geq0$ and $q>0.$  Then, by convexity of $N$,  the triangle $\Delta=((0,0),(p,q),(1,0))$ is contained within the Newton polygon and thus should have zero or one internal point.  Pick's formula relates the number of internal points $I$ to the area $A$ and integer perimeter length $P$ of an integer polygon:
\begin{equation}
I=A-\frac{1}{2}P+1.
\end{equation}
For the triangle $\Delta$ we have $I\leq1, A=\frac{1}{2}q,$ and $P=3.$ Thus Pick's formula implies $q\leq 3.$
This leaves three possibilities $(p,q)=(0,1), (1,2),$ or $(2,3),$ since we allow only $(1,1)$ as the internal point of the Newton polygon.

 There are {\em nine} such polygons: one hexagon, two pentagon, five quadrilaterals, and one triangle as in Figure~\ref{GI_l1}. 
The second pentagon  (Figure~\ref{Fig:PentS}) is related to the first pentagon (Figure~\ref{Fig:Pent}) by the transformation $\left(\begin{smallmatrix} 
1 & 0  \\ 
1 &  -1
\end{smallmatrix}\right)$ followed by the upward shift by one unit.  Now, turning to the quadrilaterals, we look at their $GL(2,\mathbb{Z})$ invariant quantities.  The areas spanned by pairs of their adjacent edges are $(1,1,1,1), (\frac{1}{2}, 1, \frac{3}{2}, 1), (1,\frac{1}{2}, 1, \frac{3}{2}), (\frac{3}{2}, 1, \frac{1}{2}, 1),$ and $(1, \frac{3}{2}, 1, \frac{1}{2})$ respectively for the parallelogram, symmetric, curious, ambitious, and obnoxious quadrilaterals of Fig.~\ref{GI_l1}.  This indicates that parallelogram is distinct, while the other four might be $GL(2,\mathbb{Z})$ equivalent.
Indeed, the symmetric quadrilateral is transformed 
\begin{itemize}
\item into the curious quadrilateral by a shift down by one followed by $\left(\begin{smallmatrix} 
1 & -1  \\ 
1 & 0 
\end{smallmatrix}\right)$ transformation,  
\item into the ambitious quadrilateral by a shift by $(-2,-2)$ followed by $\left(\begin{smallmatrix} 
0 & -1  \\ 
1 &  -2
\end{smallmatrix}\right)$ transformation, and
\item into the obnoxious quadrilateral by a shift by $(-1,0)$ followed by $\left(\begin{smallmatrix} 
-1 & 1  \\ 
-2 &  1
\end{smallmatrix}\right).$
\end{itemize}

Thus for $l=1$ case we have five distinct Newton polygons: a hexagon, a pentagon, a parallelogram, a symmetric quadrilateral, and a triangle.

\subsubsection{Complete List}
The three master polygons are those in Figs.~\ref{Fig:l3_333}, \ref{Fig:l2_2222}, and \ref{Fig:l4}.  All the others can be obtained by deleting some of the perimeter points of these three.  We organize them in Table~\ref{Tab:Relations} according to their integer perimeter length.  We also put the triangle of Figure~\ref{Fig:Triangle} in a more elegant form applying the $\left(\begin{smallmatrix} 
2 & -1  \\ 
1 &  0
\end{smallmatrix}\right)$ transformation.

\begin{table}[htbp]
\begin{center}
\begin{tabular}{cccccccccc}
$p=9$&&&&&\includegraphics[width=0.1\textwidth]{GI_l3_333}&&&&
\\  &&&&&$\downarrow$&&&&
\\
$p=8$&\includegraphics[width=0.1\textwidth]{GI_l2_2222}&&&&\includegraphics[width=0.1\textwidth]{GI_l3_3212}&&&&\includegraphics[width=0.15\textwidth]{GI_l4}
\\
 & &$\searrow$&&$\swarrow$&&$\searrow$&&$\swarrow$&
\\
$p=7$& & &\includegraphics[width=0.1\textwidth]{GI_l2_21112}& & & &\includegraphics[width=0.1\textwidth]{GI_l3_3112}&&
\\
 & &$\swarrow$&$\searrow$&&$\swarrow\searrow$&$\swarrow$&&$\searrow$&
\\
$p=6$&\includegraphics[width=0.1\textwidth]{GI_l1_Hex}& &&\includegraphics[width=0.1\textwidth]{GI_l2_2112}&&\includegraphics[width=0.1\textwidth]{GI_l2_21111}&&&\includegraphics[width=0.1\textwidth]{GI_l3_312}
\\
 & &$\searrow$& & & $\searrow$\ $\swarrow$& & &$\swarrow$&
\\
$p=5$& & &\includegraphics[width=0.1\textwidth]{GI_l1_Pent}&&& &\includegraphics[width=0.1\textwidth]{GI_l2_2111}  &&
\\
 & &$\swarrow$&&$\searrow$&&$\swarrow$&&$\searrow$&
\\
$p=4$&\includegraphics[width=0.1\textwidth]{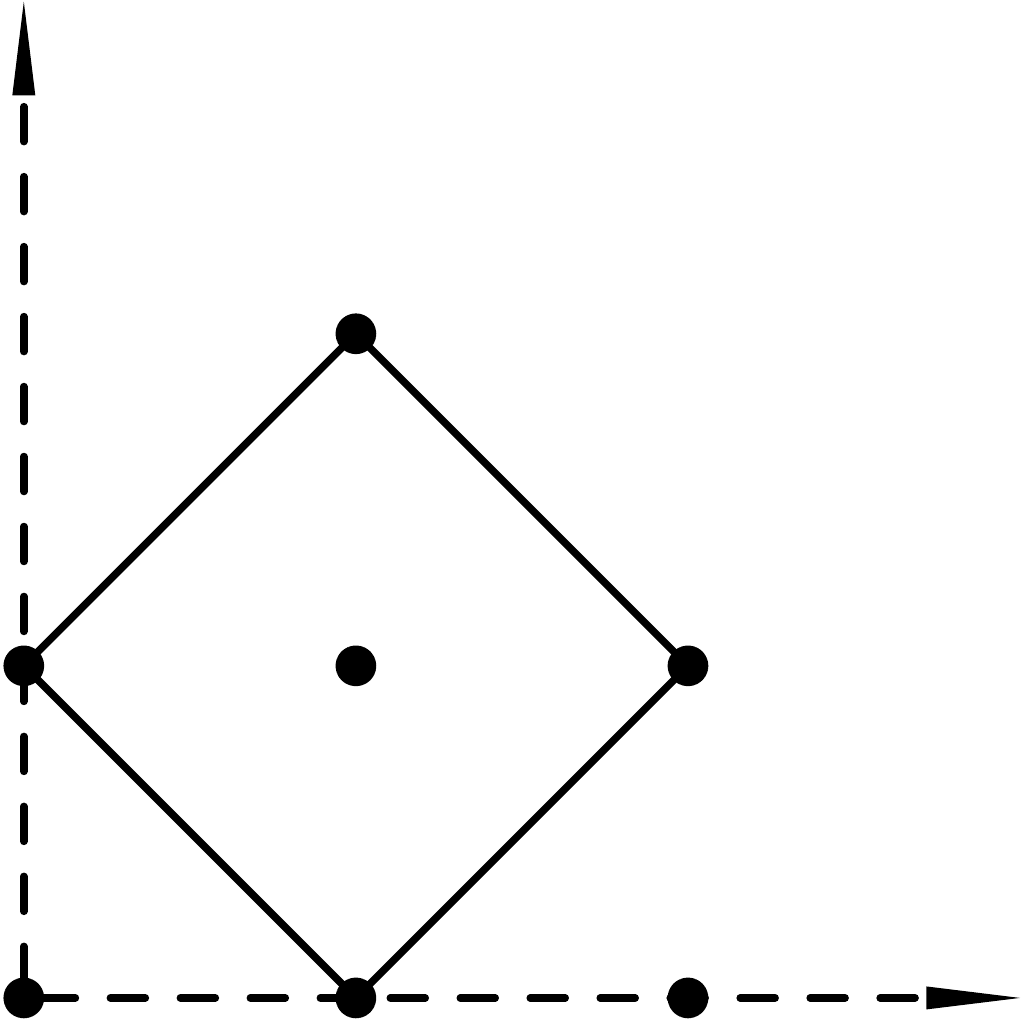}&&&&\includegraphics[width=0.1\textwidth]{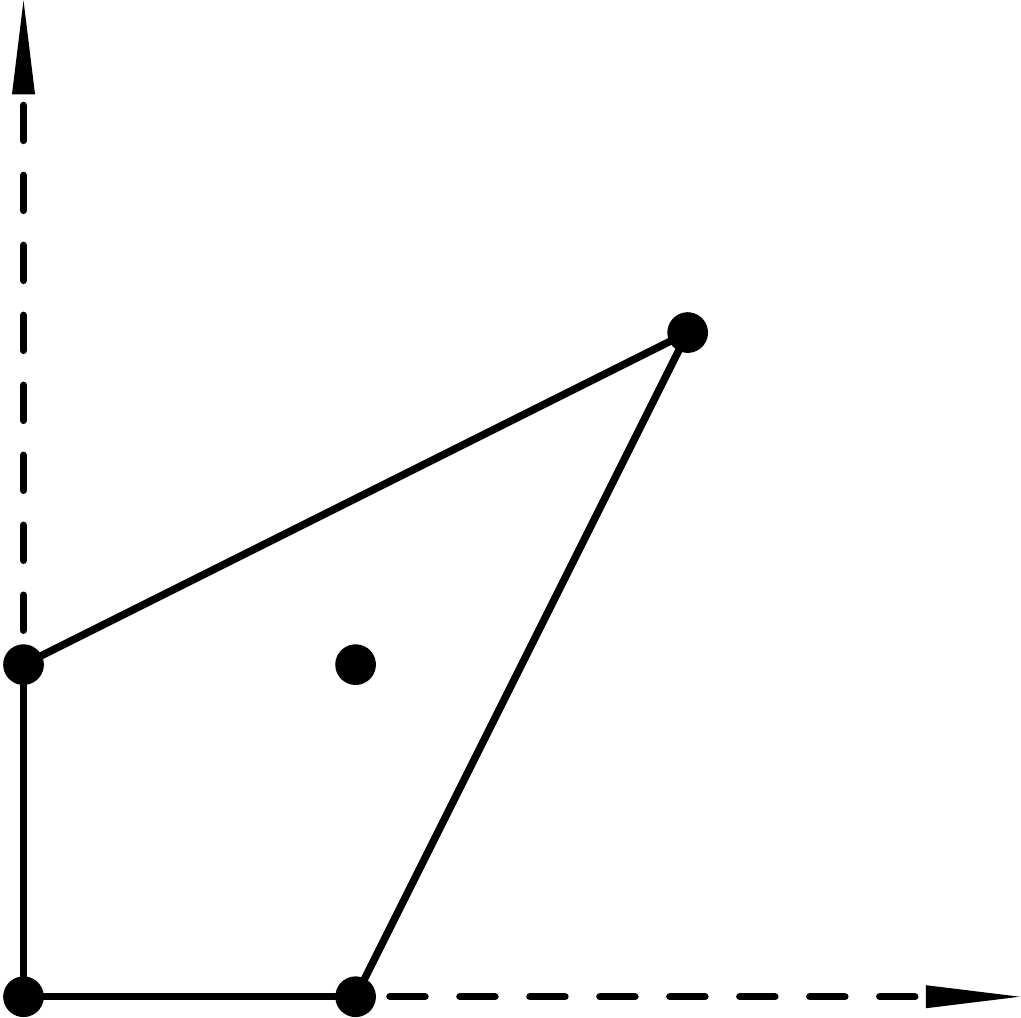}&&&&\includegraphics[width=0.1\textwidth]{GI_l2_211}
\\
&&&&&$\downarrow$&&&&
\\
$p=3$&&&&&  \includegraphics[width=0.1\textwidth]{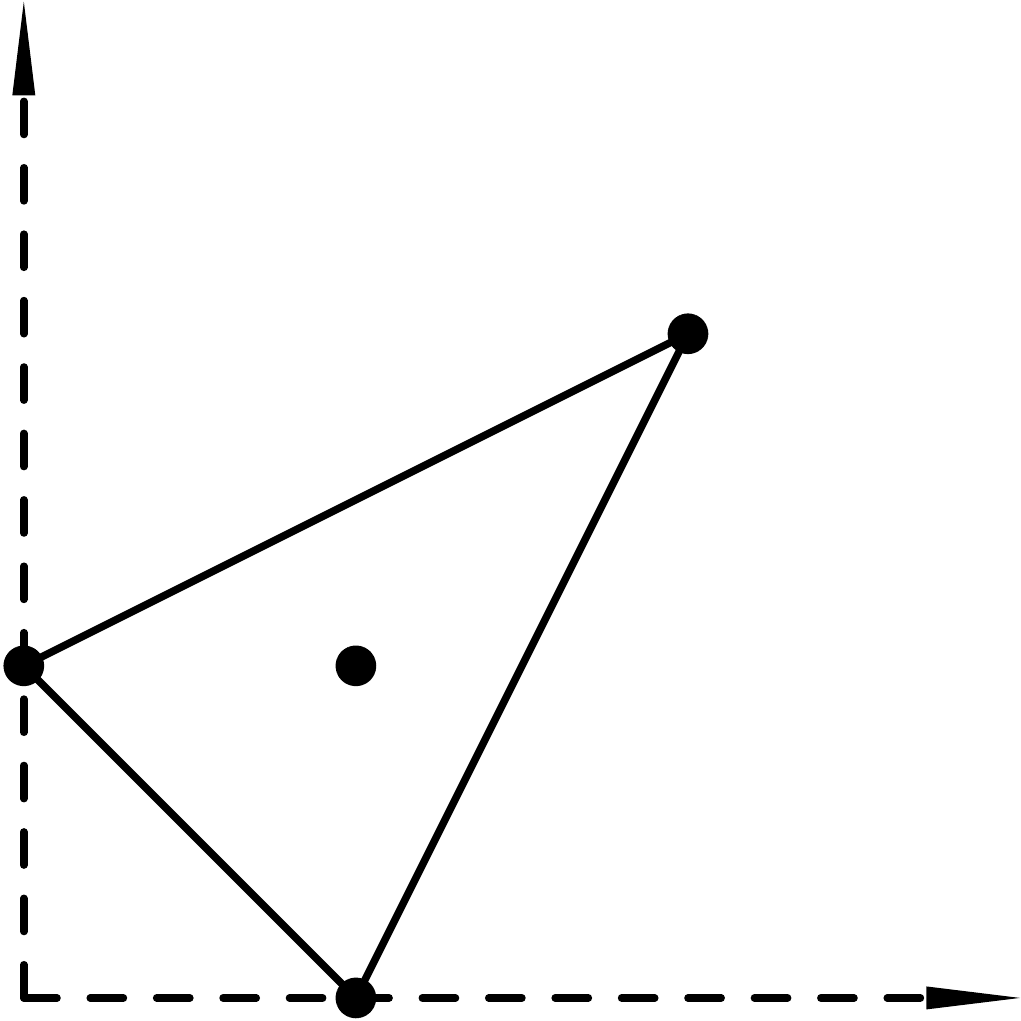} &&&&
\end{tabular}
\caption{Relations between moduli spaces.}
\label{Tab:Relations}
\end{center}
\end{table}

There are sixteen cases in total.  This is exactly the celebrated list of reflexive polygons that are significant in numerous fields, see e.g. \cite{12} for some insightful relations. Each has its own $GL(2,\mathbb{Z})$ class of monowalls with the same moduli space. 

There might be some other equivalence that would establish that some of these spaces are isometric.  Clearly, any isometric spaces would have equal number of deformation parameters, so such an isometry would only relate spaces on the same line of Figure~\ref{Tab:Relations}.  As demonstrated in \cite{Cherkis:2012qs}, number of the moduli space deformations is the number of relevant parameters in the monowall problem and it equals to $\#{\rm Perim}(N)-3$, integer perimeter length of $N$ minus three.  In the next section we formulate relation between monopole walls with different, non $GL(2,\mathbb{Z})$ related, Newton polygons that induces isometry of their moduli spaces.

\subsection{Additional Equivalence}
Given any abelian monopole wall $(a,\phi)$ we consider a map acting on all monopole walls:
\begin{equation}
\label{MonTrans}
(A,\Phi)\mapsto(A+a\mathbb{I},\Phi+\phi\mathbb{I}),
\end{equation}
where $\mathbb{I}$ is the identity matrix.  The abelian spectral curve $\Sigma_x^{(a,\phi)}$ is given by some rational function of $s$, so that $\Sigma_x^{(a,\phi)}: t=P(s)/Q(s).$  Then, if the original spectral curve of $(A,\Phi)$ was given by $G_{(A,\Phi)}(s,t)=0$, the spectral curve $\Sigma_x^{(A+a,\Phi+\phi)}$ of $(A+a\mathbb{I},\Phi+\phi\mathbb{I})$ is given by $G_{(A+a,\Phi+\phi)}(s,t Q(s)/P(s))=0.$  What is equally important is that the trace part of any monopole wall is completely determined by the boundary and the singularity data and is independent of its moduli.  As it lies in the center of the algebra, it plays no role in the moduli space metric computation, thus the transform of Eq.~\eqref{MonTrans}, while changing the spectral curve, acts isometrically on the moduli spaces.

The spectral curve of Figure~\ref{Fig:l2_2222}, for example, has the form $L(s)+M(s)t+R(s) t^2=0,$ with $L(s), M(s),$ and $R(s)$ some quadratic polynomials.  Adding a $U(1)$ monowall with the spectral curve $t=R(s)$ amounts to the substitution $t\rightarrow t/R(s).$  Adding any $U(1)$ solution does not alter the moduli space metric, while the resulting spectral curve is now given by $L(s)R(s)+M(s)t+t^2=0$ with its Newton polygon of Figure~\ref{Fig:l4}.  Thus the spectral curve $L(s)+M(s)t+R(s) t^2=0$ is mapped to $L(s)R(s)+M(s)t+t^2=0$ via the substitution $t\rightarrow t/R(s).$ This is the {\em total transform}, it maps a monopole wall with $n_+$ positive and $n_-$ negative singularities to a monopole wall with $(n_++n_-)$ only negative singularities.

If $r$ is one of the roots of $R(s)$ then adding a single Dirac monowall with negative singularity at $s=r$ amounts to making a substitution $t\rightarrow t/(s-r).$ It puts the spectral curve in the form $L(s)(s-r)+M(s) t+\frac{R(s)}{s-r} t^2=0$ with the Newton polygon of Figure~ \ref{Fig:l3_3212}.  We call this a {\em partial transform}.

We conclude that all moduli spaces of monopole walls corresponding to the Newton polygons of line $p=8$ of Table~\ref{Tab:Relations} are isometric.  

Now we apply the same argument to each line of that table.  
\begin{table}[htbp]
  \centering
\null\hfill   \subfloat[$E_6$]{\label{Fig:8}\includegraphics[width=0.17\textheight]{GI_l3_333}}\hfill
      \subfloat[$E_0=1$]{\label{Fig:2}\includegraphics[width=0.17\textheight]{GI_l1_Tri}} \hfill\null
   
\null\hfill     \subfloat[$E_5={\rm Spin}(10)$]{\label{Fig:7}\includegraphics[width=0.17\textheight]{GI_l2_2222}}
\hfill   \subfloat[$E_1=SU(2)$]{\label{Fig:3a}\includegraphics[width=0.17\textheight]{GI_l1_Rombus}}
\quad    \subfloat[$\widetilde{E}_1=U(1)$]{\label{Fig:3ap}\includegraphics[width=0.17\textheight]{GI_l1_QuadSym}}

\null\hfill     \subfloat[$E_4=SU(5)$]{\label{Fig:6}\includegraphics[width=0.17\textheight]{GI_l2_21112}}
\hfill       \subfloat[\hbox{$E_2=SU(3)\times U(1)$}]{\label{Fig:4}\includegraphics[width=0.17\textheight]{GI_l1_Pent}} 
\hfill\null   
   
   \subfloat[$E_3=SU(3)\times SU(2)$]{\label{Fig:5}\includegraphics[width=0.18\textheight]{GI_l2_21111}}
 \caption{All monopole walls with four moduli and distinct moduli spaces.}
  \label{FinalList}   
\end{table}

For $p=7$ line the argument is exactly the same with $L(s)$ and $M(s)$ quadratic and $R(s)$ linear, so all of these spaces are isometric to each other. 

For $p=6$ the spectral curve for the Newton polygon $(2,1,1,2)$ (second in  $p=6$ line  of Table~\ref{Tab:Relations}) has $L(s)$ quadratic and $M(s)$ and $R(s)$ linear; the above transformation maps it to the spectral curve of the Newton polygon $(3,1,2)$ (the fourth one on $p=6$ line of Table~\ref{Tab:Relations}).  

The spectral curve of the hexagon, on the other hand, has the form $L_1(s)+M_1(s)t+s N_1(s)t^2=0$ and, via the transformation $t\rightarrow t/N_1(s),$ it is mapped to a curve with the Newton polygon $(2,1,1,1,1)$ (the third one on line $p=6$ of Table~\ref{Tab:Relations}).  

It remains to relate the two pairs.  We choose to consider $(2,1,1,1,1)$ and $(2,1,1,2).$ To begin with, we interchange the two axes by applying $\left(\begin{smallmatrix} 
0 & 1 \\ 
1 &  0
\end{smallmatrix}\right). $  The spectral curves of the resulting Newton polygons are respectively $L_1(s)+M_2(s)t+R_1(s)t^2=0$ and $P_2(s)+Q_2(s)t+t^2=0$ and they are related by the total transform.

Thus all moduli spaces corresponding to perimeter six, $p=6,$ Newton polygons are also isometric to each other.

For $p=5$ we already established the equivalence of the two pentagons \ref{Fig:Pent} and \ref{Fig:PentS}.  The spectral curve of the latter has the form $L_1(s)+M_1(s)t+sN_1(s)t^2=0$ and by the total transform $t\rightarrow t/N_1(s)$ is mapped to the spectral curve $L_1(s)N_1(s)+M_1(s)t+st^2=0$ with the Newton polygon $(2,1,1,1)$ of Figure~\ref{Fig:l2_2111}.  Thus all $p=5$ Newton polygons have isometric moduli spaces.

So far all Newton polygons with a given integer perimeter length had isometric moduli spaces.  In other words, each line of Table~\ref{Tab:Relations}, besides $p=4$ line, corresponds to one distinct family of moduli spaces.  For $p=4$ the situation is different.  Of the three Newton polygons on $p=4$ line of Table~\ref{Tab:Relations} the first and last are equivalent via the total transform.  The middle one -- the symmetric quadrilateral -- is distinct.

We end up with a list of only eight monopole wall spaces of dimension four.  These appear in Table~\ref{FinalList}.   

This list relates to the classification of five-dimensional superconformal field theories with one-dimensional Coulomb branch of vacua of \cite{Seiberg:1996bd,Douglas:1996xp,Morrison:1996xf}.  Table~\ref{FinalList} lists the global symmetry groups of the corresponding theories.  This relation is not coincidental, as we explain in the following section.

\section{Relation to Gauge Theories and Calabi-Yau Moduli Spaces}\label{Sec:GTHCY}
\subsection{Five-dimensional Theories}
In \cite{Seiberg:1996bd} Seiberg identified superconformal five-dimensional field theories with $E_n, n\leq 8,$ global symmetries. Heterotic string theory view of these theories appeared in \cite{Ganor:1996xd}.   For low values of $n$ these global symmetry groups are:
$
E_1=SU(1),
E_2=SU(1)\times U(1),
E_3=SU(3)\times SU(2),
E_4=SU(5),$ and
$
E_5={\rm Spin}(10);
$
while $E_6, E_7,$ and $E_8$ are the exceptional ones.  
In \cite{Douglas:1996xp,Morrison:1996xf} two more theories were added to this list with 
$
\widetilde{E}_1=U(1)$ 
and
$
E_0=1.
$
Relevant deformations of $E_{N_f+1}$ theory are interpreted as the supersymmetric $SU(2)$ gauge theories with $N_f$ quarks\footnote{As Table~\ref{FinalList} and the geometric engineering \cite{Douglas:1996xp} indicate, the number of quarks relates to the integer perimeter length $p$ of the Newton polygon via $N_f=p-4.$} with masses $m_i, i=1,\ldots,N_f.$ 

Even though these five-dimensional theories are non-renormalizable and should rather be viewed as an intermediate effective description of any of the string realizations mentioned below, they give a good description of the moduli space of vacua. 
Namely, according to  \cite{Seiberg:1996bd}, they have one-dimensional Coulomb branch with the metric
\begin{align}\label{Eq:linear}
\left(t_0+16\phi-\sum_{i=1}^{N_f}\big(|\phi-m_i|+|\phi+m_i|\big)\right)d\phi^2,
\end{align}
while their Higgs branches are isometric to the moduli space of $E_n, SO(2k),$ or $SU(m)$ instantons (with the gauge groups determined by the remaining global symmetry at the point where the Higgs branch is intersecting the Coulomb branch).
A  general classification of such theories with higher-dimensional Coulomb branches appeared in \cite{Intriligator:1997pq}.  

When one of the five space-time directions is compact, the Coulomb branch doubles its dimension.  The additional  dimensions correspond to the eigenvalues of the vacuum expectation value of the holomony around the compact direction.  Such gauge theories on $\mathbb{R}^{1,3}\times S^1$ were solved in \cite{Nekrasov:1996cz} with the full quantum-corrected metric on the Coulomb branch given in terms of special geometry.  One loop asymptotic analysis of five-dimensional gauge theory with two periodic directions was carried out in \cite{arXiv:1107.2847} in complete agreement with Eq.~\eqref{Eq:linear}.

	There is a  relation, discovered by Seiberg and Witten \cite{Seiberg:1996nz}, between quantum vacua of super-Yang-Mills with eight real supercharges in three dimensions and classical monopoles on $\mathbb{R}^3.$ For super-QCD with $n$ quarks \cite{Cherkis:1997aa}, the appropriate monopoles are those with $n$ Dirac singularities.  In both cases, realizing the relevant gauge theory via the Chalmers-Hanany-Witten brane configuration \cite{Chalmers:1996xh,Hanany:1996ie}  makes the relation to the dynamics of monopoles transparent.  Along the similar lines \cite{Cherkis:2000ft}, super-QCD with eight supercharges on $\mathbb{R}^3\times S^1$ is related to periodic monopoles with Dirac singularities.  Thus, from this Chalmers-Hanany-Witten point of view, it is not surprising that the doubly periodic monopoles with singularities relate to super-QCD on $\mathbb{R}^3\times T^2.$  For more detailed reasoning with concrete brane configurations and the string theory duality chain see \cite{Cherkis:2012qs} or the diagrams of Sec.~\ref{Sec:Limits} below.

\subsection{Calabi-Yau Moduli Spaces}
The five-dimensional theories of \cite{Seiberg:1996bd} can be realized in string theory either by considering a D4-brane probe in type I$'$ string theory on $\mathbb{R}/\mathbb{Z}_2$ with $N_f$ D8-branes \cite{Ganor:1996xd} or via  geometric engineering of \cite{Katz:1996fh} by compactifying M theory on a Calabi-Yau manifold with a smooth four-cycle  $\mathbb{S}$ (of complex dimension two) shrinking to a point \cite{Douglas:1996xp, Seiberg:1996bd}.  Equally  relevant to our monowall picture is the fact that the same theories appear as effective theories on M theory five-brane wrapped on a curve  in $\mathbb{C}^*\times\mathbb{C}^*$ (the same curve as the monowall spectral curve) and as theories on $(p,q)$-networks of five-branes  \cite{Aharony:1997ju,Aharony:1997bh} (see \cite{Bergman:2013aca} for a recent discussion).  

In the geometric engineering picture \cite{Katz:1996fh}, the shrinking surface inside a Calabi-Yau space has to be a del Pezzo surface and it is  Gorenstein.  The local Calabi-Yau geometry is that of the total space of the canonical bundle of $\mathbb{S}$.   All Gorenstein toric del Pezzo surfaces are in one-to-one correspondence with reflexive convex polygons. (In fact this holds in general dimension \cite{BJ}.) Thus, it is not surprising that our intermediate result above in Figure~\ref{Tab:Relations}, is similar to the `del Pezzo tree'  \cite[Fig. 1]{CKP}.  All $E_n$ theories are geometrically engineered by compactifying M theory on a local Calabi-Yau that is a canonical bundle to a del Pezzo surface.  Whenever this del Pezzo is toric, it corresponds to a monopole wall, with the toric diagram of the former being the Newton polygon of the latter. 

This correspondence extends to the level of the moduli spaces. The map between the moduli and parameters of the gauge theory and M theory and string theory descriptions is explored in detail in \cite{Douglas:1996xp, arXiv:1107.2847}.  Viewing the del Pezzo surface $\mathbb{S}$ as a fibration over a projective line $\mathbb{P}^1_B,$ the generic fiber is $\mathbb{P}^1_f$ (see \cite{Douglas:1996xp}) and  each singular fiber is a pair of $\mathbb{P}^1$'s intersecting at a point.
Monowall parameters correspond to 1) the size of the base $\mathbb{P}^1_B$ (which corresponds to the  coupling of the five-dimensional quantum theory) and 2) the difference of the sizes of the two $\mathbb{P}^1$'s in each special fiber (which correspond to the masses of the five-dimensional theory matter multiplets).  The monowall modulus corresponds to the size of the generic fiber $\mathbb{P}^1_f.$  So far, considering the Calabi-Yau geometries (with fixed parameters) the K\"ahler moduli space is one-dimensional. The corresponding monowall has four moduli. What are the remaining three periodic moduli?

Since we are interested in a five-dimensional theory on $\mathbb{R}^{1,2}\times T^2$, the relevant M theory compactification to engineer it is on the direct product of a Calabi-Yau space with a two-torus.  For the two-torus being a direct product of two circles, $T^2=S^1_S\times S^1_R, $ the remaining three moduli are 
$\int_{\mathbb{P}^1_f\times S^1_S} C_3,
\int_{\mathbb{P}^1_f\times S^1_R} C_3,$ and $\int_{\mathbb{S}\times T^2} C_6.$
We can view any one of the first two, together with the fiber size modulus to give the complexification of the 
Calabi-Yau K\"ahler structure moduli space.  The remaining two are coordinates on a torus  fibration over it. Selecting  either $\int_{\mathbb{P}^1_f\times S^1_S} C_3$ or 
$\int_{\mathbb{P}^1_f\times S^1_R} C_3$ to complexify the size of $\mathbb{P}^1$ gives the choice of two complex structures on the same moduli space. These are the same two complex structures on the moduli space that emerge from the spectral curve of a monowall in the $x$- or $y$-direction.  Since the moduli space is hyperk\"ahler, the moduli come in multiples of four, the deformation parameters, however, come in triplets $(M,p,q).$  Let us identify these in terms of the  Calabi-Yau geometry. $M$ corresponds to the difference in sizes of the two $\mathbb{P}^1$'s, say $\mathbb{P}^1_A$ and $\mathbb{P}^1_B$ of each singular fiber.  While $p$ and $q$ correspond to the difference in $C_3$ fluxes:
$\int_{(\mathbb{P}^1_A-\mathbb{P}^1_B)\times S^1_S} C_3$ and $\int_{(\mathbb{P}^1_A-\mathbb{P}^1_B)\times S^1_R} C_3.$

Compactifying M theory on one of the two circles, we are left with a type IIA theory on the same Calabi-Yau space product with the remaining circle.  The size of $\mathbb{P}^1_f$ is the real modulus that is complexified by adding as imaginary component $\int_{\mathbb{P}^1_f} B^{NS}_2.$  The torus fiber coordinates are   $\int_{\mathbb{P}^1_f\times S^1} C_3^{RR}$ and $\int_{\mathbb{S}\times S^1} C_5^{RR}.$  

Equivalently, \cite{arXiv:1107.2847}, the same moduli space can be realized as the moduli space of type IIB string theory on the same Calabi-Yau space with the noncompact modulus being the size of the fiber $\mathbb{P}^1_f$ and  three periodic moduli  $\int_{\mathbb{P}_f}C^{RR}_2, \int_{\mathbb{P}^1_f}B^{NS}_2$, and $\int_{\mathbb{S}} C^{RR}_4.$

In the mirror \cite{Katz:1997eq}, type IIA version, the relevant Calabi-Yau space $W$ is written directly in terms of the spectral curve of the monowall. Type IIA theory is compactified on $W\times S^1.$ If the monowall spectral curve is $\Sigma_x: \{(s,t)\in\mathbb{C}^*\times\mathbb{C}^*\, |\, G(s,t)=0\},$ then the mirror Calabi-Yau space is $W:\{(s,t,u,v)\in\mathbb{C}^*\times\mathbb{C}^*\times\mathbb{C}\times\mathbb{C}\, |\, uv=G(s,t)\}.$  In our case $\Sigma_x$ is genus one with a number of punctures.  Choose  generators of $\pi_1(\Sigma_x)$ so that some correspond to the punctures while  $\alpha$ and $\beta$ are the remaining two generators, corresponding to the genus.  The compact Lagrangian three-cycles of $W$, $\Gamma_\alpha$ and $\Gamma_\beta$, correspond to the cycles $\alpha$ and $\beta$, as in \cite{Katz:1996fh}.
Now, it is the complex structure moduli space of $W$ that is relevant; it is the space of curves $\Sigma_x$ and it forms the base of our monowall moduli space. While the fiber is the intermediate Jacobian of $W$ with coordinates $\int_{\Gamma_\alpha} C^{RR}_3$ and $\int_{\Gamma_\beta} C^{RR}_3$.

\subsection{String Theory Dualities and the Two Spectral Curves}\label{Sec:Limits}

There are three distinct limits to consider, each has its interpretation as a monopole, quantum field theory, and a spectral curve degeneration.  Each poses new interesting problems for monowall and field theory interpretation. 
String theory dualities  relate these tree points of view via the following diagram:


\resizebox{0.95\hsize}{!}{
$\xymatrix{
 \shortstack{\hfill\txt{M5-brane wrapped on}\\ \hfill\txt{$\Sigma_x$ spectral curve}\\ 
 {\crug{a}. \begin{tabular}{l|ccccccccccc} 
	  M    & 0 &1 &2 &3 &\crug{4}&\crug{5}&6 &7 &8 &9 &\raisebox{.6pt}{\textcircled{\raisebox{-.4pt} {$\mbox{\fontsize{9}{10}\selectfont $10$}$}}} \\ \hline
	{\footnotesize 2} M5 & x &x &x &x &x         & x       &   &   &    &   &    \\
	M5 & x &x &x &   &            &x       & x &  &    &   &x\\ 
	\end{tabular}}
	\ar@/^2pc/[dr]^{S^1_M=S^1_{10}}}
	\POS+LD\ar@/_3pc/[dddd]+UL_{S^1_M=S^1_5} & \\
\shortstack{\hfill\txt{(p,q)-network or} \\ \hfill\txt{5D QFT on $\mathbb{R}^3\times T^2$}.\\
{\crug{b}. \begin{tabular}{l|cccccccccc} 
  IIB    & 0 &1 &2 &3 &\crug{4}&\crug{5}&6 &7 &8 &9  \\ \hline
{\footnotesize 2} NS5 & x &x &x &x &x         & x       &   &   &    &         \\
{\footnotesize k} D5 & x &x &x &   & x           & x        & x &  &    &  \\ 
\end{tabular}}}\ar@{<->}[r]^{T_4}\ar@{<->}_(0.4)S[d]
&
{\crug{f}. \begin{tabular}{l|cccccccccc} 
 IIA     & 0 &1 &2 &3 &\crug{4}&\crug{5}&6 &7 &8 &9  \\ \hline
{\footnotesize 2} NS5 & x &x &x &x &x         & x       &   &   &    &         \\
{\footnotesize k} D4 & x &x &x &   &            &  x       & x &  &    &  \\ 
\end{tabular}}\ar@{<->}[d]_{ST_5}\\
\shortstack{\hfill\txt{(q,p)-network or}\\
\hfill\txt{EM dual 5D QFT on $\mathbb{R}^3\times T^2$}\\ 
{\crug{c}. \begin{tabular}{l|cccccccccc} 
  IIB    & 0 &1 &2 &3 &\crug{4}&\crug{5}&6 &7 &8 &9  \\ \hline
{\footnotesize 2} D5 & x &x &x &x &x         & x       &   &   &    &         \\
{\footnotesize k} NS5 & x &x &x &   & x      & x        & x &  &    &  
\end{tabular}}} 
\ar@{<->}[d]_(0.45){ST_{45}}
& 
\shortstack{\hfill\txt{Monowall}\\
{\crug{g}. \begin{tabular}{l|cccccccccc} 
  IIB    & 0 &1 &2 &3 &\crug{4}&\crug{5}&6 &7 &8 &9  \\ \hline
{\footnotesize 2} D5 & x &x &x &x &x         & x       &   &   &    &         \\
{\footnotesize k}  D3 & x &x &x &   &            &         & x &  &    &  \\ 
\end{tabular}}}
\ar@{<->}[d]_{T_4}
\POS+LD\ar@{<->}[ddl]+RU_{T_5}
\\
\shortstack{\hfill\txt{Nahm Transformed Monowall}\\
{\crug{d}. \begin{tabular}{l|cccccccccc} 
  IIB    & 0 &1 &2 &3 &\crug{4}&\crug{5}&6 &7 &8 &9  \\ \hline
{\footnotesize 2} D3 & x &x &x &x &          &         &   &   &    &         \\
{\footnotesize k} D5 & x &x &x &   &x           &x       & x &  &    &  \\ 
\end{tabular}}}\ar@{<->}[d]_{T_4}
&
\shortstack{\hfill\txt{D4 wrapped on $\Sigma_y$}\\
{\crug{h}. \begin{tabular}{l|cccccccccc} 
 IIA     & 0 &1 &2 &3 &\crug{4}&\crug{5}&6 &7 &8 &9  \\ \hline
{\footnotesize 2} D4 & x &x &x &x &         & x       &   &   &    &         \\
{\footnotesize k} D4 & x &x &x &   &   x         &          & x &  &    &  \\ 
\end{tabular}}}\\
\shortstack{\hfill\txt{D4 wrapped on $\Sigma_x$}\\
{\crug{e}. \begin{tabular}{l|cccccccccc} 
 IIA     & 0 &1 &2 &3 &\crug{4}&\crug{5}&6 &7 &8 &9  \\ \hline
{\footnotesize 2} D4 & x &x &x &x &x         &        &   &   &    &         \\
{\footnotesize k} D4 & x &x &x &   &            & x         & x &  &    &  \\ 
\end{tabular}}}
 & \shortstack{\hfill\txt{M5-brane wrapped on}\\ \hfill\txt{$\Sigma_y$ spectral curve}\\
 {\crug{i}.	\begin{tabular}{l|ccccccccccc} 
	  M    & 0 &1 &2 &3 &\crug{4}&\crug{5}&6 &7 &8 &9 &\raisebox{.6pt}{\textcircled{\raisebox{-.4pt} {$\mbox{\fontsize{9}{10}\selectfont $10$}$}}} \\ \hline
	{\footnotesize 2} M5 & x &x &x &x &         &x        &   &   &    &   & x   \\
	M5 & x &x &x &   & x           &       & x &  &    &   &x\\ 
	\end{tabular}}}\ar[u]^{S^1_M=S^1_{10}}
\label{Diag:Branes}
}
$
}

\ 

In  string theory brane configurations $\crug{b}, \crug{c}, \crug{d},$ and $\crug{g}$ above relevant gauge theory emerges in the world-volume of the D5-branes \cite{Polchinski:1994fq,Witten:1995im}. Before we proceed with the discussion of various limits, let us recall the standard string theory duality relations and their M theory origin \cite{Townsend:1995kk,Witten:1995ex,Sen:1994fa,Schwarz:1995dk,Schwarz:1995du,Schwarz:1995jq,Schwarz:1995ap,Aspinwall:1995fw}. They are important in understanding all scales, couplings, and sizes involved:
\begin{itemize}
\item
Given M theory with Planck scale $l_{pl}=L$ compactified on a two torus  $S^1_R\times S^1_r,$ there are two ways of describing it as a type IIA string theory.  Each requires selecting an M theory circle. Identifying the first circle $S^1_R$ as the M theory circle, $S^1_M=S^1_R,$ for example, acquire an equivalent description as the type IIA string theory on the remaining $S^1_{r'},$ \cite{Townsend:1995kk, Witten:1995ex}. We denote this  relation by 
$$
\xymatrix{\text{M theory} && \text{Type IIA String Theory}\\
l_{pl}=L,\ S^1_R\times S^1_r\ar^{S^1_M=S^1_R}[rr]&& \stackrel{S^1_{r'}}{\scriptstyle g=\left(\frac{R}{L}\right)^{\sfrac{3}{2}}}, l=\frac{L^{\sfrac{3}{2}}}{R^{\sfrac{1}{2}}}}$$  
Here $g=R^{\frac{3}{2}}$ is the string theory coupling, $l=L^{\sfrac{3}{2}}/R^{\sfrac{1}{2}}$ is the string scale, while the string theory circle radius is $r'=r.$ 
\item
Under the T duality, the type IIA string theory on $S^1_r$, with string coupling $g=\lambda,$ and string scale $l=a$ is equivalent to the type IIB string theory on $S^1_{r'}$ with $r'=a^2/r,$ same string scale $l=a,$ and string coupling $g=\lambda a/r:$ 
	$$\xymatrix{\stackrel{S^1_r}{\scriptscriptstyle g=\lambda, l=a}\ar@{<->}[rr]|T&&\stackrel{S^1_{a^2/r}}{\scriptscriptstyle g=\lambda\frac{a}{r}, l=a}}.$$
\item
S duality $\xymatrix{\ar@{<.>}|S[rr]&&}$ inverts the string coupling and changes the string scale $a$ to $a\sqrt{g},$   \cite{Sen:1994fa}.  In string frame it implies:
$$\xymatrix{\stackrel{S^1_r}{\scriptscriptstyle g=\lambda, l=a}\ar@{<.>}[rr]|S&&\stackrel{S^1_{r}}{\scriptstyle g=\frac{1}{\lambda}, l=a\sqrt{\lambda}}}.$$
\item
Symbol $\xymatrix{\ar@{:}|\times[r]&}$ in $\xymatrix{S^1_r\times S^1_{r'}\ar@{:}|\times[r]&S^1_{r'}\times S^1_r}$  signifies the interchange of the two $S^1$ factors. 
\end{itemize}
Schematically we assemble these facts in the relation  
\begin{align}
\begin{bmatrix} \text{M theory}\\ R\\ \\ r\\ \\ L\end{bmatrix}
\xrightarrow{S^1_M=S^1_R}
\begin{bmatrix} \text{IIA} \\ g^{A}=\left(\frac{R}{L}\right)^{3/2}\\ \\ l^{A}=\frac{L^\frac{3}{2}}{R^\frac{1}{2}}\\ \\ r^A=r\end{bmatrix}
\xrightarrow{T_{r^A}}
\begin{bmatrix}\text{IIB} \\ g^{B}=g^{A}\frac{l^A}{r_A}\\ \\ l^B=l^A \\ \\ r^B=\frac{(l^A)^2}{r^A} \end{bmatrix}
\xrightarrow{S}
\begin{bmatrix}\text{IIB} \\ \tilde{g}^{B}=\frac{1}{g^{B}}\\ \\ \tilde{l}^B=l^B\sqrt{g^{B}} \\ \\ \tilde{r}^B=r^B \end{bmatrix}
\end{align}

Another important fact to note is that the Yang-Mills coupling on a Dp-brane is determined in terms of the string scale $l$ and the string coupling $g$ by the relation $g^2_{YM}=l^{p-3}g,$ \cite{Witten:1995ex}.  In our case, the most relevant is the D5-brane with $g^2_{YM}=l^{2}g.$ 

Consider M theory on a tree-torus $T^3=S^1_A\times S^1_B\times S^1_C$ (that is a product of three circles of respective radii $A, B,$ and $C$) with an M five-brane wrapped on $\Sigma\times S^1_B\times\mathbb{R}^{1,2}\subset T^3\times \mathbb{R}^2_{3,6}\times \mathbb{R}^3\times\mathbb{R}^{1,2},$ where $\Sigma\subset S^1_A\times S^1_C\times\mathbb{R}_3\times\mathbb{R}_6\eqsim\mathbb{C}^*\times\mathbb{C}^*.$ This allows for three type IIA six type IIB descriptions. The sequence of dualities relating these (and most of the equivalent brane configurations  on page~\pageref{Diag:Branes}) is captured by the following diagram\newline
\resizebox{1\hsize}{!}{
$
\scriptsize
\xymatrix@!=2.5pc{
\text{\scriptsize M theory on}& &&&
\text{\normalsize \textcolor{blue}{\crug{a}}}
S^1_A\times S^1_B\times S^1_C, l_{pl}=L 
	\ar[llldd]+(5,7)_(0.5){S^1_M=S^1_A}
\ar@{-->}[dd]^(0.5){S^1_M=S^1_B}
\ar[rrrdd]+(-5,7)^{S^1_M=S^1_C}
&&&\\
&  &&&&&&\\
\txt{\scriptsize IIA on $T^2$\\ at string\\ coupling $g$\\ and string scale $l$} & 
\stackrel{S^1_{B}\times S^1_{C}}{\scriptscriptstyle g=\left(\frac{A}{L}\right)^{\sfrac{3}{2}}, l=\frac{L^{\sfrac{3}{2}}}{A^{\sfrac{1}{2}}}}
\ar@{<->}[dd]_{T_B} 
\ar@{<->}[ddrr]+(4,7)^{T_C}^(1){\normalsize \textcolor{blue}{\crug{c}}}
&&&
{\normalsize \textcolor{blue}{\crug{e}}}
\stackrel{S^1_{A}\times S^1_{C}}{\scriptscriptstyle g=\left(\frac{B}{L}\right)^{\sfrac{3}{2}},l=\frac{L^{\sfrac{3}{2}}}{B^{\sfrac{1}{2}}}} 
\ar@{<-->}[ddll]+U_{T_A}_(1){\normalsize \textcolor{blue}{\crug{d}}} 
\ar@{<-->}[ddrr]+U^{T_C}^(1){\normalsize \textcolor{blue}{\crug{g}}} 
&&&
\text{\normalsize \textcolor{blue}{\crug{f}}}
\stackrel{S^1_{A}\times S^1_{B}}{\scriptscriptstyle g=\left(\frac{C}{L}\right)^{\sfrac{3}{2}},l=\frac{L^{\sfrac{3}{2}}}{C^{\sfrac{1}{2}}}}
\ar@{<->}[ddll]+(-4,7)_{T_A}_(1){\normalsize \textcolor{blue}{\crug{b}}}
\ar@{<->}[dd]_{T_B}
\\
&  &&&&&&\\
\txt{\scriptsize IIB on $T^2$\\ at string \\ coupling  $g$\\ and string scale $l$} & 
\stackrel{S^1_{\frac{L^3}{AB}}\times S^1_{C}}{\scriptscriptstyle g=\frac{A}{B}, l=\frac{L^{\sfrac{3}{2}}}{A^{\sfrac{1}{2}}}}  \ \ \ar@/_2pc/@{<-->}[r]_S
&\stackrel{S^1_{\frac{L^3}{AB}}\times S^1_{C}}{\scriptscriptstyle g=\frac{B}{A}, l=\frac{L^{\sfrac{3}{2}}}{B^{\sfrac{1}{2}}}}
&
**[r] \stackrel{S^1_{B}\times S^1_{\frac{L^3}{AC}}}{\scriptscriptstyle g=\frac{A}{C}, l=\frac{L^{\sfrac{3}{2}}}{A^{\sfrac{1}{2}}}} 
&&
**[l] \stackrel{S^1_{\frac{L^3}{AC}}\times S^1_{B}}{\scriptscriptstyle g=\frac{C}{A}, l=\frac{L^{\sfrac{3}{2}}}{C^{\sfrac{1}{2}}}}
&
\stackrel{S^1_{A}\times S^1_{\frac{L^3}{BC}}}{\scriptscriptstyle g=\frac{B}{C}, l=\frac{L^{\sfrac{3}{2}}}{B^{\sfrac{1}{2}}}} \ar@/_2pc/@{<-->}[r]_S\ 
&\ \ 
\stackrel{S^1_{A}\times S^1_{\frac{L^3}{BC}}}{\scriptscriptstyle g=\frac{C}{B}, l=\frac{L^{\sfrac{3}{2}}}{C^{\sfrac{1}{2}}}}
}
$
}
%
We mark the corresponding brane configurations of the duality table of page \pageref{Diag:Branes} by respective circled letters. For example, M theory five-brane wrapped on the spectral curve $\Sigma_x$ marked  by $\crug{a}$ resides in M theory on top of this diagram.  The monowall configuration on the D5-brane world-volume of $\crug{g}$ is in the second from the right type IIB theory at the bottom row. The five-dimensional theory of $\mathbb{R}^3\times T^2$ in the world-volume of the $(p,q)$-five-brane network $\crug{c}$ is in third from the left the type IIB theory in the bottom row of the diagram. 

The complete picture of the action of $T$ and $S$ duality on M theory on a three torus is represented by the following diamond diagram.  \newline
\resizebox{1\hsize}{!}{
$
\scriptsize
\xymatrix@!=2pc 
@R=2pc
@C=3pc
{
&\txt{\tiny M theory\\ \tiny on $T^3$}&\txt{\tiny IIA on $T^2$ at\\\tiny  string coupling $g$\\ \tiny and string scale $l$} & \txt{\tiny IIB on $T^2$ at\\ \tiny string  coupling $g$\\ \tiny and string scale $l$}& \txt{\tiny IIB on $T^2$ at\\  \tiny string  coupling $g$\\ \tiny and string scale $l$}& \txt{\tiny IIA on $T^2$ at\\ \tiny string coupling $g$\\ \tiny and string scale $l$}&\txt{\tiny M theory\\ \tiny on $T^3$} \\
&&&**[l]\stackrel{\cdot\times S^1_{\frac{L^3}{AB}}\times S^1_{C}}{\scriptscriptstyle g=\frac{A}{B}, l=\frac{L^{\sfrac{3}{2}}}{A^{\sfrac{1}{2}}}}
\ar@{:}|\times[r]
\ar@{<.>}@/^1.8pc/[ddd]|(0.4)S
&
**[r]\stackrel{\cdot\times S^1_{C}\times S^1_{\frac{L^3}{AB}}}{\scriptscriptstyle g=\frac{A}{B}, l=\frac{L^{\sfrac{3}{2}}}{A^{\sfrac{1}{2}}}}
&&\\
&&\stackrel{\cdot\times S^1_{B}\times S^1_{C}}{\scriptscriptstyle g=\left(\frac{A}{L}\right)^{\sfrac{3}{2}}, l=\frac{L^{\sfrac{3}{2}}}{A^{\sfrac{1}{2}}}}
\ar@{<->}[ur]|(0.6){T_B} 
\ar@{<->}[dr]|(0.37){T_C}^(0.63){\normalsize \textcolor{blue}{\crug{c}}}
&
&&\stackrel{\cdot\times S^1_{\frac{L^3}{AC}}\times S^1_{\frac{L^3}{AB}}}{\scriptscriptstyle g=\frac{L^{\sfrac{3}{2}}A^{\sfrac{1}{2}}}{BC}, l=\frac{L^{\sfrac{3}{2}}}{A^{\sfrac{1}{2}}}}
\ar@{<->}[ul]|(0.6){T_B} 
\ar@{<->}[dl]|(0.4){T_C}
& \\
\txt{\tiny Dual\\ \tiny 5D QFT}&&&
**[l]
\stackrel{\cdot\times S^1_{B}\times S^1_{\frac{L^3}{AC}}}{\scriptscriptstyle g=\frac{A}{C}, l=\frac{L^{\sfrac{3}{2}}}{A^{\sfrac{1}{2}}}}
\ar@{:}|(0.4)\times[r]
&\stackrel{\cdot\times S^1_{\frac{L^3}{AC}}\times S^1_{B}}{\scriptscriptstyle g=\frac{A}{C}, l=\frac{L^{\sfrac{3}{2}}}{A^{\sfrac{1}{2}}}} 
\POS+DL\ar@{<.>}@/_/|S @(dl,ur)
[ddddl]+UR
&&\\
\txt{\tiny Nahm Dual \\ \tiny Monopole}&&&
**[l]\stackrel{S^1_{\frac{L^3}{AB}}\times\cdot\times S^1_{C}}{\scriptscriptstyle g=\frac{B}{A}, l=\frac{L^{\sfrac{3}{2}}}{B^{\sfrac{1}{2}}}}
\ar@{}[l]|(0.6){\normalsize \textcolor{blue}{\crug{d}}} 
\ar@{:}|\times[r]
&
**[r]\stackrel{S^1_{C}\times\cdot\times S^1_{\frac{L^3}{AB}}}{\scriptscriptstyle g=\frac{B}{A}, l=\frac{L^{\sfrac{3}{2}}}{B^{\sfrac{1}{2}}}}
&&\\
\txt{\tiny Spectral\\ \tiny Curves} 
&
**[l]{\begin{array}{c} l_{pl}=L, \\S^1_A\\\times\\ S^1_B\\\times\\ S^1_C\\ \\
{\normalsize \textcolor{blue}{\crug{a}}}\end{array}} 
\ar[uuur]|{S^1_M=S^1_A}
\ar[dddr]|{S^1_M=S^1_C}^(0.85){\normalsize \textcolor{blue}{\crug{f}}}
\POS+<1pc,0pc>\ar@{->}[r]^(0.35){S^1_M=S^1_B}
& 
**[r]\stackrel{S^1_{A}\times\cdot\times S^1_{C}}{\scriptscriptstyle g=\left(\frac{B}{L}\right)^{\sfrac{3}{2}}, l=\frac{L^{\sfrac{3}{2}}}{B^{\sfrac{1}{2}}}}
\ar@{<->}[ur]|(0.6){T_A}
\ar@{}[u]^(0.3){\normalsize \textcolor{blue}{\crug{e}}}
\ar@{<->}[dr]|(0.37){T_C}^(0.7){\normalsize \textcolor{blue}{\crug{g}}}
&
&
&
**[l]{\normalsize \textcolor{blue}{\crug{h}}}
\stackrel{S^1_{\frac{L^3}{BC}} \times\cdot\times S^1_{\frac{L^3}{AB}}}{\scriptscriptstyle g=\frac{L^{\sfrac{3}{2}}B^{\sfrac{1}{2}}}{AC}, l=\frac{L^{\sfrac{3}{2}}}{B^{\sfrac{1}{2}}}}
\ar@{<->}[ul]|(0.63){T_A} \ar@{<->}[dl]|(0.35){T_C}
&
{\begin{array}{c} l_{pl}=\frac{L^2}{\sqrt[3]{ABC}},\\ S^1_{\frac{L^3}{BC}}\\ \times\\ S^1_{\frac{L^3}{AC}}\\ \times\\ S^1_{\frac{L^3}{AB}}\\ \\{\normalsize \textcolor{blue}{\crug{i}}} \end{array}} 
\ar[uuul]|{S^1_M=S^1_A}
\ar[dddl]|{S^1_M=S^1_C}
\POS-<1pc,0pc>\ar@{->}[l]_(0.3){S^1_M=S^1_B}\\
\txt{\tiny Monopole}&&&
**[l]\stackrel{S^1_{A}\times\cdot\times S^1_{\frac{L^3}{BC}}}{\scriptscriptstyle g=\frac{B}{C}, l=\frac{L^{\sfrac{3}{2}}}{B^{\sfrac{1}{2}}}}
\ar@{:}|\times[r]
&
**[r]\stackrel{S^1_{\frac{L^3}{BC}}\times\cdot\times S^1_{A}}{\scriptscriptstyle g=\frac{B}{C}, l=\frac{L^{\sfrac{3}{2}}}{B^{\sfrac{1}{2}}}}
\POS+LD\ar@{<.>}@/_1.2pc/[ddd]|(0.5)S
&&\\
\txt{\tiny 5D QFT}&&&
**[l]\stackrel{S^1_{\frac{L^3}{AC}}\times S^1_{B}\times\cdot}{\scriptscriptstyle g=\frac{C}{A}, l=\frac{L^{\sfrac{3}{2}}}{C^{\sfrac{1}{2}}}}
\ar@{:}|\times[r]
&**[r]\stackrel{S^1_{B}\times S^1_{\frac{L^3}{AC}}\times\cdot}{\scriptscriptstyle g=\frac{C}{A}, l=\frac{L^{\sfrac{3}{2}}}{C^{\sfrac{1}{2}}}}
&&\\
&&
\stackrel{S^1_{A}\times S^1_{B}\times\cdot}{\scriptscriptstyle g=\left(\frac{C}{L}\right)^{\sfrac{3}{2}}, l=\frac{L^{\sfrac{3}{2}}}{C^{\sfrac{1}{2}}}}
\ar@{<->}[ur]|(0.57){T_A}_(0.8){\normalsize \textcolor{blue}{\crug{b}}}
 \ar@{<->}[dr]|(0.43){T_B}&
&
&\stackrel{S^1_{\frac{L^3}{BC}} \times S^1_{\frac{L^3}{AC}}\times\cdot}{\scriptscriptstyle g=\frac{L^{\sfrac{3}{2}}C^{\sfrac{1}{2}}}{AB}, l=\frac{L^{\sfrac{3}{2}}}{C^{\sfrac{1}{2}}}}
\ar@{<->}[ul]|(0.63){T_A} 
\ar@{<->}[dl]|(0.4){T_B}
&\\
&&&
\stackrel{S^1_{A}\times S^1_{\frac{L^3}{BC}}\times\cdot}{\scriptscriptstyle g=\frac{C}{B}, l=\frac{L^{\sfrac{3}{2}}}{C^{\sfrac{1}{2}}}}
\ar@{:}|\times[r]
&\stackrel{S^1_{\frac{L^3}{BC}}\times S^1_{A}\times\cdot}{\scriptscriptstyle g=\frac{C}{B}, l=\frac{L^{\sfrac{3}{2}}}{C^{\sfrac{1}{2}}}}
&&
}
$
}

The full U-duality group \cite{Hull:1995mz}  acting on M theory on a $d$-dimensional torus $T^d$ is $E_{d(d)}$ (see \cite{Obers:1998fb} for references and the broad picture).  In our case of M theory on $T^3$ the U-duality group is $E_{3(3)}=SL(3,\mathbb{Z})\times SL(2,\mathbb{Z}).$  The first component, $SL(3,\mathbb{Z}),$ is the modular group of the torus $T^3$ contains elements interchanging the rows of the above diagram.  S element of the second,  $SL(2,\mathbb{Z}),$ factor acts on the diagram as a reflection with respect to the vertical axis.

Note, that the complete description of the original monowall or the full five-dimensional theory with two finite periodic directions involves two spectral curves $\Sigma_x$ and $\Sigma_y$ in $\mathbb{C}^*\times\mathbb{C}^*.$ These appear as M five-brane curves in, respectively, configurations $\crug{a}$ and $\crug{i}.$

Now we have a few interesting limits to mention:
\begin{description}
\item[Tropical limit:]
Tracing the above diamond diagram, a monowall is realized on the brane configuration $\crug{g}$ in the world-volume of the D5-brane.  The Yang-Mills coupling $g_{YM}$ on this brane satisfies $g^2_{YM}=l^2 g=L^3/C.$  We are holding it fixed.  Under $T$ duality on the second circle it becomes a D4-brane of configuration $\crug{e}$ wrapped on a curve $\Sigma_x\subset S^1_A\times S^1_C\times\mathbb{R}_3\times\mathbb{R}_6\simeq\mathbb{C}^*\times\mathbb{C}^*.$  In M theory $\crug{a}$ this is the M five-brane with world-volume $\mathbb{R}^{1,2}_{0,1,2}\times\Sigma_x\times S^1_B\subset
\mathbb{R}^{1,2}_{0,1,2}\times S^1_A\times S^1_B\times S^1_C\times\mathbb{R}_3\times\mathbb{R}_6\times\mathbb{R}^3_{7,8,9}.$ The tropical limit is $A=C=h\rightarrow 0$.  To hold the Yang-Mills coupling of the monowall fixed we need $L^3\sim h.$

Tracing the diamond diagram, the monowall $\crug{g}$ has periods $A=h$ and $1/B$ and in this limit one of the circles degenerates to zero size and the monowall becomes the Hitchin system on a cylinder.  The five-dimensional theory $\crug{b}$, on the other hand, is now on $\mathbb{R}^{1,2}_{0,1,2}\times S^1_{1/h}\times S^1_{B}$, and in this limit becomes the four-dimensional Seiberg-Witten theory with one periodic direction and finite coupling $g_{YM}^2=L^3/A$.

Note, that in this limit the spectral curve $\Sigma_y,$ on which the M five-brane of $\crug{i}$ is wrapped, remains regular $\Sigma_y\subset S^1_{1/B}\times S^1_{1/B}\times\mathbb{R}_3\times\mathbb{R}_6\simeq\mathbb{C}^*\times\mathbb{C}^*,$ while the middle circle $S^1_{B'}=S^1_{L^3/(AC)}=S^1_{1/h}$ opens up becoming $\mathbb{R}$ and the Planck scale remains finite $l_{Pl}\sim B^{-\frac{1}{3}}.$ This is the only remaining Seiberg-Witten curve of the theory in this limit, as the other curve $\Sigma_x$ turns into a tropical curve in this limit.

\item[5D field theory limit:]  If we are interested in exploring the five-dimensional quantum theory of $\crug{b}$ without any periodic directions, we are to send $B\rightarrow\infty$ and $\frac{L^3}{AC}\rightarrow\infty.$ This can be  achieved within the above tropical limit by sending $B\rightarrow\infty.$ Since $S^1_B$ was the spectator circle, i.e. it did not involve the spectral curve, the curve $\Sigma_x$ still tends to the tropical limit.  What is new, is that the other spectral curve $\Sigma_y$ involves $S^1_{A'}\times S^1_{C'}=S^1_{1/B}\times S^1_{1/B}$ becomes tropical as well.  This is the reason why our analysis of monowalls in the tropical limit reproduces the Seiberg and Morrison results of \cite{Morrison:1996xf} and the asymptotic metric of \cite{Seiberg:1996bd}.

\item[3D monopole limit:]
The decompactification limit of the monowall torus of $\crug{g}$ is $A\rightarrow\infty$ and $\frac{L^3}{BC}\rightarrow\infty,$ while holding the Yang-Mills coupling on the D5-brane fixed: $g^2_{YM}=L^3/C={\rm Const}.$  This implies $L^3\sim C$ while $A\rightarrow\infty$ and $B\rightarrow 0.$

The five-dimensional space of the field theory $\crug{b}$ degenerates to a three-dimensional space, since both circles in $S^1_{L^3/(AC)}\times S^1_B$ shrink to zero size. The limiting theory is the ${\cal N}=4$ three-dimensional super-QCD of \cite{Seiberg:1996nz}.  The monowall in this limit becomes a monopole in $\mathbb{R}^3$ and the moduli space becomes ALF in complete agreement with \cite{Seiberg:1996nz,Chalmers:1996xh,Hanany:1996ie,Cherkis:2000ft}.
\end{description}

\section{Phase Space of a Monowall}\label{Sec:PhaseSpace}
\subsection{Litvinov-Maslov's Dequantization or Tropical Geometry}

Following \cite{LM}, consider the positive real line $\mathbb{R}_+$ with the standard operations $+:(u,v)\mapsto u+v$ and $\cdot: (u,v)\mapsto u\cdot v.$ The exponential map $e:\mathbb{R}\rightarrow\mathbb{R}_+$  induces operations $\oplus$ and $\odot$ on the full real line $\mathbb{R},$ namely, letting $u=e^a$ and $v=e^b,$ we have 
\begin{align}
+: (e^a,e^b)&\mapsto e^a+e^b=e^{a\oplus b},\\
\cdot:(e^a, e^b)&\mapsto e^a e^b=e^{a\odot b}.
\end{align}
So $a\oplus b=\ln\left(e^a+e^b\right)$ and $a\odot b=a+b.$  More generally, letting $u=e^{\frac{a}{h}}$ and $v=e^\frac{b}{h}$ induces the following operations on $\mathbb{R}:$
\begin{align}
a\oplus_h b&=h \ln\left(e^\frac{a}{h}+e^\frac{b}{h}\right),\\
a\odot_h b&=a+b.
\end{align}
Let us denote the real line $\mathbb{R}$ with these operations by $S_h,$ then $(S_h,\oplus_h,\odot_h)$ is a semiring.  For any positive values of $h$ the semiring $(S_h,\oplus_h,\odot_h)$ is isomorphic  to $(\mathbb{R}_+,+,\cdot)$ with conventional addition and multiplication (under the exponential isomorphism given above sending $a\in S_h$ to $\exp(a/h)$).  What makes the operations $\oplus_h$ and $\odot_h$ interesting, however, is that they have a good limit as $h$ vanishes, namely
\begin{align}
a\oplus_0 b&=\lim_{h\rightarrow 0}h \ln\left(e^\frac{a}{h}+e^\frac{b}{h}\right)=\max (a, b),\\
a\odot_0 b&=a+b.
\end{align}
The semiring $(S_0,\oplus_0,\odot_0)$ differs from $(\mathbb{R}_+,+,\cdot)$, in particular, it is idempotent since $a\oplus_0 a=a.$

Litvinov and Maslov formulated a broad correspondence principle in which $(S_h,\oplus_h,\odot_h)$ with nonzero $h$ is viewed as a quantum object, while $(S_0,\oplus_0,\odot_0)$ is its classical limit; thus the name `dequantization'.  According to this principle various statements over $(S_h,\oplus_h,\odot_h)$ have corresponding statements over $(S_0,\oplus_0,\odot_0).$

For some given Newton polygon $N,$ let $F(u,v)=\mathop{\sum}_{(m,n)\in{N}} F_{m,n} u^m v^n$ be its corresponding family of real polynomials and let us look at the  family of curves $F(u,v)=0$  in $\mathbb{R}_+^2$ with the coefficients $F_{m,n}$ parameterizing this family and explore its dequantization.  To begin with,  we let $F_{m,n}=\exp(f_{m,n}/h)$ and $F(u,v)=\exp(f(a,b)/h),$ then in $S_0$ this algebraic relation becomes $f(a,b)=\max_{(m,n)\in N}\left\{m a+n b+f_{m,n} \right\}.$  An analogue of a real algebraic variety given by $F(u,v)=0$ is the set where the function $f(a,b)=\max_{(m,n)}\left(m a+n b+f_{m,n}\right)$ is not differentiable\footnote{See \cite{IV} for the exact description and the relation to Ragsdale conjecture and Hilbert's 16th problem.}, it is called the tropical curve. In fact, a more elegant form of dequantization by  using hyperfields (see \cite{Viro10} and \cite{Viro11}).

The function $f(a,b)=\max_{(m,n)}\left(m a+n b+f_{m,n}\right)$ is linear on every connected component of the complement of the tropical curve.  A geometric way of viewing this is by considering an arrangement of planes $c=m a+n b+f_{m,n}$ in three-space $\mathbb{R}^3$ with coordinates $(a,b,c).$  Given such a plane arrangement there is a hull or {\em crystal}  which is formed by the closure of the set of all points above all of these planes $\mathcal{C}=\{(a,b,c) | c\geq m a+n b+f_{m,n}\}.$  

Our spectral curves are complex, so a dequantization of $\mathbb{C}$ is in order.  Hyperfields again provide appropriate language, however, for our limited purposes here, the dequantization of the real line as described above will be sufficient.

\subsection{Amoebas, Melting Crystals, and the K\"ahler Potential}
To see clearly what is happening in the limit $\h\rightarrow0$ it is useful to introduce an amoeba \cite{GKZ} associated to a curve $\Sigma=\{(s,t) | s\in\mathbb{C}^*, t\in\mathbb{C}^*, \sum_{(m,n)\in{N}} G_{m,n} s^m t^n=0\}.$  If we view $\mathbb{C}^*\times\mathbb{C}^*$ as a direct product of two cylinders  $\mathbb{C}^*\times\mathbb{C}^*=(\mathbb{R}\times S^1)\times(\mathbb{R}\times S^1)=\mathbb{R}^2\times T^2,$ then we can project the curve $\Sigma\subset \mathbb{C}^*\times\mathbb{C}^*$ onto the first factor in $\mathbb{R}^2\times T^2$ via
\begin{align}
{\rm Log}:\, &\mathbb{C}^*\times\mathbb{C}^*\rightarrow\mathbb{R}^2\\
       & (s,t)\mapsto(\log|s|, \log|t|).
\end{align}
The amoeba $\mathcal{A}$ of the curve $\Sigma$ is the image of this curve under this map: $\mathcal{A}={\rm Log}(\Sigma).$

A slightly different approach to amoebas and tropical varieties is via the Ronkin function \cite{Ronkin}, defined by averaging $\log|G(s,t)|$ of the polynomial $G(s,t)=\sum_{(m,n)\in{N}} G_{m,n} s^m t^n$ along the torus factor:
\begin{align}
R(a,b)=\int_{|x|=a, |y|=b} \log|G(s,t)|\frac{ds}{s}\frac{dt}{t}.
\end{align}
In the vicinity of any point $(a,b)$ such that the torus $T^2_{(a,b)}$ does not intersect the curve $\Sigma$ this function is linear in $a$ and $b$, so one can define the amoeba $\mathcal{A}$ as the set of points where the Ronkin function is not linear.  Moreover, each connected component of the complement of the amoeba is a convex region of the $(a,b)$-plane $\mathbb{R}^2.$ It corresponds to one of the integer points $(m,n)$ of the Newton polygon of $G(s,t),$ and the Ronkin function is linear on this connected component with the slope equal to $(m,n).$

Now let us keep track of the sizes of the two cylinders in the above picture.  $\mathbb{C}^*\times\mathbb{C}^*=\mathbb{R}\times S^1_{\hat{x}}\times\mathbb{R}\times S^1_y,$ where the two circles $S^1_{\hat{x}}$ and $S^1_y$ are parameterized respectively by  coordinates $\hat{x}$ and $y$ each with the same period $2\pi h.$   Then
\begin{align}
s&=e^{(a+i\hat{x})/h}&&\text{and}& 
t&=e^{(b+i y)/h}.
\end{align}
For the curve $G(s,t)=0$ define a rescaled Ronkin function 
\begin{align}
r(a,b):=h R(a,b)=h\int_0^{2\pi h} \int_0^{2\pi h} \log|G(e^{(a+i\hat{x})/h},e^{(b+i y)/h})|\frac{d\hat{x}}{h}\frac{dy}{h}.
\end{align}
In these $(a,b)$ coordinates the region of nonlinearity of $r(a,b)$ shrinks as $h$ is sent to zero, and the tropical variety defined above can be viewed as a classical limit of the amoeba.  It is called the {\em skeleton of the amoeba} $\mathcal{A}.$  Moreover, in the limit 
$\lim_{h\rightarrow 0} r(a,b)=\max_{(m,n)\in{N}} (m a+n b+g_{m,n}),$ where 
$\left|G_{m,n}\right|=\exp\left(\frac{g_{m,n}}{h}\right).$  

Consider the region $H$ above the graph of the function $r(a,b)$
\begin{align}
H=\left\{(a,b,c) | c\geq r(a,b) \right\}.
\end{align}
In the limit $h=0$ this region is exactly the {\em crystal} $\mathcal{C}$ of the plane arrangement $$\left\{ c=m a+n b+g_{m,n} \, \big| \, (m,n)\in{N}\right\},$$ consisting of the closure of all points above all of these planes. For nonzero $h$ the region $H$ is that  same crystal $\mathcal{C}$ with melted corners -- {\em a melted crystal}.  

We shall distinguish the planes corresponding to perimeter points of the Newton polygon and those corresponding to the internal points.  For each perimeter point $(m,n)\in\partial{N}$ with the $G(s,t)$ coefficient $G_{m,n}$ we call the plane $c=m a+n b+\log \left|G_{m,n}\right|$ the $(m,n)$-{\em perimeter plane} or just the {\em perimeter plane}.  For an internal point $(m,n)\in{\rm Int}\, {N}$ we call the plane $c=m a+n b+\log \left|G_{m,n}\right|$ an internal plane.  As defined above for $G(s,t)$, the {\em crystal} $\mathcal{C}$ is the domain above all perimeter and internal planes:
\begin{align}
\mathcal{C}=\left\{(a,b,c) | c\geq m a+n b+\log \left|G_{m,n}\right|, (m,n)\in{N}\right\}.
\end{align}
The {\em complete crystal} $\mathcal{C}_{per}$ is the domain above all of the perimeter planes only:
\begin{align}
\mathcal{C}_{per}=\left\{(a,b,c) | c\geq m a+n b+\log \left|G_{m,n}\right|, (m,n)\in\partial{N}\right\}.
\end{align}
The {\em melted crystal} $H$ is the domain above the Ronkin function:
\begin{align}
H=\{(a,b,c) | c\geq R(a,b) \}.
\end{align}
These definitions and the convexity of the Ronkin function imply that $H\subset\mathcal{C}\subset\mathcal{C}_{per}.$  Let us call the volume of $\mathcal{C}_{per}\setminus H$ -- the region above the perimeter planes and under the Ronkin function --  the {\em melted volume} and denote it by $V_{melt}$, and the volume of $\mathcal{C}_{per}\setminus\mathcal{C}$ the {\em regularized volume} and denote it by $V_{reg}$.  Since $H\subset\mathcal{C}$, $V_{melt}>V_{reg}$ while both are functions of the coefficients $G_{m,n}$ of the polynomial $G(s,t).$

We can now formulate the conjecture that the leading part of the K\"ahler potential on the moduli space of a monopole wall with the Newton polygon ${N}$ is given by the melted volume $V_{melt}.$  The perimeter coefficients are the parameters, and they are held fixed. The internal coefficients $G_{m,n}$ constitute half of the moduli space coordinates and $V_{melt}$ is a function of their absolute values.  This structure of the asymptotic K\"ahler potential implies  existence of asymptotic isometries of the moduli space metric.

\subsection{The Low-dimensional Test}
Let us illustrate the conjecture in the case of a four-dimensional monowall moduli space.  In this case the Newton polygon ${N}$ has only one internal point; by shifts let us arrange this internal point to be at the origin, so that its coefficient, the modulus, is $G_{0,0}.$  The corresponding plane $L_{0,0}$ is horizontal, positioned at the hight $\phi:=\log|G_{0,0}|,$ which is one of the monowall moduli.  Let us denote the remaining three periodic moduli by $\theta_1={\rm Arg}\,G_{0,0}, \theta_2,$ and $\tau.$

As is the case for other monopole moduli spaces with only four moduli, asymptotically the moduli space metric approaches one with a triholomorphic isometry.  Any such asymptotic metric has the form
\begin{align}\label{Eq:Asym}
U d\phi^2+\frac{(d\tau+\omega)^2}{U},
\end{align}
where $U$ is a harmonic function of $\phi,\theta_1,\theta_2$ and $\omega$ is a one-form on the same space. For monopoles on $\mathbb{R}^3$ the function $U$ behaves as $1/r;$ for periodic monopoles it behaves as $\log r.$ In our case of doubly periodic monopoles, with only one noncompact coordinate, $U$ has to be linear and thus has the form $U=c_0+c_1\phi.$  The one form $\omega$ is dual to $U$, i.e. $dU=*d\omega$, which in our case implies e.g. $\omega=c_1\theta_1 d\theta_2.$ Since the $\tau$-circle is to form a fibration over the two-torus of $\theta_1$ and $\theta_2$,  the constant $c_1$ has to be integer.  

According to the conjecture 
\begin{align}
U=\frac{d^2}{d\phi^2} V_{melt},
\end{align}
up to terms exponentially small in $\phi.$  Now let us focus on the leading behavior of $V_{melt}$ for large $\phi:$

1. In the limit of a large internal coefficient the Ronkin function is well approximated by the maximum of the tropical planes; thus in this limit $V_{melt}\simeq V_{reg}.$

2. The cross section $\mathcal{C}_{per}\cap L_{0,0},$ which is the horizontal face of the crystal $\mathcal{C}$, at high modulus approaches a $\log|G_{0,0}|$ multiple of the polar polygon 
\begin{align}\label{Eq:def}
{N}^{\vee}=\{v | (v,w)>-1, \forall w\in{N}\}.
\end{align}
To be exact, at some sufficiently large value of $\phi=\log|G_{0,0}|$ the cross section $\mathcal{C}_{per}\cap L_{0,0}$ is a polygon bounded by lines parallel to the lines in the definition \eqref{Eq:def} of the polar polygon ${N}^{\vee}.$  As we increase $\phi$ and, accordingly, move up the plane $L_{0,0}$, these lines are moved away from the center at  constant rates.  Thus, the larger the value of $\phi$ the less significant the initial positions of the lines become.  Asymptotically, the cross section approaches $\phi {N}^{\vee}.$

3. One other way of seeing the previous statement is by first considering the situation when all the perimeter planes are passing through the origin.  In this case ${\cal C}_{per}$ is a cone, and the volume of that cone below $L_{0,0}$ is exactly $V_{reg}=\frac{1}{3} {\rm Area}_{GKZ}({N}^\vee) \left(\log|C_{0,0}|\right)^3.$ Now, moving one of the perimeter planes up or down by a finite distance changes $V_{reg}$ by amount at most quadratic in $\phi,$ therefore 
\begin{align}
V_{reg}=\frac{1}{3} {\rm Area}_{GKZ}({N}^\vee) \left(\log|G_{0,0}|\right)^3+\mathcal{O}\left((\log|G_{0,0}|)^2\right).
\end{align}
In these expressions we are using the conventions of \cite{GKZ} with the area of a basic simplex $[(0,0),(1,0),(1,0)]$ being equal to 1 instead of 1/2.  Thus ${\rm Area}_{GKZ}$ is twice the conventional area ${\rm Area}:$ ${\rm Area}_{GKZ}=2{\rm Area}.$

Let us add to this two more geometric facts:\newline
Pick's formula
\begin{align}
\frac{1}{2}{\rm Area}_{GKZ}({N}^\vee)={\rm Area}({N}^\vee)={\rm Int}({N}^\vee)+\frac{1}{2} {\rm Perim}({N}^\vee)-1
\end{align}
and
the perimeter relation for reflexive polygons\footnote{See \cite{12} for three intriguing explanations of this fact.} \newline
\begin{align}
{\rm Perim}({N})+{\rm Perim}({N}^\vee)=12.
\end{align}
Thus, in our case with a single internal point, we find that ${\rm Area}_{GKZ}({N}^\vee)={\rm Perim}({N}^\vee)=12-{\rm Perim}({N}),$ and the asymptotic metric of Eq.~\eqref{Eq:Asym} has 
\begin{align}
U=\frac{d^2}{d\phi^2} V_{melt}=2{\rm Area}_{GKZ}(N^\vee)\phi+O(\phi^0)=(24-2{\rm Perim}({N}))\phi+O(\phi^0).
\end{align}

Now we have all of the  needed geometric ingredients in order to compare to the gauge theory computations of \cite{Seiberg:1996bd} and \cite{arXiv:1107.2847}, we recall the relation between the Newton polygon integer perimeter length ${\rm Perim}({N})$ and the number $N_f$ of quarks in the corresponding $SU(2)$ gauge theory: $N_f={\rm Perim}({N})-4.$  This gives
\begin{align}
U=(16-2 N_f)\phi+\text{const},
\end{align}
in perfect agreement with Seiberg's one loop gauge theory result of Eq.~\eqref{Eq:linear}.

\subsection{Secondary Fan and the Phase Space}
For a given monowall, its moduli space depends on the boundary data consisting of the asymptotic behavior at $z\rightarrow\pm\infty$ and the positions of the positive and negative singularities.  We call these parameters.  Parameters are distinct from moduli: if moduli parameterize $L^2$ deformations and give coordinates on the moduli space, the parameters determine the shape of the moduli space and correspond to perturbations of the solution that are not $L^2$.   For generic values of parameters the moduli space is smooth, however, as we change these parameters the space can degenerate and undergo some drastic changes.  We would like to describe the structure of the space of parameters.  In particular, we would like to understand the phase structure, i.e. to describe the walls on which degenerations can occur and the connected domains within which the moduli spaces are diffeomorphic (though not necessarily isometric).  Here we restrict our discussions to the tropical limit, relying heavily on the beautiful combinatorics and geometry of \cite{GKZ}.

\begin{figure}[htbp]
\begin{center}
\centering
\includegraphics[width=0.15\textwidth]{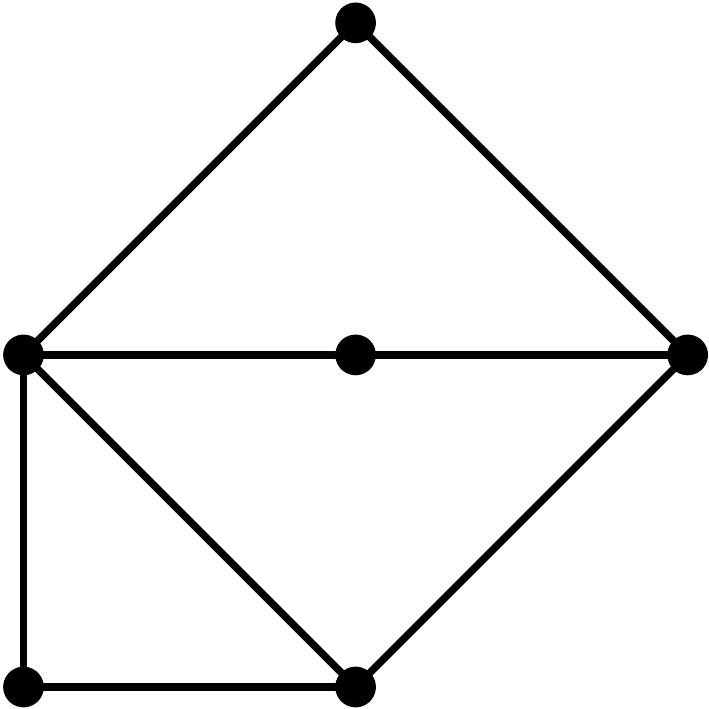}\qquad\qquad
\includegraphics[width=0.15\textwidth]{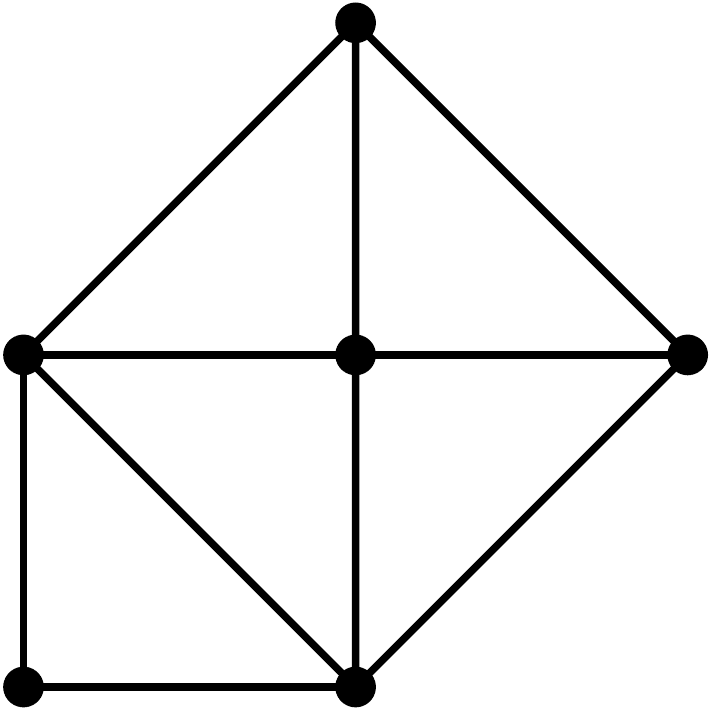}\qquad\qquad
\includegraphics[width=0.15\textwidth]{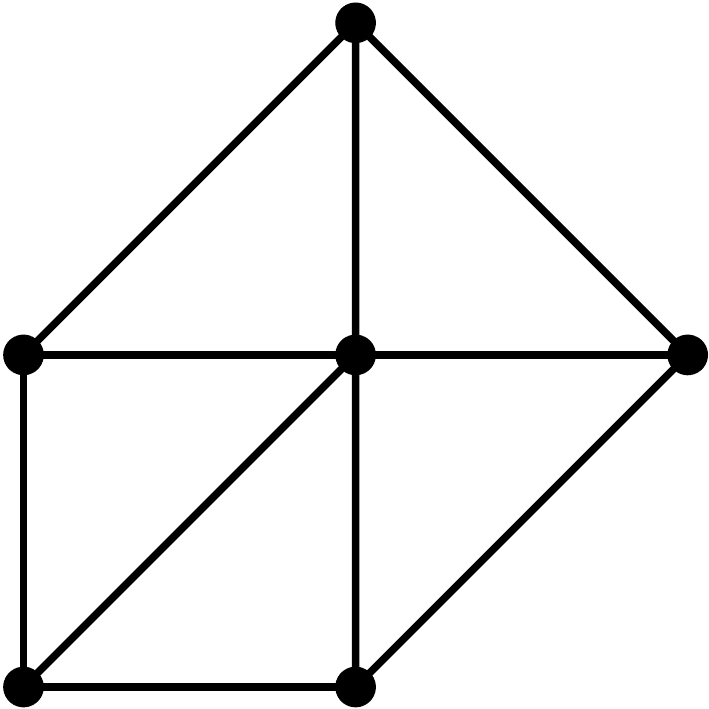} \\
\includegraphics[width=0.15\textwidth]{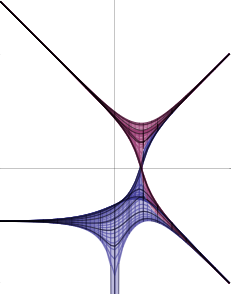}\qquad\qquad
\includegraphics[width=0.15\textwidth]{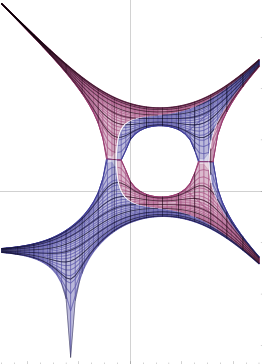}\qquad\qquad
\includegraphics[width=0.15\textwidth]{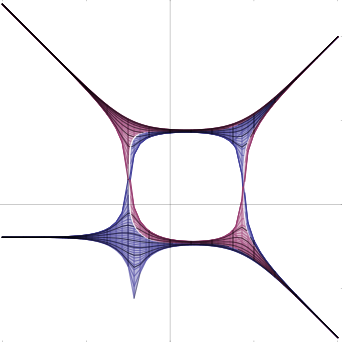}
\caption{Some regular triangulations and Corresponding Amoebas}
\label{AmoebasTriang}
\end{center}
\end{figure}

Near the tropical limit the amoeba is very close to its skeleton.  A generic skeleton, in turn, is dual to some regular triangulation of the Newton polygon.  A triangulation is regular if it can be constructed in the following architectural manner. Use the Newton polygon as a floor plan and erect columns of various hights at each integer point of the Newton polygon.  Next, through a canvas over these columns and stretch it down by its perimeter.  As a result you have built a tent over the Newton polygon; the roof of this tent is piecewise linear (with some columns supporting it and, perhaps, some not even reaching the roof) with linear planar pieces joining at edges.  Projection of these edges gives a {\em regular subdivision} of the Newton polygon.  Generically, this subdivision is a triangulation (unless the tops of four of more columns lie in a common plane).  Such triangulations are called {\em regular triangulations}. Every regular subdivision is dual to a skeleton of some amoeba, and vice versa.   Few examples are given in Figure~\ref{AmoebasTriang}.

As we vary the moduli (the internal column heights), holding the parameters (the perimeter column heights) fixed, we can go through a series of triangulations, each describing a region of the moduli space.

\begin{figure}[htbp]
  \centering
   \subfloat[(5,1,4,4,1,0)]
   {\label{Fig:Trianga}\includegraphics[width=0.15\textwidth]{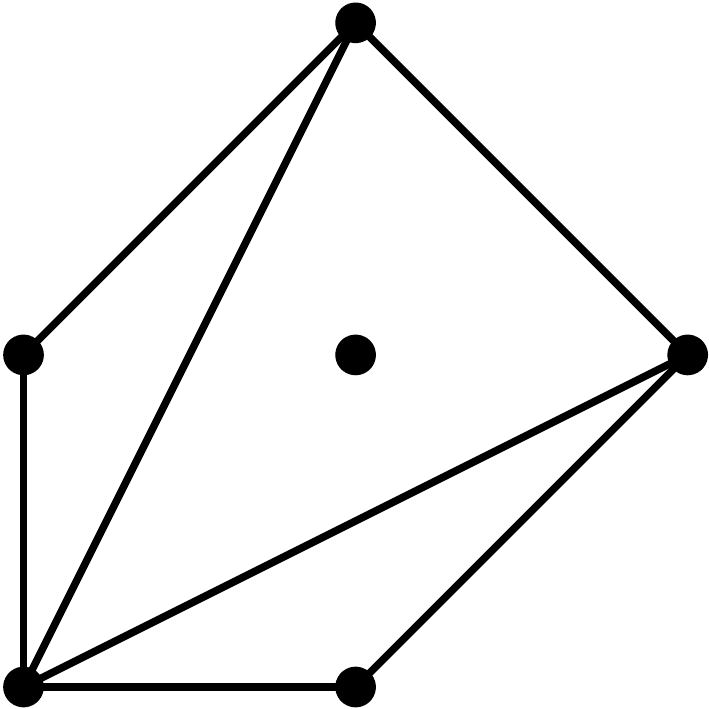}}\qquad
   \subfloat[(3,4,2,5,1,0)]{\label{Fig:Triangb}\includegraphics[width=0.15\textwidth]{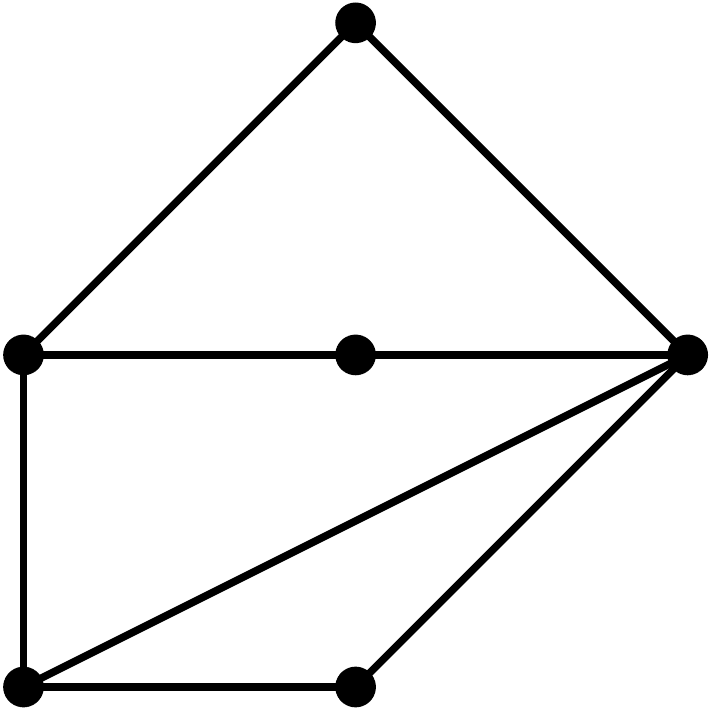}}\qquad
   \subfloat[(1,5,2,4,3,0)]{\label{Fig:Triangc}\includegraphics[width=0.15\textwidth]{GI_l1_Pent_triangc}}\qquad
   \subfloat[(1,3,4,2,5,0)]{\label{Fig:Triangd}\includegraphics[width=0.15\textwidth]{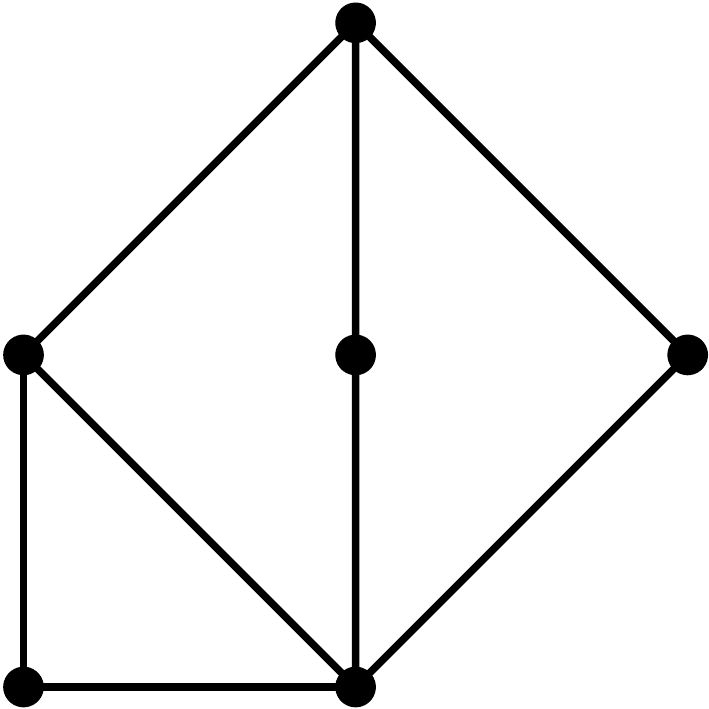}}\qquad
   \subfloat[(3,1,5,2,4,0)]{\label{Fig:Triange}\includegraphics[width=0.15\textwidth]{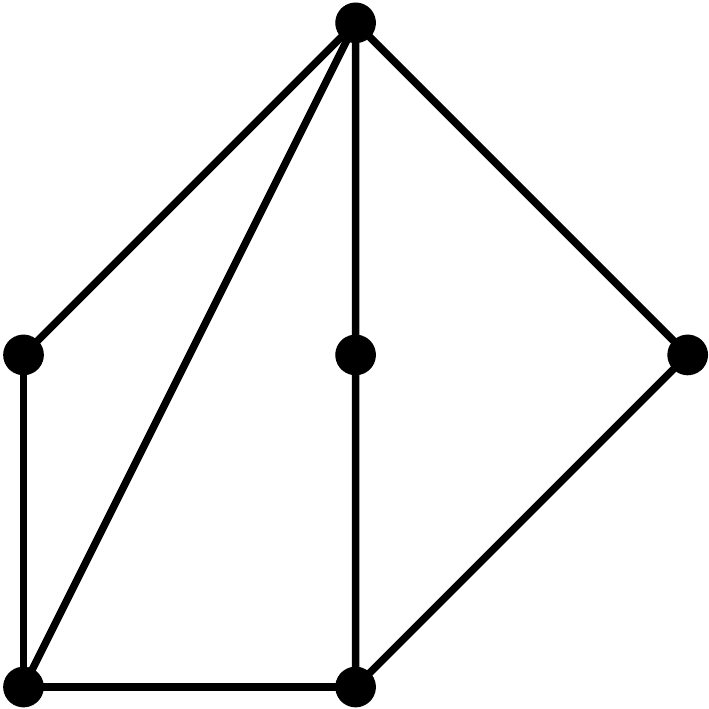}}\\
    \subfloat[(1,3,2,2,3,4)]
   {\label{Fig:Triangf}\includegraphics[width=0.15\textwidth]{GI_l1_Pent_triangf}}\qquad
   \subfloat[(4,1,3,3,1,3)]{\label{Fig:Triangg}\includegraphics[width=0.15\textwidth]{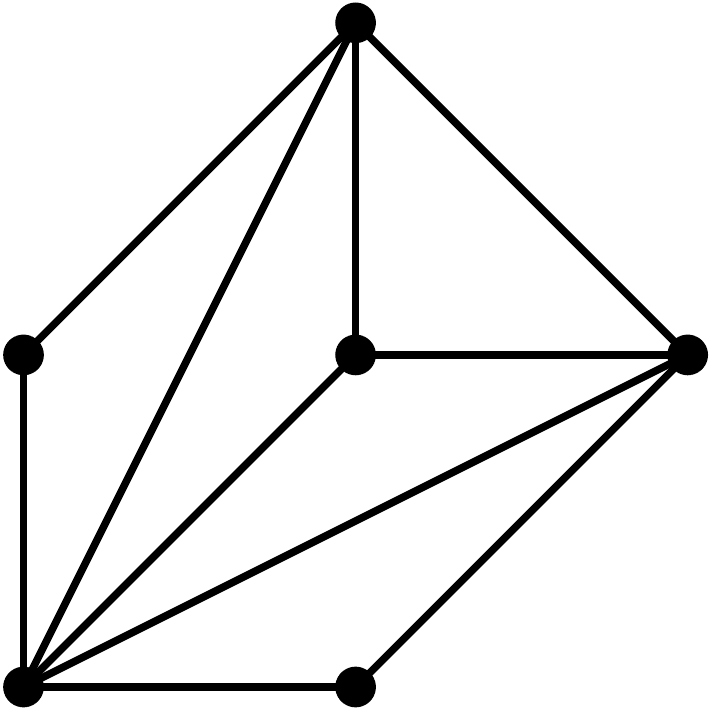}}\qquad
   \subfloat[(3,1,3,2,2,4)]{\label{Fig:Triangh}\includegraphics[width=0.15\textwidth]{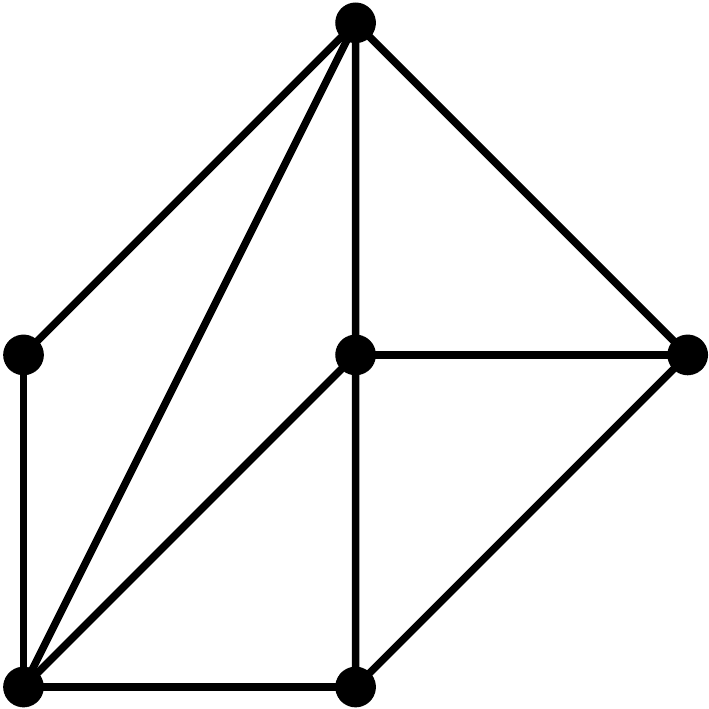}}\qquad
   \subfloat[(2,2,2,2,2,5)]{\label{Fig:Triangi}\includegraphics[width=0.15\textwidth]{GI_l1_Pent_triangi}}\qquad
   \subfloat[(3,2,2,3,1,4)]{\label{Fig:Triangj}\includegraphics[width=0.15\textwidth]{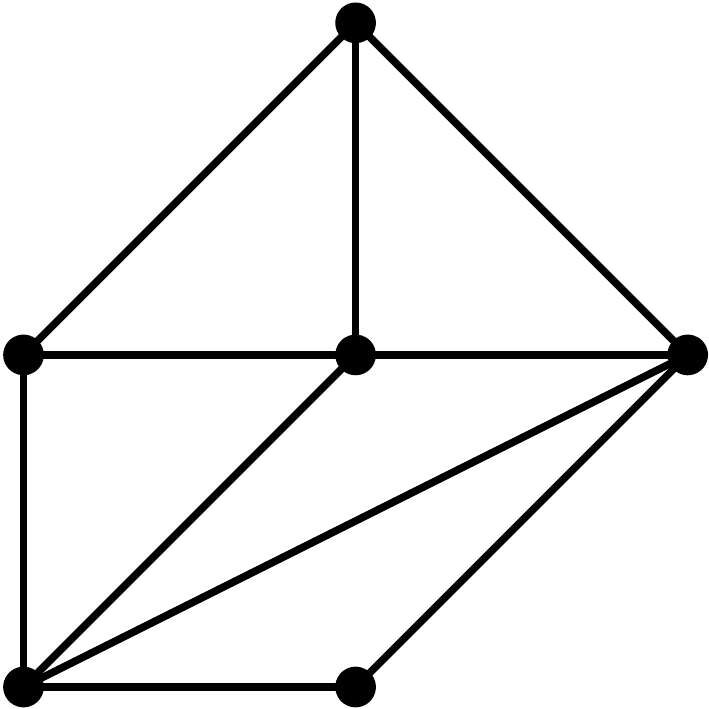}}
 \caption{All regular triangulations of $E_2$ with their five-vector.}
  \label{E2Triangulations}
\end{figure}

For example, the Newton polygon of the $E_2$ monowall admits ten regular triangulations listed in Figure~\ref{E2Triangulations}.  Triangulations listed in the first line are using only the perimeter points as vertices.  Such triangulations are special; we  call them {\em associahedral triangulations}.

\begin{figure}[htbp]
\begin{center}
\includegraphics[width=0.2\textwidth]{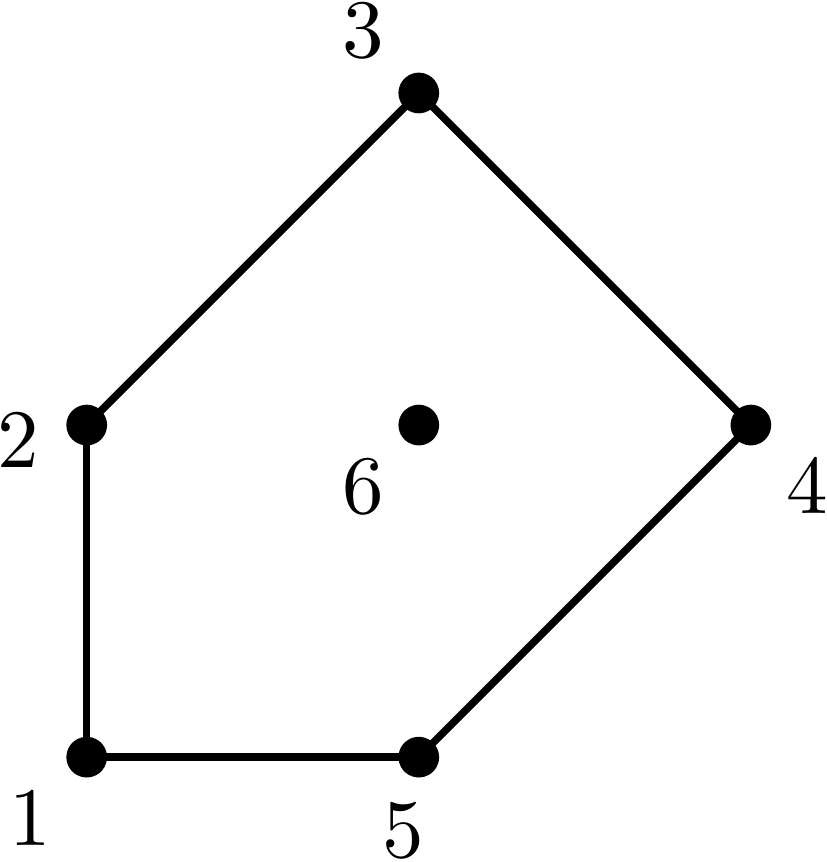}
\caption{Newton polygon of $E_2$ monowall with labelled vertices}
\label{Numbering}
\end{center}
\end{figure}

What is the relation between the (tropical) spectral polynomial and the corresponding triangulation?  The answer to this question is given in \cite{GKZ} in terms of the secondary polyhedron $\Sigma(N)$ that is constructed as follows.  Given a Newton polygon, consider all of its regular triangulations.  Each triangulation $Tr$ determines a vector $\vec{v}_{Tr}$  in a $d$-dimensional space, where $d$ is the number of integer points of the Newton polygon $N.$  Namely, if $e_1,\ldots,e_d$ are the integer points of $N$, then the $i$-th component $v_i$ of the vector $v_{Tr}$  equals to the area covered by all of the triangles of $Tr$ for which  $e_i$ is a vertex.  (For simplicity, to keep all vector coordinates integer, the area of a basic simplex is taken to be 1 instead of $\frac{1}{2}$.) 
\begin{align}
v_i=\sum_{{\Delta\in Tr\atop  v_i\in{\rm Vert} \Delta}}{\rm Area}_{GKZ}(\Delta)
\end{align}
 For example, for the vertex numbering of Figure~\ref{Numbering}, the six-vectors of the triangulation are given under each triangulation in Figure~\ref{E2Triangulations}.

\begin{figure}[htbp]
\begin{center}
  \centering
   \subfloat[Vertex Rising]
   {\label{Fig:Trianga}
\includegraphics[width=0.1\textwidth]{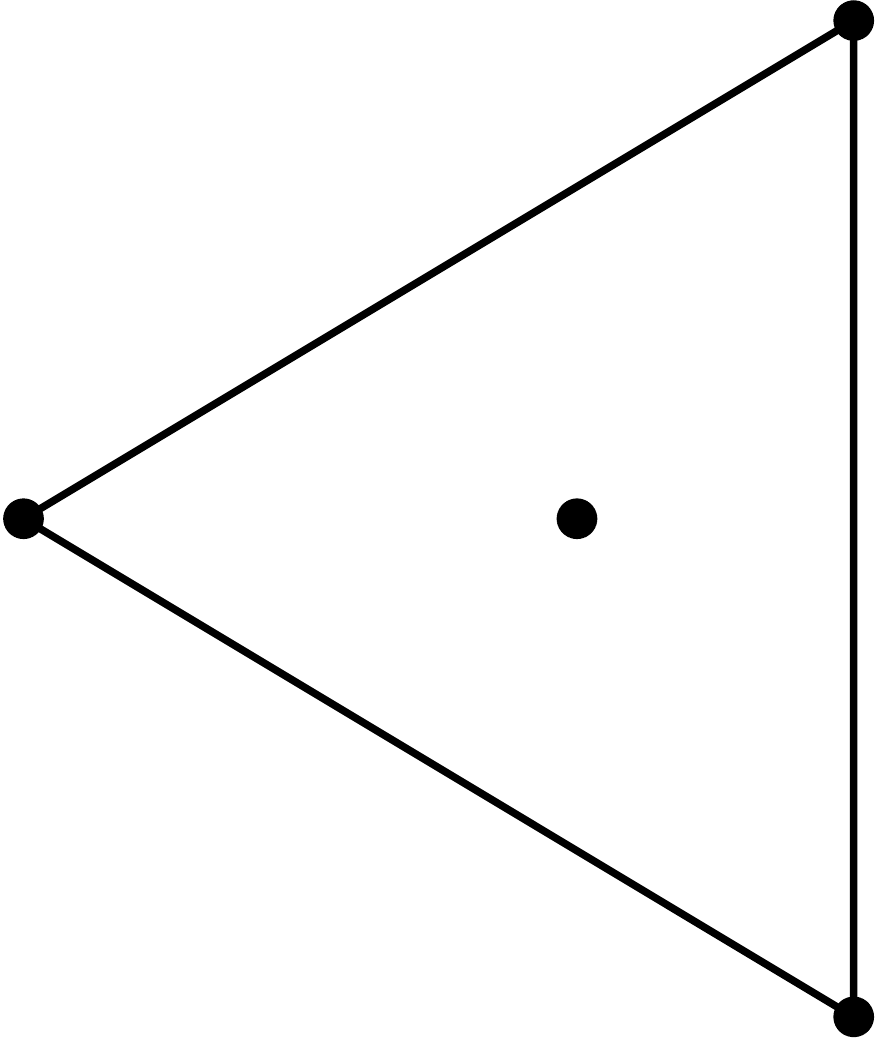}\ 
\includegraphics[width=0.1\textwidth]{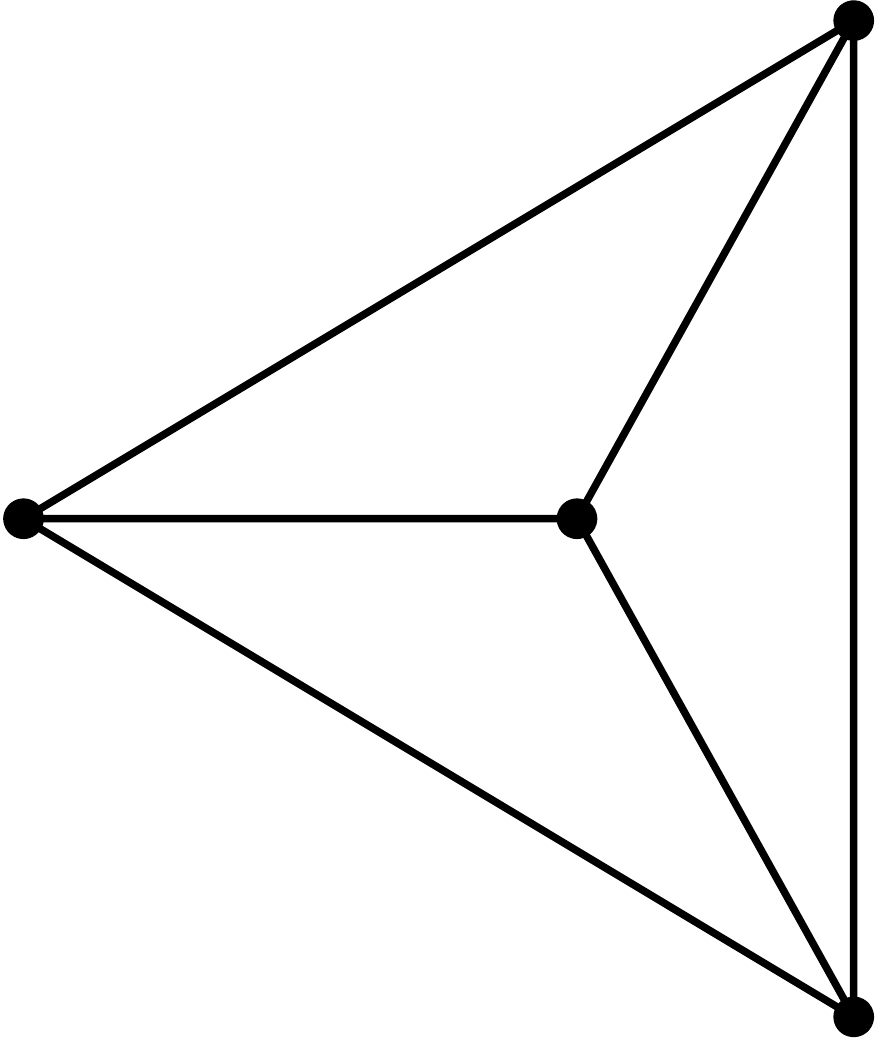}}\qquad
   \subfloat[Flop]
   {\label{Fig:Trianga}
\includegraphics[width=0.1\textwidth]{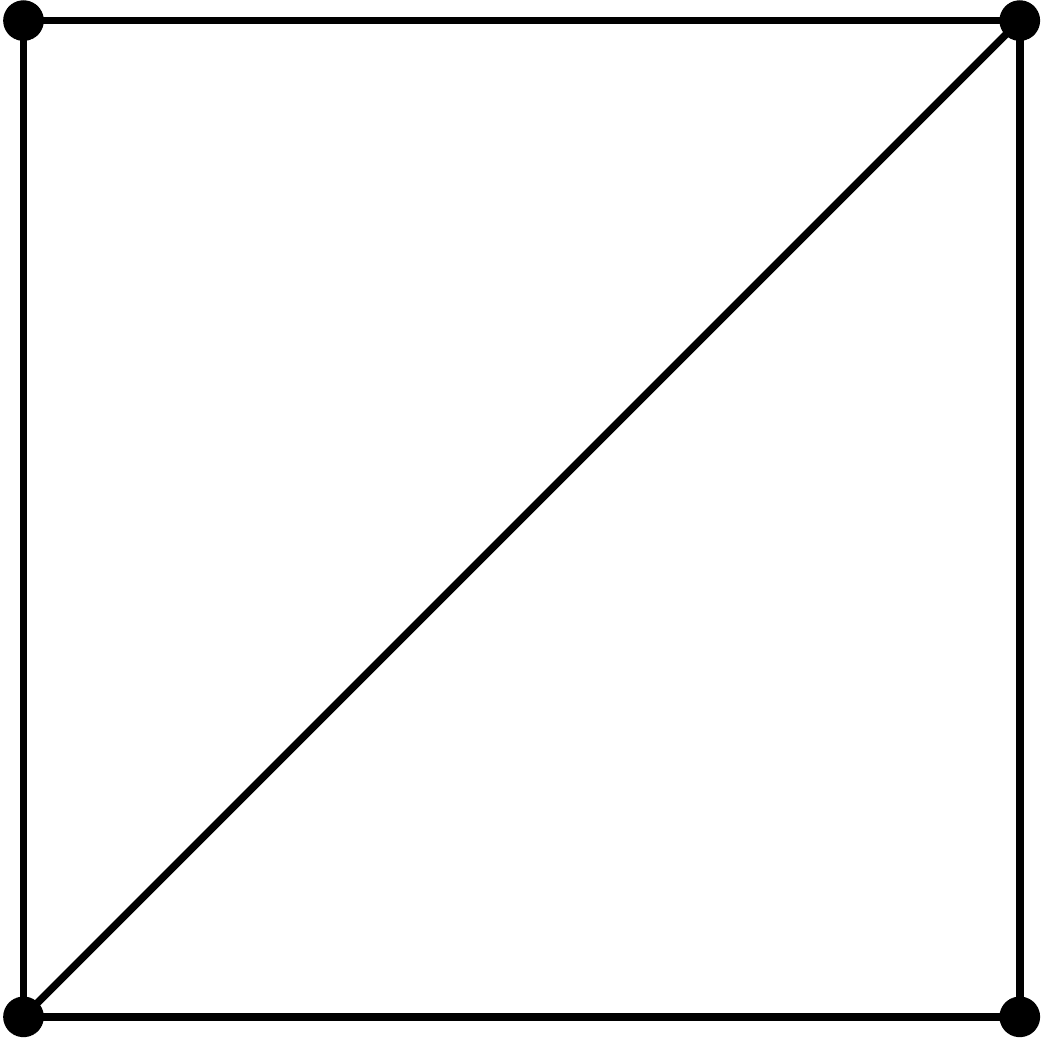} \ 
\includegraphics[width=0.1\textwidth]{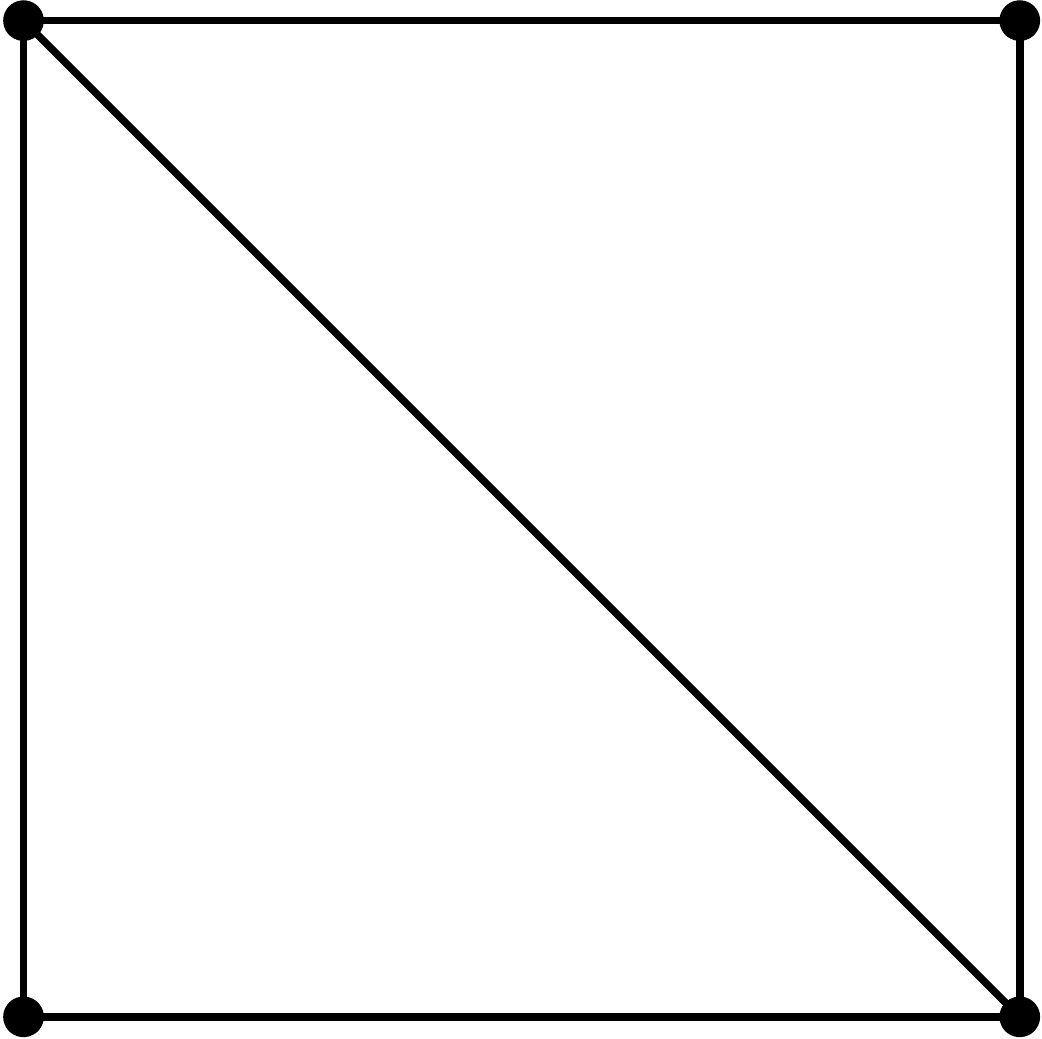}}
\caption{Two possible changes relating neighbouring regular triangulations.}
\label{Flops}
\end{center}
\end{figure}

Clearly, the vectors are constrained, namely
\begin{enumerate}
\item
the sum of vector coordinates is thrice the area of $N:$  $\sum_{i=1}^d v_i=3 {\rm Area}\, N,$ since every triangle area contributes tree times.
\item
the sum of vector components weighted by the corresponding integer point  $\sum_{i=1}^d e_i v_i$ is $\frac{2}{3}{\rm Area}(N)$ times of the center of mass of $N.$
\end{enumerate}
Thus all $\vec{v}_{Tr}$ are lying in a $d-3$ hyperspace.  Moreover, they are vertices of a convex polyhedron $\Sigma(N)$ called the secondary polyhedron of $N.$  The edges of this polyhedron connect triangulations related by two kinds of possible transitions of Figure~\ref{Flops}.

\begin{figure}[htbp]
\begin{center}
\includegraphics[width=\textwidth]{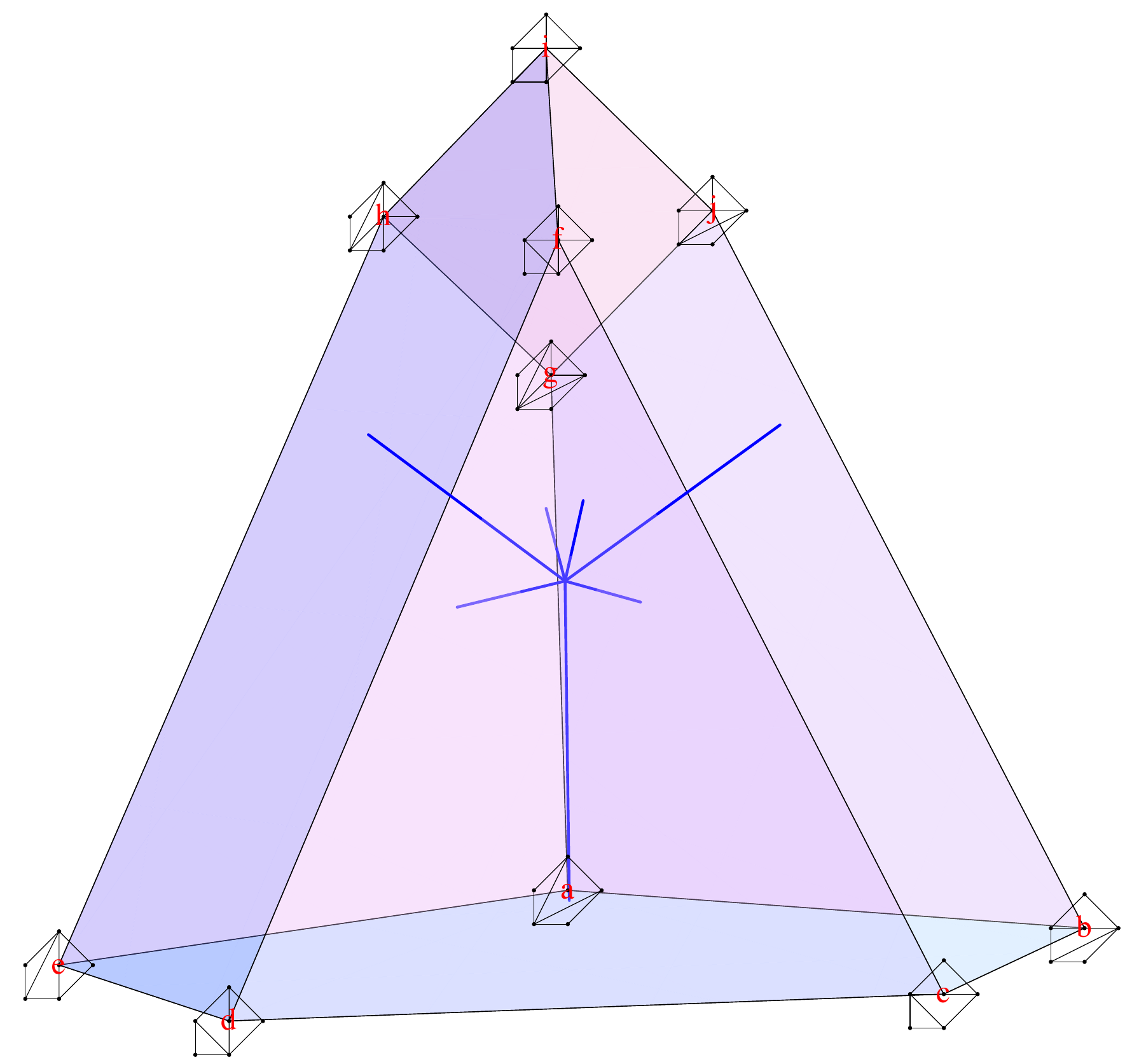}
\caption{The secondary polyhedron (of the monowall with Newton polygon of Figure~\ref{Numbering}) together with the rays of its normal fan.  The vertices correspond to the regular triangulations of Figure~\ref{E2Triangulations}.}
\label{E2Secondary}
\end{center}
\end{figure}

For example, the secondary polyhedron of $E_2$ monowall is shown in Figure~\ref{E2Secondary}.
The biggest face of the secondary polyhedron $\Sigma(N)$ is formed by the vertices corresponding to the triangulations that do not use any of the internal points of $N$ - the associahedral triangulations \cite{BGS}.  In other words this face is the secondary polyhedron of the perimeter of $N$, $\Sigma({\rm Perim}(N)).$  It is called the associahedron of ${\rm Perim}(N)$, thus we call this the {\em associahedral face}.     

Now, consider the normal fan ${\rm Fan}(N)$ of $\Sigma(N)$.  It is called the {\em secondary fan}. Its rays are outward normals to the faces of $\Sigma(N)$ and it divides the space of all coefficients $|G_{m,n}|$ into cones, each cone of maximal dimension is labelled by the corresponding triangulation.  Two  cones share a face is the corresponding vertices in $\Sigma(N)$ are connected by its edge.  All coefficient values forming a vector in a given cone $\sigma_{Tr}$ have the tropical curve dual to the cone's corresponding triangulation $Tr$ of $N.$ 

What is important to us is that we have two kids of coefficients: 1. the perimeter coefficients, that are the parameters, and 2. the internal points coefficients, that are the moduli.  Starting with some point we keep the parameters fixed, while varying the moduli.  Thus, as we change the moduli, we shall be crossing various cones.  Our goal is to identify the phases, i.e. to divide the space of parameters into domains, such that within each domain the sequence of cones we cross as we vary the moduli is the same.  Any two points in parameter space belong to the same phase, only if every point on the interval connecting them has the same sequence of cones crossed when we vary the moduli.

\begin{figure}[htbp]
\begin{center}
\includegraphics[width=0.7\textwidth]{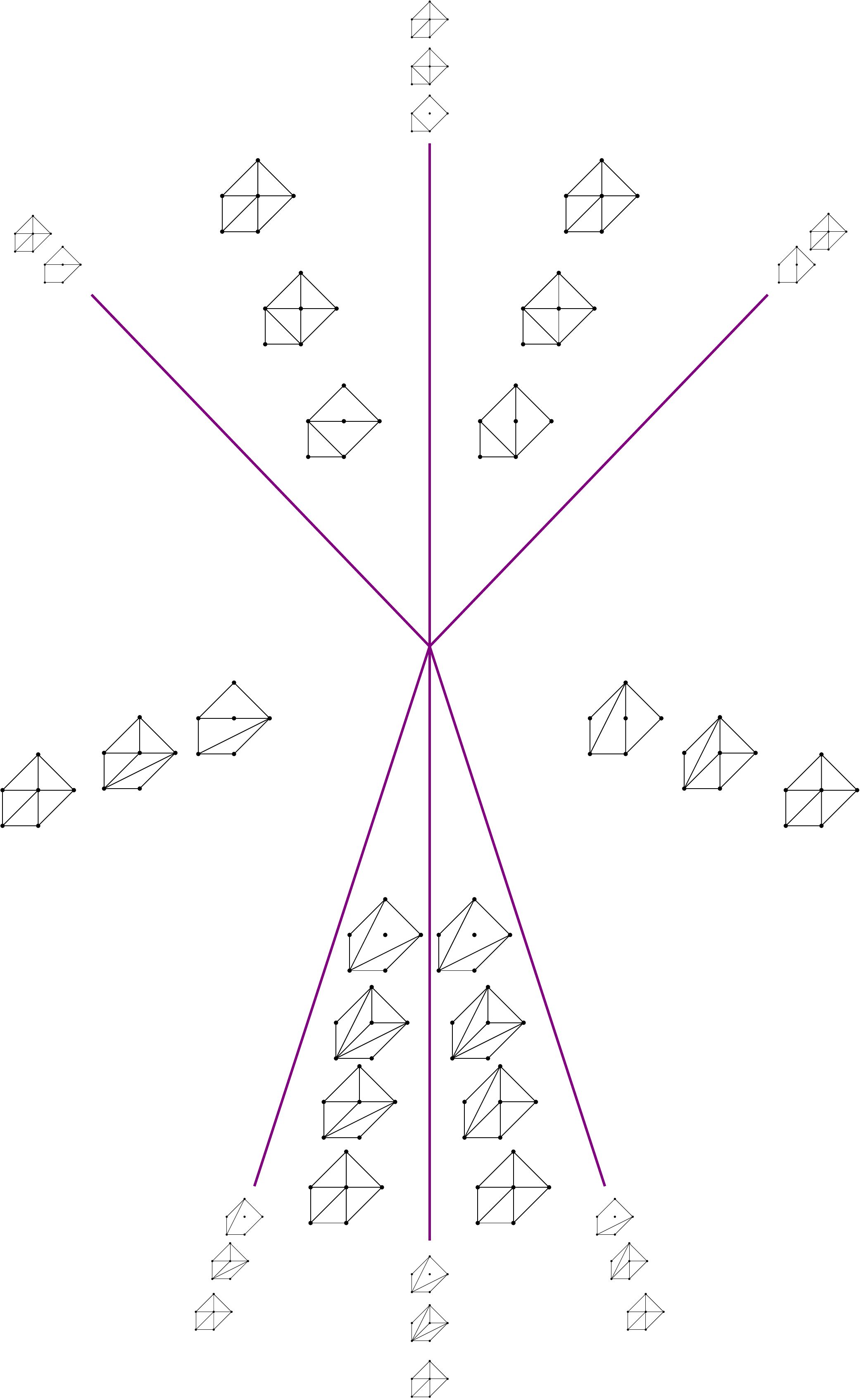}
\caption{This is the phase diagram ${\rm Ph}(N)$ of the $E_2$ monopole wall. The sequence of triangulations  within each cone is to be read outwards, so that it corresponds to increasing the modulus.}
\label{PhaseDiagram}
\end{center}
\end{figure}

From the geometric picture above it is clear that we should project the secondary fan ${\rm Fan}(N)$ onto the associahedral plane along the moduli subspace.  The resulting projection of ${\rm Fan}(N)$ is a new fan ${\rm Ph}(N).$ We call it {\em the phase fan} of $N.$  Each point in a given cone of ${\rm Ph}(N)$ corresponds to the same sequence of triangulations given by the cones of ${\rm Fan}(N)$ above it.  For the $E_2$ example this projection is given by the vitruvian  diagram in Figure~\ref{PhaseDiagram}.

\subsection{Comparison}
\begin{figure}[htbp]
\begin{center}
  \centering
   \subfloat[$D_1$]
   {\label{Fig:D1}
\includegraphics[width=0.1\textwidth]{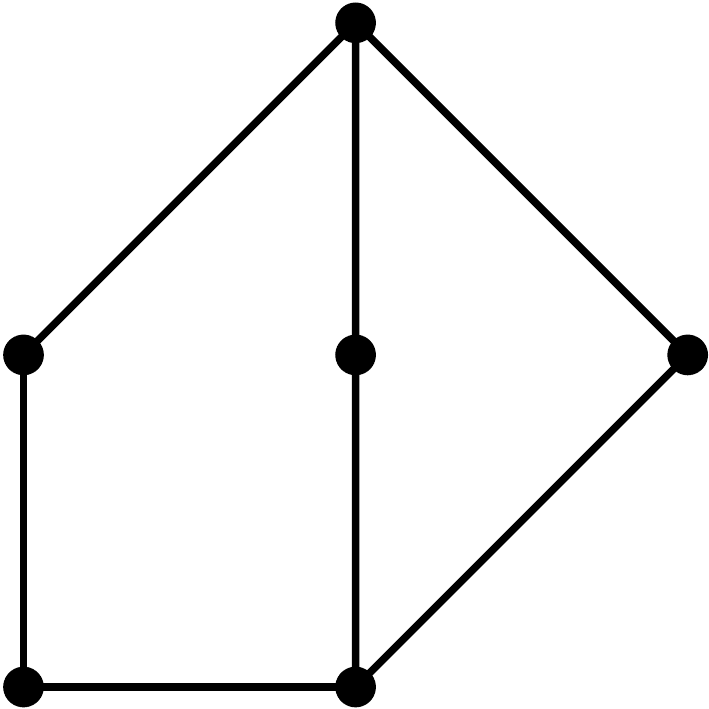}}\qquad
   \subfloat[$D_0$]
   {\label{Fig:D0}
\includegraphics[width=0.1\textwidth]{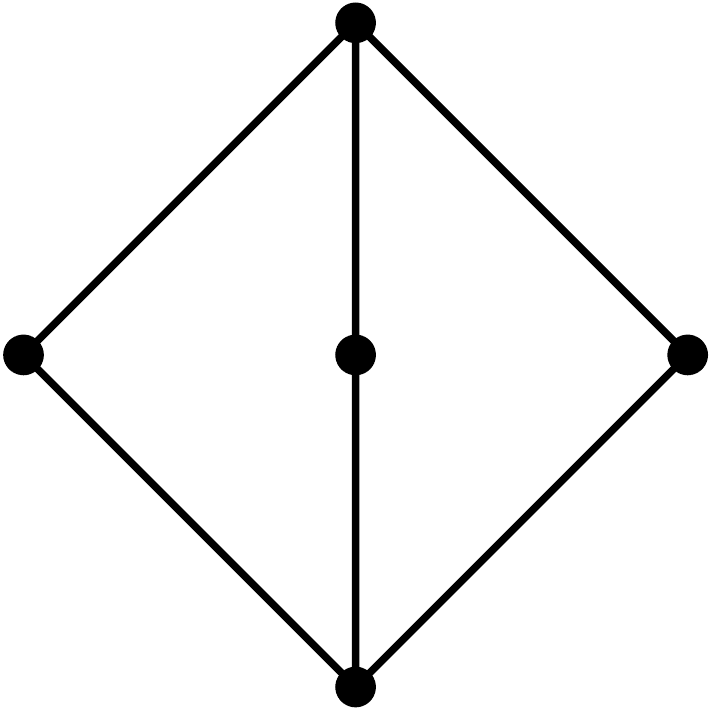}} \qquad 
   \subfloat[$\widetilde{D}_0$]
   {\label{Fig:D0tilde}
\includegraphics[width=0.1\textwidth]{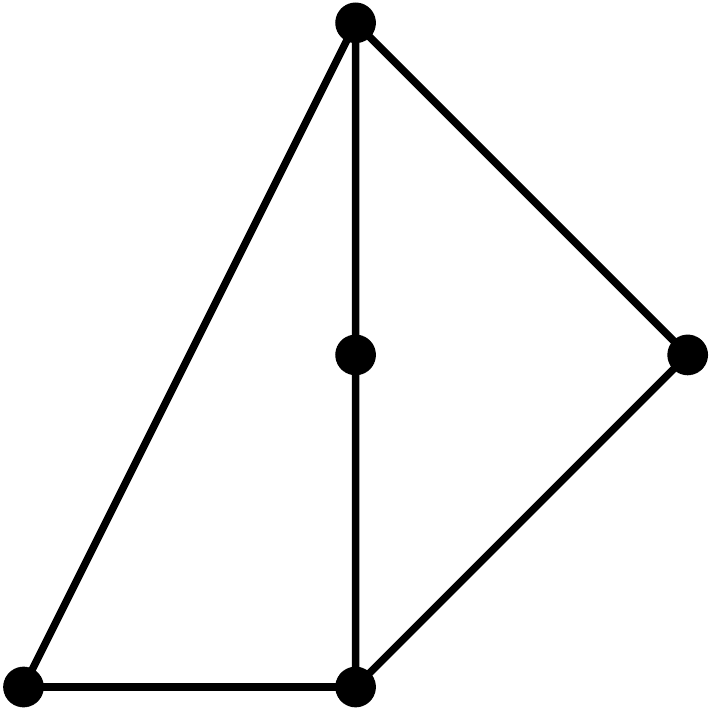}}
\caption{D-type diagrams in the $E_2$ monowall phase space.}
\label{Dtype}
\end{center}
\end{figure}

Let us now turn to the Seiberg and Morrison results of \cite{Morrison:1996xf} comparing the phase structure of the five-dimensional supersymmetric $E_n$ field theories with extreme transitions in Calabi-Yau manifolds.  Our phase diagram of Figure \ref{PhaseDiagram} reproduces the $E_2$ field theory diagram \cite[Fig.1]{Morrison:1996xf}. \footnote{For the exact comparison, we recall that,  instead of $(t_0,m)$ coordinates, \cite[Fig.1]{Morrison:1996xf} is drawn in $(m_0,m)$ coordinates, with $m_0=t_0-2|m|.$}  The upward ray is indeed governed by the $E_1$ Newton sub-polygon of $E_2$, while the direct downward ray by that of $E_0$ sub-polygon. The slanted downward rays correspond to $\tilde{E}_1.$  It is straightforward to identify $D_1$ diagram in Figure~\ref{Fig:D1} 
with $SU(2)$ theory at finite coupling with a massless fundamental multiplet, by considering its dual as a (p,q)-brane network. Sending the mass to infinity amounts to moving  the horizontal ray in the network downwards, removing the left upper point of $D_1$ and leaving the $D_0$ diagram of Figure~\ref{Fig:D0}.  Similarly,  removing that ray upwards, removing the left lower point of $D_1$ and leaves the $\tilde{D}_0$ diagram of Figure~\ref{Fig:D0tilde}.  The asymptotic rays of the (p,q)-network correspond to the same theory as $D_0$ but at a different Chern-Simons level.
Thus ${\rm Ph}(N)$ describing the phase structure of a monowall at the same time describes the phase structure of the five-dimensional supersymmetric field theories.  

What is the relation to the del Pezzo diagram \cite[Fig.2]{Morrison:1996xf}?  The phase structure ${\rm Ph}(N)$ emerged by projecting the secondary fan ${\rm Fan}(N)$ of Figure~\ref{E2Secondary} onto the associahedral plane.  Some of the cones of the secondary fan, namely those that are downwards directed in Figure~\ref{E2Secondary}, correspond to the vertices of the associahedral face.  Call these the {\em associahedral cones} of ${\rm Fan}(N).$  The del Pezzo diagram  \cite[Fig.2]{Morrison:1996xf} is the subfan of ${\rm Fan}(N)$ consisting of all of its non-associahedral cones.  In terms of the secondary polygon, diagram  \cite[Fig.2]{Morrison:1996xf} and its cross-section   \cite[Fig.3]{Morrison:1996xf} are dual to the graph formed by the top (i.e. non-associahedral) skeleton of the secondary polytope.

To summarize, for a monowall with the Newton polygon $N$, the phase structure of its moduli space (in tropical limit) is given by the projection of ${\rm Fan}(N)$ on the associahedral subspace along the moduli subspace (such as e.g. \cite[Fig.1]{Morrison:1996xf}), i.e. along the directions corresponding to the internal points of $N.$  In order to obtain the local Calabi-Yau extremal transitions diagram (such as \cite[Fig.2]{Morrison:1996xf}), restrict ${\rm Fan}(N)$ to a subfan consisting of cones that are not associahedral.


\section{Acknowledgments}
This work was supported in part by a grant from the Simons Foundation (\#245643 to SCh).
The author is grateful to  Oleg Viro for illuminating discussions, to John Schwarz for comments on the manuscript, to the Simons Center for Geometry and Physics  and to the Caltech Particle Theory group for hospitality.  


\end{document}